\documentclass[12pt]{article}
\usepackage{amsmath}
\usepackage{graphicx}
\usepackage{natbib}

\usepackage{url} % not crucial - just used below for the URL

\usepackage[ruled,vlined]{algorithm2e}
\SetAlFnt{\small}
\usepackage{amsfonts}
\usepackage{amsmath}
\usepackage{bbm}
\usepackage{floatrow}
\usepackage{subcaption}
\newcommand{\EE}{\mathbb{E}}
\newcommand{\Cov}{\mbox{Cov}}

\newcommand{\Var}{\mbox{Var}}
\newcommand{\bZ}{\mathbf{Z}}
\newcommand{\bo}{\mathbf{0}}
\newcommand{\bx}{\mathbf{x}}
\newcommand{\by}{\mathbf{y}}
\newcommand{\bz}{\mathbf{z}}

\newtheorem{prop}{Proposition}

\newtheorem{lem}{Lemma}
\newtheorem{cor}{Corollary}

\def\T{ {\mathrm{\scriptscriptstyle T}} }
\def\dif{\mathrm d}
\def\me{\mathrm e}
\def\diag{\mathrm{diag}}
\def\tr{\mathrm{tr}}

\begin{document}

\def\spacingset#1{\renewcommand{\baselinestretch}%
{#1}\small\normalsize} \spacingset{1.5}

%%%%%%%%%%%%%%%%%%%%%%%%%%%%%%%%%%%%%%%%%%%%%%%%%%%%%%%%%%%%%%%%%%%%%%%%%%%%%%

\begin{titlepage}

\begin{center}
{\Large Hamiltonian Assisted Metropolis Sampling}

\vspace{.1in} Zexi Song\footnotemark[1] \& Zhiqiang Tan\footnotemark[1]

\vspace{.1in}
\today
\end{center}

\footnotetext[1]{Department of Statistics, Rutgers University. Address: 110 Frelinghuysen Road,
Piscataway, NJ 08854. E-mails: zexisong@stat.rutgers.edu, ztan@stat.rutgers.edu.}

\paragraph{Abstract.}

Various Markov chain Monte Carlo (MCMC) methods are studied to improve upon random walk Metropolis sampling, for simulation from complex distributions.
Examples include Metropolis-adjusted Langevin algorithms, Hamiltonian Monte Carlo,
and other recent algorithms related to underdamped Langevin dynamics.
We propose a broad class of irreversible sampling algorithms, called Hamiltonian assisted Metropolis sampling
(HAMS), and develop two specific algorithms with appropriate tuning and preconditioning strategies.
Our HAMS algorithms are designed to achieve two distinctive properties, while using an augmented target density with momentum as an auxiliary variable.
One is generalized detailed balance, which induces an irreversible exploration of the target.
The other is a rejection-free property, which allows our algorithms to perform satisfactorily with relatively large step sizes.
Furthermore, we formulate a framework of generalized Metropolis--Hastings sampling, which not only highlights our construction of HAMS at a more
abstract level, but also facilitates possible further development of irreversible MCMC algorithms.
We present several numerical experiments, where the proposed algorithms are found to consistently yield superior results among existing ones.

%\vspace{-.2in}

\end{titlepage}

\newpage
\spacingset{1.5} % DON'T change the spacing!

%%%%%%%%%%%%%%%%%%%%%%%%%%%%%%%%%%%%%%%%%%%%%%%%%%%%%%%%%%%%%%%%%%%%%%%%%%%%%%%%%%%%%%%%%%
%%%---------------------------------------------------------------------------------------
%%%%%%%%%%%%%%%%%%%%%%%%%%%%%%%%%%%%%%%%%%%%%%%%%%%%%%%%%%%%%%%%%%%%%%%%%%%%%%%%%%%%%%%%%%

\section{Introduction}
\label{sec:intro}

In various statistical applications, it is desired to generate observations from a probability density
$\pi(x)$, referred to as the target distribution.
The density function $\pi(x)$ is often defined such that an unnormalized density function $\tilde \pi(x) \propto \pi(x)$ can be readily evaluated,
but the normalizing constant $\int \tilde \pi(x)\,\dif x$ is intractable due to high-dimensional integration.
A prototypical example is posterior sampling for Bayesian analysis, where
the product of the likelihood and prior is an unnormalized posterior density.
For such sampling tasks, a useful methodology is Markov chain Monte Carlo (MCMC),
where a Markov chain is simulated such that the associated stationary distribution coincides with the target $\pi(x)$.
Under ergodic conditions, observations from the Markov chain can be considered an approximate sample from $\pi(x)$.
See for example \cite{Liu2001} and \cite{Brooks2011}.

One of the main workhorses in MCMC is Metropolis--Hastings sampling \citep{Metropolis1953,Hastings1970}.
Given current variable $x_0$, the Metropolis--Hastings  algorithm generates $x^*$ from a proposal density
$x^*\sim Q(x^*| x_0)$, and then accepts $x_1 = x^*$ as the next variable with probability
\begin{equation}
  \rho (x^*|x_0) = \min\left\{1, \frac{\pi(x^*) Q(x_0|x^*) }{\pi(x_0) Q(x^*|x_0)} \right\}, \label{eq:accept-prob}
\end{equation}
or rejects $x^*$ and set $x_1=x_0$, where $\pi(x^*)/\pi(x_0)$ can be evaluated as $\tilde\pi(x^*) / \tilde\pi(x_0)$
without requiring the normalizing constant.
The update from $x_0$ to $x_1$ defines a Markov transition $K(x_1 | x_0)$, depending on both the proposal density and the acceptance-rejection step,
such that reversibility is satisfied: $\pi(x_0) K(x_1 | x_0) = \pi(x_1) K(x_0 | x_1)$. This condition is also
called detailed balance, originally in physics.
As a result, the Markov chain defined by the transition kernel $K$ is reversible and admits $\pi(x)$ as a stationary distribution.

The Metropolis--Hastings algorithm is flexible in allowing various choices of the proposal density $Q$. A simple choice, known as
random walk Metropolis (RWM), is to add a Gaussian noise to $x_0$ for generating $x^*$.
However, RWM may perform poorly for sampling from complex distributions.
To tackle this issue, various MCMC methods are developed by exploiting gradient information in the target density $\pi(x)$.
A common approach is to use discretizations of physics-based continuous dynamics as proposal schemes,
while staying within the framework of Metropolis--Hastings sampling.
One group of algorithms include preconditioned Metropolis-adjusted Langevin algorithm (pMALA) \citep{Besag1994, Roberts1996}
and preconditioned Crank-Nicolson Langevin (pCNL) \citep{Cotter2013},
related to (overdamped) Langevin diffusion.
Another popular algorithm is Hamiltonian Monte Carlo (HMC),
which introduces a momentum variable and uses a leapfrog discretization of the deterministic Hamiltonian dynamics as the proposal scheme
combined with momentum resampling \citep{Duane1987,Neal2011}.
A subtle point is that the momentum can be artificially negated at the end of leapfrog to ensure reversibility.

There are also various MCMC methods, designed by simulating irreversible Markov chains which converge to the target distribution (sometimes with
auxiliary variables). One group of algorithms include guided Monte Carlo (GMC) \citep{Horowitz1991,Ottobre2016} and
the underdamped Langevin sampler (UDL) \citep{Bussi2007}, related to the underdamped Langevin dynamics.
Another group of algorithms includes irreversible MALA \citep{Ma2018} and non-reversible parallel tempering \citep{syed-etal-2019},
related to lifting with a binary auxiliary variable \citep{Gustafson1998,Vucelja2016}.
A third group of algorithms include the bouncy particle \citep{Bouchard-Cote2018} and Zig-Zag samplers \citep{Bierkens2019}, using Poisson jump processes.

The contribution of this article can be summarized as follows. First, we propose a broad class of irreversible sampling algorithms, called Hamiltonian assisted Metropolis sampling (HAMS),
and develop two specific algorithms, HAMS-A/B, with appropriate tuning and preconditioning strategies.
Our HAMS algorithms use an augmented target density (corresponding to a Hamiltonian) with momentum as an auxiliary variable.
Each iteration of HAMS consists of
a proposal step depending on the gradient of the Hamiltonian, and an acceptance-rejection step using an acceptance probability different from the usual formula (\ref{eq:accept-prob}).
%and successively update the momentum variable together with the original (location) variable.
The two steps are designed to achieve generalized detailed balance and a rejection-free property discussed below.
Second, we formulate a framework of generalized Metropolis--Hastings sampling, which not only highlights our construction of HAMS as a special case,
but also facilitates possible further development of irreversible MCMC algorithms.
Third, we present several numerical experiments, where the proposed algorithms are found to consistently yield superior results among existing ones.

Compared with existing algorithms, there are two important properties which are {\it simultaneously} satisfied by our HAMS algorithms.
The first is generalized detailed balance (or generalized reversibility), where the backward transition is related to the forward transition after negating the momentum.
This condition is known in the study of continuous dynamics in physics \citep{Gardiner1997},
but seems to receive insufficient treatment in the MCMC literature, where the acceptance-rejection step is also crucial for proper sampling from a target distribution.
By generalized detailed balance, the momentum can be accepted without sign negation, which induces an irreversible exploration of the target.
Second, our algorithms satisfy a rejection-free property, that is, the proposal is always accepted at each iteration, in the case where
the target distribution is standard normal. By preconditioning, the rejection-free property can also be achieved when the target distribution
is normal with a pre-specified variance $\Sigma$.
A similar motivation can be found in the construction of pCNL algorithm \citep{Cotter2013}.
From our experiments, this property allows our algorithms to perform satisfactorily with relatively large step sizes.

\vspace{.1in}\textbf{Notation.}
Assume that a target density $\pi(x)$ is defined on $\mathbb R^k$. The potential energy function $U(x)$ is defined such that $\pi(x) \propto \exp\{-U(x)\}$ as in physics.
Denote the gradient of $U(x)$ as $\nabla U(x)$. The (multivariate) normal distribution
with mean $\mu$ and variance $V$ is denoted as $\mathcal N(\mu,V)$, and the density function as $\mathcal N(\cdot |\mu,V)$.
Whenever possible, we treat a probability distribution and its density function interchangeably.
Write $\bo$ for a vector or matrix with all $0$ entries,
and $I$ for an identity matrix of appropriate dimensions.

%%%%%%%%%%%%%%%%%%%%%%%%%%%%%%%%%%%%%%%%%%%%%%%%%%%%%%%%%%%%%%%%%%%%%%%%%%%%%%%%%%%%%%%%%%
%%%---------------------------------------------------------------------------------------
%%%%%%%%%%%%%%%%%%%%%%%%%%%%%%%%%%%%%%%%%%%%%%%%%%%%%%%%%%%%%%%%%%%%%%%%%%%%%%%%%%%%%%%%%%

\section{Related methods}
\label{sec:review}

We describe several MCMC algorithms, related to our work, for sampling from a target distribution $\pi(x)$.
Throughout, we write the current variable as $x_0$, a proposal as $x^*$, and the next variable as $x_1$ after the acceptance-rejection step.
Denote as $\Sigma$ a constant variance matrix used as an approximation to the variance of the target $\pi(x)$.

Random walk Metropolis sampling generates a proposal $x^*$ by directly adding a Gaussian noise to $x_0$ and then performs
acceptance or rejection.

\vspace{.05in}\textit{Random walk Metropolis sampling} (RWM). \vspace{-.1in}
\begin{itemize}\addtolength{\itemsep}{-.1in}
  \item Generate $x^* = x_0 + \epsilon Z$, where $Z \sim \mathcal N(\bo, \Sigma)$ and $\epsilon>0$ is a tunable step size.
  \item Set $x_1 = x^*$ with acceptance probability $\rho(x^* |x_0) = \min(1,\pi(x^*)/\pi(x_0))$ by (\ref{eq:accept-prob}), \\
  or set $x_1=x_0$ with the remaining probability.
\end{itemize}
RWM does not exploit gradient information, and may be slow in exploring the target $\pi(x)$. On the other hand,
RWM is operationally low-cost, without gradient evaluation.

The preconditioned Metropolis-adjusted Langevin algorithm (pMALA) generates a proposal $x^*$ by moving along the gradient from current $x_0$ \citep{Roberts1996}.
Hence pMALA is more directed and encourages exploration to high density regions.

\vspace{.05in}\textit{Preconditioned Metropolis-adjusted Langevin algorithm} (pMALA). \vspace{-.1in}
\begin{itemize}\addtolength{\itemsep}{-.1in}
  \item Generate $x^* = x_0 - \frac{\epsilon^2}{2}\Sigma \nabla U(x_0)  + \epsilon Z$, where $Z \sim \mathcal N(\bo, \Sigma)$ and $\epsilon>0$ is a step size.
  \item Set $x_1 = x^*$ with probability (\ref{eq:accept-prob}), where $Q(x^*|x_0) = \mathcal N(x^*|x_0 - \frac{\epsilon^2}{2}\Sigma \nabla U(x_0), $ $\epsilon^2\Sigma)$, \\
        or set $x_1 = x_0$ with the remaining probability.
\end{itemize}

The preconditioned Crank-Nicolson Langevin (pCNL) algorithm is originally designed for posterior sampling with a latent Gaussian field model \citep{Cotter2013}.
The target density is $\pi(x)\propto \exp\{-U(x)\} \propto \exp\{\ell(x)\}\mathcal N(x|\bo,C)$, a product of a likelihood function and a normal prior with variance $C$.
For easy comparison, we use a parameterization in terms of the step size $\epsilon$ and the potential gradient, $\nabla U(x) = - \nabla \ell(x) + C^{-1} x$.

\vspace{.05in}\textit{Preconditioned Crank-Nicolson Langevin} (pCNL). \vspace{-.1in}
\begin{itemize}\addtolength{\itemsep}{-.1in}
  \item Sample $Z \sim \mathcal N(\bo, C)$ and compute
  \begin{align}
  x^* & = \sqrt{1-\epsilon^2} x_0 +\frac{\epsilon^2}{1+\sqrt{1-\epsilon^2}} C\nabla \ell(x_0) + \epsilon Z \nonumber \\
  & = x_0 - \frac{\epsilon^2}{1+\sqrt{1-\epsilon^2}} C\nabla U(x_0) + \epsilon Z . \label{eq:pCNL}
  \end{align}
  \item Set $x_1 = x^*$ with probability (\ref{eq:accept-prob}), where $Q(x^*|x_0) = \mathcal N(x^*|x_0 - \frac{\epsilon^2}{1+\sqrt{1-\epsilon^2}}C \nabla U(x_0),\epsilon^2 C)$, \\
        or set $x_1 = x_0$ with the remaining probability.
\end{itemize}

It is interesting to compare pMALA and pCNL. On one hand, pCNL is close to pMALA with the preconditioning matrix $\Sigma$ chosen to be $C$, as the step size $\epsilon \to 0$
and hence $\frac{\epsilon^2}{1+\sqrt{1-\epsilon^2}} \to \frac{\epsilon^2}{2}$ in (\ref{eq:pCNL}).
On the other hand, as $\epsilon$ stays away from 0,
the coefficient $\frac{\epsilon^2}{1+\sqrt{1-\epsilon^2}}$ associated with the potential gradient in pCNL can differ considerably from $\frac{\epsilon^2}{2}$ in pMALA.
As discussed in \cite{Cotter2013}, a simple advantage of this difference is that when the likelihood gradient $\nabla \ell$ is dropped, the resulting proposal from (\ref{eq:pCNL})
becomes  $x^* = \sqrt{1-\epsilon^2} x_0 +\epsilon Z$, which is invariant and reversible with respect to the prior $\mathcal N(\bo,C)$.
In this case, the proposal $x^*$ is accepted with probability 1 in pCNL, but not in pMALA.
To achieve such a rejection-free property also plays an important role in our work.

From the preceding discussion, it seems straightforward to define a modified pMALA algorithm, by replacing the update coefficient $\frac{\epsilon^2}{2}$ with
 $\frac{\epsilon^2}{1+\sqrt{1-\epsilon^2}}$ in pMALA. Equivalently, this algorithm can also be obtained from pCNL, by replacing the prior variance $C$
 by a general preconditioning matrix $\Sigma$, which can be specified as an approximation to the variance of the target distribution $\pi(x)$,
 instead of being fixed as the prior variance $C$.
 As a result, the modified pMALA algorithm is rejection-free (i.e., the proposal $x^*$ is always accepted) when the target density is $\mathcal N(\bo, \Sigma)$.
 To our knowledge, such an extension of pMALA and pCNL appears not explicitly studied before.
In Section \ref{subsec:special}, we obtain the modified pMALA algorithm as a boundary case of the proposed HAMS algorithms.

\vspace{.05in}\textit{Modified preconditioned Metropolis-adjusted Langevin algorithm} (pMALA*). \vspace{-.1in}
\begin{itemize} \addtolength{\itemsep}{-.1in}
  \item Generate $x^* = x_0 - \frac{\epsilon^2}{1+\sqrt{1-\epsilon^2}}\Sigma \nabla U(x_0)  + \epsilon Z$, where $Z \sim \mathcal N(\bo, \Sigma)$.
  \item Set $x_1 = x^*$ with probability (\ref{eq:accept-prob}), where $Q(x^*|x_0) = \mathcal N(x^*|x_0 - \frac{\epsilon^2}{1+\sqrt{1-\epsilon^2}}\Sigma \nabla U(x_0),\epsilon^2\Sigma)$, \\
        or set $x_1 = x_0$ with the remaining probability.
\end{itemize}

We also point out that modified pMALA is distinct from a related gradient-based algorithm in \cite{Titsias2018}, which is proposed in the context of posterior sampling
with the target density $\pi(x)\propto \exp\{-U(x)\} \propto \exp\{\ell(x)\}\mathcal N(x|\bo,C)$. The associated proposal scheme (without preconditioning) can be written as
\begin{align}
x^*  = \frac{2}{\delta}\tilde C x_0 +\tilde C \nabla \ell (x_0)  + Z = x_0 -\tilde C \nabla U(x_0)  + Z, \quad Z \sim \mathcal N(\bo,  \frac{2}{\delta} \tilde C^2 + \tilde C), \label{eq:proposal-TP}
\end{align}
where $\tilde C  = (\frac{2}{\delta}I +C^{-1})^{-1}$ and $\nabla U(x_0) = - \nabla \ell(x_0) + C^{-1} x_0$.
When the prior variance $C$ is an identity matrix (i.e., $C=I$), the proposal scheme (\ref{eq:proposal-TP}) reduces to
\begin{align*}
x^* = x_0 - \frac{\delta}{2+\delta} \nabla U(x_0) + Z, \quad Z \sim \mathcal N(\bo, \frac{\delta(\delta+4)}{(\delta+2)^2} I),
\end{align*}
which is equivalent to the proposal scheme in pCNL and in modified pMALA with $\Sigma=I$, after matching $\epsilon^2= \frac{\delta(\delta+4)}{(\delta+2)^2}$.
However, except for this coincidence, the algorithm of \cite{Titsias2018} based on (\ref{eq:proposal-TP}) as well as its preconditioned version
in general differ from modified pMALA above.
In fact, modified pMALA can also be derived using auxiliary variables, but invoking a different Taylor expansion
to approximate the target density from \cite{Titsias2018}. See the Supplement Section \ref{sec:auxiliary} for further discussion on auxiliary variables and second-order schemes.

The following methods require augmenting the sample space to include a momentum variable $u\in\mathbb R^k$,
which is assumed to be normally distributed, $u\sim\mathcal N(\bo,M)$. The variance $M$ is also called
a mass matrix, and the quantity $u^\T M^{-1}u/2$
represents the kinetic energy in physics. The joint target density of $(x,u)$ becomes
\begin{align}
\pi(x,u)\propto \exp\{-H(x,u)\} = \exp\{-U(x) - \frac{1}{2} u^\T M^{-1}u\}, \label{eq:aug-density}
\end{align}
where $H(x,u) = U(x) + \frac{1}{2}u^\T M^{-1} u$, called a total energy or Hamiltonian.
For sampling from an augmented target distribution $\pi(x,u)$, Hamiltonian  Monte Carlo
generates a proposal by first redrawing a momentum variable and then performing a series of deterministic updates,
based on molecular dynamics (MD) simulations such that the Hamiltonian $H(x,u)$ is approximately preserved \citep{Duane1987,Neal2011}.

\vspace{.05in} \textit{Hamiltonian Monte Carlo} (HMC). \vspace{-.1in}
\begin{itemize}\addtolength{\itemsep}{-.1in}
  \item Sample $u^*\sim \mathcal N(\bo,M)$, reset $u_0=u^*$, and set $x^* = x_0$.
  \item For $i$ from 1 to $nleap$, repeat:\\
            $u^* \gets u^* - \frac{\epsilon}{2}\nabla U(x^*)$, \quad
            $x^* \gets x^* + \epsilon M^{-1} u^*$, \quad
            $u^* \gets u^* - \frac{\epsilon}{2}\nabla U(x^*)$.
  \item Set $(x_1,u_1) = (x^*, u^*)$ with probability $\min(1,\exp(H(x_0,u_0) - H(x^*,u^*)))$\\
        or set $(x_1,u_1) = (x_0, -u_0)$ with the remaining probability.
\end{itemize}

The steps within the for loop are called leapfrog updates, which provide an accurate discretization of
the Hamiltonian dynamics, defined as a system of differential equations by Newton's laws of motion
such that the Hamiltonian $H(x,u)$ is preserved over time.
Although the update of $u$ can be ignored, the acceptance-rejection step above is stated such that the update of $(x,u)$ matches UDL and GMC later with $c=1$,
if momentum were not resampled.
For HMC, both the step size $\epsilon$ and the number of leapfrog steps $nleap$ need to be tuned.
For automated tuning, it seems popular to use the No-U-Turn Sampler (NUTS) proposed by \cite{Gelman2014}.
Nevertheless, HMC often requires a large number of leapfrog steps which is
computationally costly.

An important extension of the Hamiltonian dynamics is Langevin dynamics, which can be defined as a system of stochastic differential equations,
\begin{align}
\dif x_t = u_t \,\dif t, \quad \dif u_t = - \eta \,\dif x_t - \nabla U(x_t) \,\dif t + \sqrt{2\eta} \,\dif W_t, \label{eq:langevin}
\end{align}
where $\eta > 0$ is a friction coefficient and $W_t$ is the standard Brownian process. In the case of $\eta \to 0$, the Langevin dynamics
reduces to the deterministic Hamiltonian dynamics, $\dif x_t = u_t \,\dif t$ and $\dif u_t =- \nabla U(x_t) \,\dif t$.
In the high-friction limit (i.e., large $\eta$), the overdamped Langevin diffusion process is obtained:
$  \dif x_t  = -\eta^{-1} \nabla U(x_t) \,\dif t + \sqrt{2\eta^{-1}} \,\dif W_t$.
Hence (\ref{eq:langevin}) is also called underdamped Langevin dynamics.
Although Langevin dynamics has long been used in molecular simulations \citep[e.g.,][]{Berendsen1982}, there is extensive and growing research
related to Langevin dynamics in physics and chemistry \citep[e.g.,][]{Horowitz1991,Scemama2006,Bussi2007,Goga2012,Farago2013,Farago2020}
and machine learning and statistics \citep[e.g.,][]{Ottobre2016,Cheng2018,Dalalyan2018}.
In particular, the Metropolized version of the algorithm in \cite{Bussi2007} can be described as follows, to accommodate an acceptance-rejection step.

\vspace{.05in}\textit{Underdamped Langevin sampling} (UDL). \vspace{-.1in}
\begin{itemize}\addtolength{\itemsep}{-.1in}
  \item Sample $Z_1,Z_2 \sim \mathcal N(\bo,M)$ independently, and compute
  \begin{itemize}
  \item[] $u^+ = \sqrt{c}u_0 + \sqrt{1-c}Z_1$,
  \item[] $\tilde u = u^+ - \frac{\epsilon}{2}\nabla U(x_0), \quad x^* = x_0 + \epsilon M^{-1} \tilde u, \quad u^- = \tilde u - \frac{\epsilon}{2}\nabla U(x^*)$,
  \item[] $u^* = \sqrt{c}u^- + \sqrt{1-c}Z_2$,
  \end{itemize}
  where $0\le c \le 1$ is a tuning parameter and can be interpreted as $c = \me^{-\eta \epsilon/2}$.
  \item Set $(x_1,u_1) = (x^*, u^*)$ with probability $\min(1,\exp(H(x_0,u^+) - H(x^*,u^-)))$\\
   or set $(x_1,u_1) = (x_0,-u_0)$ with the remaining probability.
\end{itemize}

There are several interesting features in UDL.
First, the proposal scheme in UDL contains a (deterministic) leapfrog update, which is sandwiched by two random updates of the momentum.
Notably, the current momentum $u_0$ is partially refreshed at the beginning, where the amount of ``carryover" is controlled by the parameter $c$.
At the two extremes, $c=0$ or 1, UDL recovers pMALA or Metropolized leapfrog respectively.
When $c=0$, the first updated momentum $u^+=Z_1$ is independent of $u_0$ and the final updated momentum $u^*=Z_2$ can be ignored.
In this case, UDL reduces to HMC with one leapfrog step (after redrawing the momentum) and hence is equivalent to pMALA as discussed in \cite{Neal2011}.
When $c=1$, UDL generates a proposal by one leapfrog update and then accept or reject (with $u_0$ flipped)
based on the change in the Hamiltonian.

Second, the proposal scheme in UDL is derived in \cite{Bussi2007} by a particular choice of operator splitting in discretizing the Langevin dynamics (\ref{eq:langevin}).
Compared with other possible choices, the UDL proposal scheme is shown to satisfy a generalized formulation of detailed balance. However, as discussed later in Section~\ref{sec:extension},
whether a sampling algorithm leaves a target distribution invariant depends also on how acceptance or rejection is executed.
While \cite{Bussi2007} only mentioned that acceptance-rejection can be performed similarly as in \cite{Scemama2006},
the acceptance-rejection step above is explicitly added by our understanding. In the Appendix, we verify the validity of the UDL algorithm
in leaving the target augmented density $\pi(x,u)$ invariant, using our proposed framework of generalized Metropolis--Hastings sampling.

Third, both the two momentum updates are in the form of an order-1 autoregressive process, which leaves the momentum distribution invariant:
if $u_0 \sim \mathcal N(\bo,M)$ then $u^+ \sim \mathcal N(\bo,M)$ and, similarly, if $u^- \sim \mathcal N(\bo,M)$ then $u^* \sim \mathcal N(\bo,M)$.
As discussed in \cite{Bussi2007}, such updates using two independent noise vectors are exploited to achieve generalized detailed balance.
In fact, it is instructive to compare UDL with a related algorithm in \cite{Horowitz1991}, which uses only one noise vector per iteration as described below.
To our knowledge, it seems difficult to show that generalized detailed balance is satisfied by this algorithm, although
invariance with respect to $\pi(x,u)$ is valid because each iteration is a composition of two steps,
first $(x_0, u_0) \to (x_0, u^+)$
and then $(x_0, u^+) \to (x_1,u_1)$ by Metropolized leapfrog, and each step leaves the target $\pi(x,u)$ invariant.

\vspace{.05in}\textit{Guided Monte Carlo} (GMC). \vspace{-.1in}
\begin{itemize}\addtolength{\itemsep}{-.1in}
  \item Sample $Z_1 \sim \mathcal N(\bo,M)$, and compute
  \begin{itemize}
  \item[] $u^+ = \sqrt{c}u_0 + \sqrt{1-c}Z_1$,
  \item[] $\tilde u = u^+ - \frac{\epsilon}{2}\nabla U(x_0), \quad x^* = x_0 + \epsilon M^{-1} \tilde u, \quad u^- = \tilde u - \frac{\epsilon}{2}\nabla U(x^*)$.
  \end{itemize}
  \item Set $(x_1,u_1) = (x^*, u^-)$ with probability $\min(1,\exp(H(x_0,u^+) - H(x^*,u^-)))$\\
   or set $(x_1,u_1) = (x_0,-u^+)$ with the remaining probability.
\end{itemize}

Another interesting method is the irreversible MALA algorithm in \cite{Ma2018}.
Compared with our method using an augmented density with momentum as an auxiliary variable,
this method relies on a binary auxiliary variable to facilitate irreversible sampling,
while using discretizations of continuous dynamics in the original variable $x$ as proposal schemes.
See Section~\ref{sec:extension} and Supplement Section~\ref{sec:gMH} for further discussion.

% Lifting—A nonreversible Markov chain Monte Carlo algorithm, Marija Vucelja, American Journal of Physics 84, 958 (2016)

%%%%%%%%%%%%%%%%%%%%%%%%%%%%%%%%%%%%%%%%%%%%%%%%%%%%%%%%%%%%%%%%%%%%%%%%%%%%%%%%%%%%%%%%%%

\section{Proposed Methods}
\label{sec:proposed}

We develop our methods in several steps. We first construct proposal schemes using gradient information, then introduce modifications to derive a class of generalized reversible algorithms HAMS,
and finally study two specific algorithms, HAMS-A/B, and propose tuning and preconditioning strategies.
To focus on main ideas, consider the augmented target density (\ref{eq:aug-density}) with momentum variance $M=I$, that is,
\begin{align}
\pi(x,u) \propto  \exp(-H(x,u))= \exp(-U(x) - u^\T u/2) , \label{eq:aug-density2}
\end{align}
until Section~\ref{subsec:precondition} to discuss preconditioning.
The proposed algorithms are then placed in a more abstract framework of generalized Metropolis--Hastings sampling in Section~\ref{sec:extension}.

%----------------------------------------------------------------------------------
\subsection{Construction of Hamiltonian proposals}
\label{subsec:aux}

We provide a simple, broad class of proposal distributions, which are suitable for use in standard Metropolis--Hastings sampling from an augmented density $\pi(x,u)$.
These proposal schemes will be modified later for developing irreversible algorithms.

Given current variables $(x_0,u_0)$, a proposal $(x^*,u^*)$ can be generated as
\begin{equation} \label{eq:proposal}
  \begin{pmatrix}
    x^*\\
    u^*
  \end{pmatrix}
  =
  \begin{pmatrix}
    x_0\\
    u_0
  \end{pmatrix}
  -A
  \begin{pmatrix}
    \nabla U(x_0)\\
    u_0
  \end{pmatrix}
  +
  \begin{pmatrix}
    Z_1\\
    Z_2
  \end{pmatrix},
  \quad
  \begin{pmatrix}
    Z_1\\
    Z_2
  \end{pmatrix}\sim\mathcal N(\bo,2A - A^2),
\end{equation}
where $A$ is a $(2k)\times (2k)$ symmetric positive semi-definite (PSD) matrix and $Z_1,Z_2\in \mathbb R^k$ are Gaussian noises
independent of $(x_0,u_0)$, with $k$ the dimension of $x$ and that of $u$. We require $\bo \leq A \leq 2 I$, where inequalities between matrices are in the PSD sense.
This ensures that $2A - A^2$ is also symmetric positive semi-definite, although allowed to be singular.
The update in (\ref{eq:proposal}) takes a gradient step from the current variables $(x_0,u_0)$ and then injects Gaussian noises $(Z_1,Z_2)$. Hence the proposal scheme (\ref{eq:proposal})
is similar to that in pMALA. However, (\ref{eq:proposal}) is applied to $(x,u)$ jointly, instead of $x$ alone.

The proposal scheme (\ref{eq:proposal}) can be derived through an auxiliary variable argument related to \cite{Titsias2018},
while incorporating an over-relaxation technique as in \cite{Adler1981} and \cite{Neal1998}.
See Supplement Section~\ref{sec:auxiliary} for details.

Another important motivation for the proposal scheme (\ref{eq:proposal}) is that Metropolis--Hastings sampling using (\ref{eq:proposal}) becomes
rejection-free, while generating correlated draws, in the canonical case where the target density $\pi(x)$ is $\mathcal N(\bo,I)$,
that is, $U(x)= -x^\T x/2$ with the gradient $\nabla U(x) = x$.
In fact, the proposal scheme (\ref{eq:proposal}) in this case gives
\begin{equation} \label{eq:proposal2}
  \begin{pmatrix}
    x^*\\
    u^*
  \end{pmatrix}
  = (I - A)
  \begin{pmatrix}
    x_0\\
    u_0
  \end{pmatrix}+
  \begin{pmatrix}
    Z_1\\
    Z_2
  \end{pmatrix},
  \quad
  \begin{pmatrix}
    Z_1\\
    Z_2
  \end{pmatrix}\sim\mathcal N(\bo,2A - A^2).
\end{equation}
The update from $(x_0,u_0)$ to $(x^*,u^*)$ in (\ref{eq:proposal2}) can be seen to define
an order-$1$ vector autoregressive process, VAR(1), which is reversible and admits $\mathcal N(\bo, I)$ as a stationary distribution due to symmetry of $A$ \citep{Osawa1988}.
The stationary distribution can be easily verified: if $(x_0,u_0)\sim \mathcal N(\bo, I)$, then $(x^*,u^*)$ is normal and
the mean and variance are
\begin{align}
\EE[(x^{*\T},u^{*\T})^\T] = \bo,\quad   \Var[(x^{*\T},u^{*\T})^\T ] = (I - A)(I - A)^\T + 2A - A^2 = I . \label{eq:var-calc}
\end{align}
%If we further require $\bo < A < 2I$ with strict inequalities, then
%all the eigenvalues of $I-A$ are strictly between $-1$ and $1$, and hence the VAR(1)
%process is ergodic, converging to $\mathcal N(\bo, I)$ from any starting value.
The reversibility of (\ref{eq:proposal2}) with stationary distribution $\mathcal N(\bo, I)$ implies that
when the target density $\pi(x,u)$ is $\mathcal N(\bo, I)$,
Metropolis--Hastings sampling using the proposal scheme (\ref{eq:proposal2}) is rejection-free: the
draws $(x^*,u^*)$ are always accepted.
This can also be shown by using the proposal density, $Q(x^*,u^*|x_0,u_0) = \mathcal N(x^*,u^* | (I-A)(x^\T,u^\T)^\T, 2A - A^2)$,
and directly verifying that the acceptance probability (\ref{eq:accept-prob}) with $x$ replaced by $(x,u)$ reduces to 1.
%Finally, for the choice $A = I$, (\ref{eq:proposal2}) reduces to $(x^*,u^*)|(x_0,u_0)\sim \mathcal N(\bo, I)$ and Metropolis--Hastings reduces to perfect sampling.
%The proposals are always accepted and form independent and identically distributed draws from $\mathcal N(\bo, I)$.

% Osawa (1088), Reversibility of first-order autoregressive processes, Stochastic Processes and their Applications,  28,  61-69.

Our discussion focuses on the proposal scheme (\ref{eq:proposal}) for a Hamiltonian with momentum $u\sim \mathcal N(\bo,I)$ and the VAR(1) representation (\ref{eq:proposal2})
in the canonical case $x\sim \mathcal N(\bo,I)$, related to the normal approximation (\ref{eq:normal-approx}) with identity variance $I$ in the auxiliary variable derivation.
The development can be readily extended to handle general variance matrices, for a momentum distribution $u\sim \mathcal N(\bo,M)$ and a normal
approximation to $\pi(x)$ with variance matrix $\Sigma$.
%The VAR(1) representation and rejection-free property can be obtained under a normal target $x\sim \mathcal N(0,\Sigma)$.
%Furthermore, the auxiliary variables derivation can be applied by adjusting the normal approximation to $\pi(x)$ accordingly.
Nevertheless, as discussed in Section~\ref{subsec:precondition}, it is convenient to set $M = I$ and if an approximation of $\Var(x)$ is available,
apply linear transformation to $x$ such that the target density $\pi(x)$ can be roughly aligned with
an identity variance $\Sigma=I$.

%----------------------------------------------------------------------------------

\subsection{HAMS: a class of generalized reversible algorithms}
\label{subsec:general}

In this and subsequent sections, we exploit the class of proposals (\ref{eq:proposal}) with general choices of $A$ matrix,
to first derive a broad class of generalized reversible algorithms HAMS and then study two specific algorithms HAMS-A/B more elaborately.

For simplicity, consider the following form of $A$ matrix in (\ref{eq:proposal}),
\begin{align}
A =
\begin{pmatrix}
a_1 I & a_2 I\\
a_2 I & a_3 I
\end{pmatrix} \label{eq:matrix-A}
\end{align}
with each $I$ a $k\times k$ identity matrix and $a_1,a_2,a_3$ scalar coefficients.
We require $a_1,a_3\geq 0,a_1+a_3\leq 2$ and $a_1a_3\geq a_2^2$, which is sufficient for the constraint $\bo\leq A \leq 2I$ (in the PSD sense).
Substituting this choice of $A$ into (\ref{eq:proposal}) yields
\begin{align}
  x^* = x_0 - a_1\nabla U(x_0) - a_2 u_0 + Z_1, \label{eq:update-x}\\
  u^* = u_0 - a_2\nabla U(x_0) - a_3 u_0 + Z_2, \label{eq:update-u}
\end{align}
where $(Z^\T _1,Z^\T _2)^\T  \sim \mathcal N(\bo,2A - A^2)$ as before.
As discussed in Section~\ref{subsec:aux}, standard Metropolis-Hastings sampling using this proposal scheme is rejection-free, that is,
$(x^*,u^*)$ is always accepted, when the target density $\pi(x)$ is $\mathcal N(\bo,I)$.

\vspace{.1in}\textbf{Modification for generalized reversibility.}
We first make a modification to (\ref{eq:update-x})--(\ref{eq:update-u}) by replacing the momentum $u_0$ with $-u_0$.
Although a formal justification is to achieve generalized reversibility as shown in Proposition~\ref{prop1}, we give a heuristic motivation by
noticing that $a_2 u_0$ in (\ref{eq:update-x}) and $a_2\nabla U(x_0)$ in (\ref{eq:update-u}) are of the same sign. In contrast,
for the discretization of Hamiltonian dynamics using Euler's method:
 \[
   x^* = x_0 + \epsilon u_0, \quad u^* = u_0 - \epsilon\nabla U(x_0),
 \]
the momentum $u_0$ and gradient $\nabla U(x_0)$ are of the opposite signs. This discrepancy can be resolved by setting $u_0\mapsto - u_0$,
for which (\ref{eq:update-x})--(\ref{eq:update-u}) become
\begin{align}
  x^* &= x_0 - a_1\nabla U(x_0) + a_2 u_0 + Z_1, \label{eq:update-x2} \\
  u^* &= -u_0 - a_2\nabla U(x_0) + a_3 u_0 + Z_2, \label{eq:update-u2}
\end{align}
where $(Z^\T _1,Z^\T _2)^\T  \sim \mathcal N(\bo,2A - A^2)$ as before.

The proposal $(x^*,u^*)$ in (\ref{eq:update-x2})--(\ref{eq:update-u2}) can be accepted or rejected, similarly as in standard Metropolis--Hastings sampling
but using a different acceptance probability, which we derive through generalized detailed balance.
Rewrite the proposal scheme (\ref{eq:update-x2})--(\ref{eq:update-u2}) as
\begin{align}
& \tilde Z_1 = Z_1 - a_1\nabla U(x_0) + a_2 u_0,\quad \tilde Z_2 = Z_2 - a_2\nabla U(x_0) + a_3 u_0, \label{eq:forward-Z} \\
& x^* = x_0 + \tilde Z_1, \quad u^* = -u_0 + \tilde Z_2.  \label{eq:forward-xu}
\end{align}
Equations (\ref{eq:forward-xu})--(\ref{eq:forward-Z}) determine a forward transition from $(x_0,u_0)$ to $(x^*, u^*)$, depending on
noises $(Z_1,Z_2)$. To construct a backward transition, define new noises
\begin{align}
  Z_1^* = \tilde Z_1 - a_1\nabla U(x^*) - a_2 u^*,\quad Z_2^* = \tilde Z_2 - a_2\nabla U(x^*) - a_3 u^*. \label{eq:backward-Zdef}
\end{align}
Then (\ref{eq:backward-Zdef}) and (\ref{eq:forward-xu}) can be equivalently rearranged to
\begin{align}
& -\tilde Z_1 = -Z_1^* - a_1\nabla U(x^*)  + a_2 (-u^*),\quad -\tilde Z_2 = -Z_2^* -a_2 \nabla U(x^*) + a_3 (-u^*), \label{eq:backward-Z}\\
& x_0 = x^* + (-\tilde Z_1), \quad -u_0 = u^* +(-\tilde Z_2) .  \label{eq:backward-xu}
\end{align}
Importantly, equations (\ref{eq:backward-Z})--(\ref{eq:backward-xu}) can be seen to correspond to the {\it same} mapping as (\ref{eq:forward-Z})--(\ref{eq:forward-xu}), but
applied from $(x^*, -u^*)$ to $(x_0, -u_0)$ using the new noises $(-Z_1^*, -Z_2^*)$.
In other words, (\ref{eq:backward-Z})--(\ref{eq:backward-xu}) are obtained from (\ref{eq:forward-Z})--(\ref{eq:forward-xu})
by replacing $(x_0,u_0)$, $(x^*, u^*)$, and $(Z_1,Z_2)$ with
$(x^*, -u^*)$, $(x_0, -u_0)$, and $(-Z_1^*, -Z_2^*)$ respectively.

From the preceding discussion, the forward and backward transitions of the proposals in (\ref{eq:forward-Z})--(\ref{eq:forward-xu}) and (\ref{eq:backward-Z})--(\ref{eq:backward-xu})
can be illustrated as
\begin{align}
  \begin{pmatrix}
    x_0 \\
    u_0
  \end{pmatrix}
  \stackrel{(Z_1,Z_2)}{\longrightarrow}
  \begin{pmatrix}
    x^* \\
    u^*
  \end{pmatrix},
  \qquad
  \begin{pmatrix}
    x^* \\
    -u^*
  \end{pmatrix}
  \stackrel{-(Z_1^*,Z_2^*)}{\longrightarrow}
  \begin{pmatrix}
    x_0 \\
    -u_0
  \end{pmatrix}, \label{eq:transition-diagram}
\end{align}
where the two arrows denote the {\it same} mapping, depending on $(Z_1,Z_2)$ or $-(Z_1^*,Z_2^*)$.
For $(Z^\T_1,Z^\T_2)^\T \sim \mathcal N(\bo,2A - A^2)$, the proposal density from $(x_0,u_0)$ to $(x^*, u^*)$ is
\begin{align*}
Q ( x^*, u^* | x_0, u_0) = \mathcal N( Z_1,Z_2 | \, \bo, 2A - A^2 ) .
\end{align*}
Moreover, evaluation of the {\it same} proposal density from  $(x^*, -u^*)$ to $(x_0, -u_0)$ gives
\begin{align*}
Q (x_0, -u_0 |x^*, -u^*) = \mathcal N( -(Z_1^*, Z_2^*)| \, \bo, 2A - A^2 ),
\end{align*}
because the transition from $(x^*, -u^*)$ to $(x_0, -u_0)$ is determined by the same mapping as $(x_0, u_0)$ to $(x^*, u^*)$, only with the noises $(-Z_1^*, -Z_2^*)$ used instead of $(Z_1,Z_2)$.

By mimicking (and extending) the standard Metropolis--Hastings probability, we set $(x_1,u_1)=(x^*,u^*)$ with the acceptance probability
\begin{align}
   \rho( x^*,u^* | x_0,u_0 ) =\min\left(1, \frac{\pi(x^*, -u^*) Q( x_0, -u_0 | x^*, -u^*) }
  {\pi(x_0,u_0) Q ( x^*, u^* | x_0, u_0) }\right),   \label{mr1}
\end{align}
or set $(x_1,u_1)=(x_0,-u_0)$ with the remaining probability.
Due to the evenness of mean-zero normal distributions, the probability (\ref{mr1}) can be calculated as
\begin{align}
  \rho( x^*,u^* | x_0,u_0 ) =\min\left(1, \frac{\exp\left\{ -H(x^*,u^*)-\frac{1}{2}{\bZ^*}^\T (2A-A^2)^{-1}\bZ^* \right\}}
  {\exp\left\{ -H(x_0,u_0)-\frac{1}{2}\bZ^\T (2A-A^2)^{-1}\bZ \right\}}\right),  \label{mr1-b}
\end{align}
where $\bZ = (Z^\T_1,Z^\T_2)^\T$ and $\bZ^* = ({Z_1^*}^\T, {Z_2^*}^\T)^\T$. Note that $u_1=u^*$ upon acceptance, but $u_1=-u_0$ in the case of rejection.
The resulting transition from $(x_0,u_0)$ to $(x_1,u_1)$ can be shown to satisfy generalized detailed balance.

\begin{prop} \label{prop1}
  For an augmented density $\pi(x,u)$ in (\ref{eq:aug-density2}), let $K_0(x_1 ,u_1 | x_0,u_0)$ be the transition kernel from $(x_0,u_0)$ to $(x_1,u_1)$, defined by
  the proposal scheme (\ref{eq:forward-Z})--(\ref{eq:forward-xu}) and the acceptance probability (\ref{mr1}). Then generalized detailed balance holds for $x_1\not=x_0$:
  \begin{equation} \label{eq:gDB}
    \pi(x_0,u_0) K_0(x_1 ,u_1 | x_0,u_0) = \pi(x_1,-u_1) K_0(x_0,-u_0 | x_1, -u_1).
  \end{equation}
Furthermore, the augmented density $\pi(x,u)$ is a stationary distribution of the Markov chain defined by transition kernel $K_0$.
\end{prop}

Condition (\ref{eq:gDB}), called generalized detailed balance (or generalized reversibility),
differs from detailed balance (or reversibility) in standard Metropolis--Hastings sampling because the momentum variable is negated in
defining the backward transition.
Accordingly, the acceptance probability (\ref{mr1}) is called a generalized Metropolis--Hastings probability.
A similar concept of detailed balance is known in connection with Fokker-Planck equations in physics \citep[Section 5.3.4]{Gardiner1997}.
The momentum is called an odd variable, for which the time-reversed variable is defined with sign negation to achieve generalized detailed balance.
Such a general formulation of detailed balance is used in the derivation of underdamped Langevin sampling \citep{Bussi2007}, but
overall seems to be under-appreciated in the MCMC literature. See Section~\ref{sec:extension} for a further extension.

\vspace{.1in}\textbf{Modification for updating momentum.}
To further broaden our method, we introduce another modification to the proposal scheme (\ref{eq:forward-Z})--(\ref{eq:forward-xu}).
In fact, a potential limitation of (\ref{eq:forward-Z})--(\ref{eq:forward-xu}), compared with the popular leapfrog scheme, is that the updated momentum $u^*$ ignores the new gradient information $\nabla U(x^*)$.
To incorporate $\nabla U(x^*)$ in updating the momentum, we revise (\ref{eq:forward-xu}) with an additional term in $u^*$ as
\begin{align}
 x^* = x_0 + \tilde Z_1, \quad u^* = -u_0 + \tilde Z_2 + \phi (\tilde Z_1 + \nabla U(x_0) - \nabla U(x^*) ),  \label{eq:forward-xu2}
\end{align}
where $\phi$ is a (constant) tuning parameter, and $(\tilde Z_1,\tilde Z_2)$ remain the same as in (\ref{eq:forward-Z}).
Moreover, the update (\ref{eq:forward-xu2}) can be rearranged to
\begin{align}
x_0 = x^* + (-\tilde Z_1),  \quad -u_0 = u^* + (-\tilde Z_2) + \phi (-\tilde Z_1 + \nabla U(x^*) - \nabla U(x_0) ).  \label{eq:backward-xu2}
\end{align}
With $(Z_1^*, Z_2^*)$ still defined as (\ref{eq:backward-Zdef}), equations (\ref{eq:forward-Z}) and (\ref{eq:forward-xu2}) and equations
(\ref{eq:backward-Z}) and (\ref{eq:backward-xu2}) can be seen to be determined by the {\it same} mapping, similarly as illustrated in (\ref{eq:transition-diagram}).
The forward transition is from $(x_0,u_0)$ to $(x^*, u^*)$ depending on $(Z_1,Z_2)$, whereas
the backward transition is from $(x^*, -u^*)$ to $(x_0, -u_0)$ depending on $-(Z_1^*, Z_2^*)$.
With the modified proposal $(x^*,u^*)$, the acceptance-rejection is the same as before: set $(x_1,u_1)=(x^*,u^*)$ with probability (\ref{mr1}) or
$(x_1,u_1) = (x_0,-u_0)$ with the remaining probability. Then
generalized detailed balance remains valid for the transition from $(x_0,u_0)$ and $(x_1,u_1)$.

%\vspace{-.05in}
\begin{prop} \label{prop1-b}
  For an augmented density $\pi(x,u)$ in (\ref{eq:aug-density2}), let $K_\phi(x_1 ,u_1 | x_0,u_0)$ be the transition kernel from $(x_0,u_0)$ to $(x_1,u_1)$, defined by
  the proposal scheme (\ref{eq:forward-Z}) and (\ref{eq:forward-xu2}) and the acceptance probability (\ref{mr1}). Then generalized detailed balance holds for $x_1\not=x_0$:
  \begin{equation} \label{eq:gDB-b}
    \pi(x_0,u_0) K_\phi (x_1 ,u_1 | x_0,u_0) = \pi(x_1,-u_1) K_\phi (x_0,-u_0 | x_1, -u_1).
  \end{equation}
Furthermore, the augmented density $\pi(x,u)$ is a stationary distribution of the Markov chain defined by transition kernel $K_\phi$.
\end{prop}

%\vspace{.1in}
\textbf{General HAMS.} Using the proposal scheme and acceptance probability as in Proposition~\ref{prop1-b}
leads to a class of generalized reversible MCMC algorithms, which is called Hamiltonian assisted Metropolis sampling (HAMS) and shown in Algorithm~\ref{alg1}.

Although the modifications of the proposal scheme from (\ref{eq:update-x})--(\ref{eq:update-u}) to (\ref{eq:update-x2})--(\ref{eq:update-u2})
and then to (\ref{eq:forward-Z}) and (\ref{eq:forward-xu2}) are constructed for different purposes,
the resulting HAMS algorithm preserves the rejection-free property with a standard normal target density $\pi(x)$, which is satisfied by
standard Metropolis--Hastings sampling with proposal scheme (\ref{eq:update-x})--(\ref{eq:update-u}).
In fact, the second modification from (\ref{eq:forward-xu}) to (\ref{eq:forward-xu2}) has no effect
when $\pi(x)$ is $\mathcal N(\bo,I)$, because in this case $\tilde Z_1 + \nabla U(x_0) - \nabla U(x^*) = \tilde Z_1 + x_0 - x^* = \bo$.
The justification for the first modification is subtler. Whether rejection-free is achieved by a sampling algorithm depends on both a proposal scheme and
an associated acceptance-rejection mechanism. When $\pi(x)$ is $\mathcal N(\bo,I)$, our HAMS algorithm is rejection-free,
due to the fact the proposal scheme (\ref{eq:update-x2})--(\ref{eq:update-u2}) is used in conjunction with the generalized acceptance probability (\ref{mr1}),
not the standard Metropolis--Hastings probability.
We provide further discussion in Section~\ref{sec:extension}, where it can be seen that consideration of the rejection-free property is instrumental to
a general approach for constructing generalized reversible algorithms.

\begin{cor} \label{cor:rejection-free}
 Suppose that the target density $\pi(x)$ is  $\mathcal N(\bo,I)$. Then the generalized acceptance probability (\ref{mr1}) or equivalently (\ref{mr1-b})
 reduces to 1, and hence $(x^*, u^*)$ from the proposal scheme (\ref{eq:update-x2})--(\ref{eq:update-u2}) is always accepted under the HAMS algorithm.
\end{cor}

\begin{algorithm}[t] %[H]
\SetAlgoLined
Initialize $x_0,u_0$\\
 \For{ $t = 0,1,2,..., N_{iter}$}{
   Sample $w \sim \text{Uniform} [0,1]$ and $(Z_1,Z_2)^\T \sim N(\bo,2A -A^2) $ with $A = \begin{pmatrix}
   a_1 I & a_2 I\\
   a_2 I & a_3 I
   \end{pmatrix} $ \\

   $\tilde Z_1 = Z_1 - a_1\nabla U(x_t) + a_2 u_t$\\
   $\tilde Z_2 = Z_2 - a_2\nabla U(x_t) + a_3 u_t$\\
   Propose $x^* = x_t + \tilde Z_1$ and $u^* = -u_t + \tilde Z_2 + \phi(\tilde Z_1 + \nabla U(x_t) - \nabla U(x^*) )$\\
   $Z_1^* = \tilde Z_1 - a_1 \nabla U(x^*) - a_2 u^*$\\
   $Z_2^* = \tilde Z_2 - a_2 \nabla U(x^*) - a_3 u^*$\\
   $\rho = \exp\left\{H(x_t,u_t) - H(x^*,u^*) + \frac{1}{2}\bZ^\T (2A-A^2)^{-1}\bZ -
   \frac{1}{2} \bZ^{*\T}(2A-A^2)^{-1}\bZ^*\right\}$\\
  \eIf{$w < \min(1,\rho)$}{ $(x_{t+1},u_{t+1}) = (x^*,u^*)$ \qquad\# Accept
   }{
    $(x_{t+1},u_{t+1}) = (x_t,-u_t)$ \qquad\# Reject
  }
 }
\caption{General HAMS}  \label{alg1}
\end{algorithm}

The general HAMS involves four tuning parameters $\phi,a_1,a_2,a_3$, which need to be specified for practical implementation.
In the following sections, we develop more concrete versions of HAMS with a reduced number of tuning parameters.
As the augmented target density is $2k$ dimensional,
HAMS in general allows the noise term $(Z_1,Z_2)$ to be drawn directly from a $2k$ dimensional Gaussian distribution.
Nevertheless, there are related methods developed for simulating Langevin dynamics, using $k$ dimensional noises at each time step \citep{Farago2013,Farago2020}.
We investigate HAMS which also uses only $k$ dimensional Gaussian noises in each iteration. This requires the variance matrix $2A - A^2$ to be singular.
There are two possible choices: either $A$ itself is singular or $2 I -A$ is singular, corresponding to HAMS-A and HAMS-B in Section~\ref{subsec:hams-ab}.

%---------------------------------------------------------------------------------
\subsection{HAMS-A and HAMS-B}
\label{subsec:hams-ab}

We develop two concrete versions of HAMS with the noise variance $2A - A^2$ singular, hence using only $k$ dimensional Gaussian noises in each iteration.

\vspace{.1in}\textbf{HAMS-A.}  First, we set $A$ singular by taking $a_1 = a, a_3 = b$ and $a_2 = \sqrt{ab}$ in (\ref{eq:matrix-A}). The constraints on $A$ require that $a\ge 0,b\geq 0$ and $a+b\leq 2$.
To avoid trivial cases, we also assume that $a>0$.
The noise variance becomes
\begin{equation}
  \label{eq:var1}
  \Var  \begin{pmatrix}
    Z_1\\
    Z_2
  \end{pmatrix} = 2A - A^2 =
  \begin{pmatrix}
    a(2-a-b) I & \sqrt{ab}(2-a-b) I\\
    \sqrt{ab}(2-a-b) I & b(2-a-b) I
  \end{pmatrix}.
\end{equation}
As expected, this implies that $Z_1$ and $Z_2$ are proportional: $Z_2 = \sqrt{b/a} Z_1$.
By definitions (\ref{eq:forward-Z}), (\ref{eq:forward-xu2}), and (\ref{eq:backward-Zdef}), it can be easily verified that
$\tilde Z_2 = \sqrt{b/a}\tilde Z_1$ and $Z_2^* = \sqrt{b/a} Z_1^*$ as well. The proportionality between
$Z_1^*$ and $Z_2^*$ is important, because it ensures that both forward and backward transitions, illustrated in (\ref{eq:transition-diagram}),
can be determined using a single noise vector, $Z_1$ or $-Z_1^*$.
Hence the proposal density from $(x_0,u_0)$ to $(x^*, u^*)$ is $\mathcal N ( Z_1 | \bo, a(2-a-b)I)$
and that from $(x^*, -u^*)$ to $(x_0, -u_0)$ is $\mathcal N ( -Z_1^* | \bo, a(2-a-b)I)$.
The acceptance probability (\ref{mr1}) can be evaluated as (\ref{mr2}) below, while (\ref{mr1-b}) is not well defined.

From the preceding discussion, the HAMS algorithm can be simplified as follows, given current variables $(x_0,u_0)$:
\begin{align*}
  & \tilde Z = Z - a\nabla U(x_0) + \sqrt{ab} u_0,\quad  Z\sim \mathcal N(\bo,a(2-a-b)I),\\
  & x^* = x_0 + \tilde Z,\quad u^* = -u_0 + \sqrt{\frac{b}{a}}\tilde Z + \phi (\tilde Z + \nabla U(x_0) - \nabla U(x^*)), \\
  & Z^* = \tilde Z - a\nabla U(x^*) - \sqrt{ab} u^*.
\end{align*}
The proposal $(x^*,u^*)$ is accepted with probability
\begin{equation}
  \label{mr2}
  \min\left(1,\exp\left\{H(x_0,u_0) - H(x^*,u^*) + \frac{Z^\T Z-(Z^*)^\T Z^*}{2a(2-a-b)} \right\}\right).
\end{equation}
Except for the choice of $\phi$ derived below, this algorithm is shown as HAMS-A in Algorithm~\ref{alg2}, after a transformation $Z=\sqrt{a(2-a-b)}\zeta$ with $\zeta \sim \mathcal N(\bo,I)$.

To derive a specific choice for $\phi$, we examine the situation where the target density $\pi(x)$ deviates from standard normal.
As discussed in Section~\ref{subsec:general}, the HAMS algorithm is rejection-free, that is,
the acceptance probability (\ref{mr2}) is always 1, when the target density $\pi(x)$ is $\mathcal N(\bo,I)$.
We seek a choice of $\phi$ such that the acceptance probability can be minimally affected by the deviation of $\gamma$ from 1,
when $\pi(x)$ is $\mathcal N(\bo,\gamma^{-1} I)$. For simplicity, we study the
behavior of the quantity inside $\exp()$ in (\ref{mr2}) as $\gamma$ varies.

\begin{lem} \label{lem:hamsa}
Suppose that the target density $\pi(x)$ is $\mathcal N(\bo, \gamma^{-1} I)$. Then the quantity  inside $\exp()$ in (\ref{mr2}) can be expressed as as a quadratic form,
\begin{align*}
H(x_0,u_0) - H(x^*,u^*) + \frac{Z^\T Z-(Z^*)^\T Z^*}{2a(2-a-b)} = (x_0^\T, u_0^\T, Z^\T) G(\gamma) (x_0^\T, u_0^\T, Z^\T)^\T ,
\end{align*}
where $G(\gamma)$ is a $3 \times 3$ block matrix. For $i,j=1,2,3$, the $(i,j)$th block of $G(\gamma)$ is of the form $g_{ij}(\gamma) I$,
where $g_{ij}(\gamma)$ is a scalar, polynomial of $\gamma$, with coefficients depending on $(a,b,\phi)$.
For any $a> 0,b\geq 0$ and $a+b\leq 2$, the coefficients of the leading terms of $g_{11}(\gamma)$, $g_{22}(\gamma)$, $g_{33}(\gamma)$
are simultaneously minimized in absolute values by the choice $\phi = \sqrt{ab}/(2-a)$.
\end{lem}

It seems remarkable that a single choice of $\phi$ leads to simultaneous minimization of the absolute coefficients of the leading terms of $g_{11}(\gamma)$, $g_{22}(\gamma)$, $g_{33}(\gamma)$.
Moreover, the particular choice $\phi = \sqrt{ab}/(2-a)$ also ensures that HAMS-A reduces to leapfrog or modified pMALA in the special cases where $a+b=2$ or $b=0$, as discussed in Section~\ref{subsec:special}.

\vspace{.1in}\textbf{HAMS-B.}
For a singular $2A-A^2$, another possibility is to set $2I - A$ singular. We take $a_1 = 2-a,a_3 = 2-b$ and
$a_2 = \sqrt{ab}$ in (\ref{eq:matrix-A}), with the constraints that $a>0,b\geq0$ and $a+b\leq 2$. The noise variance is then
\begin{equation}
  \label{eq:var2}
  \Var  \begin{pmatrix}
    Z_1\\
    Z_2
  \end{pmatrix} = 2A - A^2 =
  \begin{pmatrix}
    a(2-a-b) I & \sqrt{ab}(a+b-2) I\\
    \sqrt{ab}(a+b-2) I & b(2-a-b) I
  \end{pmatrix},
\end{equation}
which implies that $Z_1$ and $Z_2$ are proportional: $Z_2 = -\sqrt{b/a} Z_1$. However,
it does not in general hold that $Z_2^* = -\sqrt{b/a} Z_1^*$, except for the choice $\phi = \sqrt{b/a}$.
Moreover, this choice of $\phi$ is the only one such that any proportionality between $(Z_1^*, Z_2^*)$ holds.
This situation is in contrast with HAMS-A, where $Z_2^* = \sqrt{b/a} Z_1^*$ automatically holds for any choice of $\phi$
and additional consideration is needed to derive a specific choice of $\phi$.

\begin{lem} \label{lem:hamsb}
For the preceding choice of $A$ in (\ref{eq:matrix-A}), it holds that $Z_2^* =r Z_1^*$ for a constant coefficient $r\in \mathbb R$ and arbitrary values $(x_0, u_0, Z_1)$
by definitions (\ref{eq:forward-Z}), (\ref{eq:forward-xu2}), and (\ref{eq:backward-Zdef}) if and only if $r=-\sqrt{b/a}$ and $\phi = \sqrt{b/a}$.
\end{lem}

\begin{algorithm}[t] %[H]
\SetAlgoLined
Initialize $x_0,u_0$\\
 \For{ $t = 0,1,2,..., N_{iter}$}{
   Sample $w \sim \text{Uniform} [0,1]$ and $\zeta \sim \mathcal N(\bo,I)$ \\
   Propose $x^* = x_t - a\nabla U(x_t) + \sqrt{ab} u_t + \sqrt{a(2-a-b)}\zeta$\\
   \If{HAMS-A}
   {
      Propose $u^* = \left(\frac{2b}{2-a} - 1\right) u_t - \frac{\sqrt{ab}}{2-a}(\nabla U(x_t) + \nabla U(x^*)) + \frac{2\sqrt{b(2-a-b)}}{2-a}\zeta$\\
      $\zeta^* = \left(1-\frac{2b}{2-a}\right) \zeta - \frac{\sqrt{a(2-a-b)}}{2-a}(\nabla U(x_t) + \nabla U(x^*)) + \frac{2\sqrt{b(2-a-b)}}{2-a}u_t$
   }
   \If{HAMS-B}
   {  Propose $u^* = u_t - \frac{\sqrt{ab}}{2-a}(\nabla U(x_t) + \nabla U(x^*))$\\
      $\zeta^* = \zeta - \frac{\sqrt{a(2-a-b)}}{2-a}(\nabla U(x_t) + \nabla U(x^*))$
   }
   $\rho = \exp\left\{H(x_t,u_t) - H(x^*,u^*) + \frac{1}{2}\zeta^\T \zeta - \frac{1}{2}(\zeta^*)^\T \zeta^*\right\}$\\
  \eIf{$w < \min(1,\rho)$}{ $(x_{t+1},u_{t+1}) = (x^*,u^*)$ \qquad\# Accept
   }{
    $(x_{t+1},u_{t+1}) = (x_t,-u_t)$ \qquad\# Reject
  }
 }
 \caption{HAMS-A/HAMS-B}  \label{alg2}
\end{algorithm}

To maintain the forward and backward transitions, illustrated in (\ref{eq:transition-diagram}), using a single noise vector, we take the only feasible choice $\phi = \sqrt{b/a}$.
Then the HAMS algorithm can be simplified as follows, given current variables $(x_0,u_0)$: \vspace{-.05in}
\begin{align*}
  & \tilde Z = Z - (2-a)\nabla U(x_0) + \sqrt{ab} u_0,\quad  Z\sim \mathcal N(0,a(2-a-b)I), \\
  & x^* = x_0 + \tilde Z,\quad u^* = u_0 + \sqrt{\frac{b}{a}}(\nabla U(x_0) + \nabla U(x^*)), \\
  & Z^* = \tilde Z - (2-a)\nabla U(x^*) - \sqrt{ab} u^*.
\end{align*}
Similarly as discussed for HAMS-A, the acceptance probability (\ref{mr1}) can be evaluated as (\ref{mr2}).
To facilitate comparison with HAMS-A, we use a reparametrization, $\tilde a = 2-a$ and $\tilde b= ab/(2-a)$,
such that $ab = \tilde a \tilde b$ and $a(2-a-b) = \tilde a (2 - \tilde a - \tilde b)$.
The transformation is one-to-one between $\{(a,b): a>0, b>0, a+b\le 2\}$ and $\{ (\tilde a, \tilde b): \tilde a>0, \tilde b >0, \tilde a+ \tilde  b \le 2\}$.
The resulting algorithm, with $(\tilde a, \tilde b)$ relabeled as $(a,b)$, is shown as HAMS-B in Algorithm~\ref{alg2}.
Then the two algorithms, HAMS-A and HAMS-B, agree in the expressions for $x^*$.

%---------------------------------------------------------------------------------
\subsection{Default choices of carryover}
\label{subsec:altparam}

While the $(a,b)$ parameterization arises naturally in our development above, the $(\epsilon,c)$
parameterization used in existing algorithms (see Section \ref{sec:review}) has a desirable interpretation, with $\epsilon$ corresponding to a
step size and $c$ the amount of carryover momentum. By matching leapfrog and modified pMALA in special cases (see Section~\ref{subsec:special}),
our HAMS algorithms can be translated into an $(\epsilon,c)$ parameterization with the following formulae:
\begin{align}
  a = \frac{\epsilon^2}{1+\sqrt{1-\epsilon^2}} = 1 - \sqrt{1-\epsilon^2},\quad b = c(2-a),\quad 0\leq \epsilon, c \leq 1.\label{eq:abepsc}
\end{align}
Because $a$ is expressed as a function of $\epsilon$ only, and $b$ given $a$ is a function of $c$ only, we also refer
to $a$ as a step size and $b$ as a carryover.

So far, the number of tuning parameters is reduced from four in general HAMS (Algorithm~\ref{alg1}) to two in HAMS-A/B (Algorithm~\ref{alg2}).
To facilitate applications, we seek to further reduce tuning by studying the lag-1 auto-covariance matrix for a HAMS chain in stationary when
the target density $\pi(x)$ is standard normal.

\begin{lem} \label{lem:b-choice}
  Suppose that the target density $\pi(x)$ is $\mathcal N(\bo,1)$, and $(x_0,u_0)\sim\mathcal N(\bo,I)$.
  Given step size $a$, the maximum modulus of the eigenvalues of the lag-1 auto-covariance
  matrix $\Cov((x_0,u_0), (x_1,u_1))$ is minimized by the following choice of $b$:
  \begin{align} \label{eq:b-choice}
    \text{HAMS-A:  } b = (\sqrt{2} - \sqrt{a})^2,\qquad \text{HAMS-B:  } b = \frac{a(2-a)}{(\sqrt{2} + \sqrt{2-a})^2}.
  \end{align}
\end{lem}

For convenience, the formulae (\ref{eq:b-choice}) can be used as the default choices of carryover $b$, given step size $a$.
On the other hand, such choices are derived under an idealized setting, where the target
density $\pi(x)$ is $\mathcal N(\bo,I)$. For the default tuning to be effective, we often need to first
apply transformations to bring $\pi(x)$ closer to $\mathcal N(\bo,I)$, which will be discussed in Section \ref{subsec:precondition}.
If such a transformation is not available for various reasons, then it is preferable to tune both $a$ and $b$
instead of using the default values in (\ref{eq:b-choice}).

%---------------------------------------------------------------------------------
\subsection{Special Cases of HAMS-A/B}
\label{subsec:special}

Recall that the constraints on the step size and carryover are $a\geq0,b\geq 0, a+b\leq 2$. In the following, we examine
three boundary cases.

The first case is when $ a + b = 2 $ (or equivalently $c=1$). For both HAMS-A and HAMS-B, the updates become deterministic from $(x_0,u_0)$ to $(x^*,u^*)$ .
To help understanding, we introduce an
intermediate variable $\tilde u$. Then the updates can be written as
\begin{align*}
& \zeta \sim \mathcal N(\bo,I),\qquad\zeta^* = - \zeta \,\text{(HAMS-A)}, \qquad\zeta^* = \zeta \, \text{(HAMS-B)},\\
& \tilde u = u_0 - \sqrt{\frac{a}{2-a}}\nabla U(x_0) = u_0 - \frac{\epsilon}{1+\sqrt{1-\epsilon^2}}\nabla U(x_0),\\
& x^* = x_0 + \sqrt{a(2-a)}\tilde u_0 = x_0 + \epsilon\tilde u,\\
& u^* = \tilde u - \sqrt{\frac{a}{2-a}}\nabla U(x^*) = \tilde u - \frac{\epsilon}{1+\sqrt{1-\epsilon^2}} U(x^*),
\end{align*}
where the Metropolis ratio is $\rho=\exp(H(x_0,u_0) - H(x^*,u^*))$. The above is similar to the leapfrog discretization of the
Hamiltonian dynamics but with step size $\epsilon/(1+\sqrt{1-\epsilon^2})$ instead of $\epsilon/2$ for
momentum updates. The proposal $(x^*,u^*)$ can be accepted or rejected (with $u_0$ flipped) based on the change in the
Hamiltonian from the update.

The second case is when $ b = 0 $ (or equivalently $c=0$). We introduce another intermediate variable $\tilde{\zeta}$ to the updates. Then HAMS-A and HAMS-B reduce to
\begin{align*}
& \zeta \sim \mathcal N(\bo,I),\qquad u^* = - u_0 \, \text{(HAMS-A)}, \qquad u^* = u_0 \,\text{(HAMS-B)}, \\
& \tilde{\zeta} = \zeta - \sqrt{\frac{a}{2-a}}\nabla U(x_0) = \zeta - \frac{\epsilon}{1+\sqrt{1-\epsilon^2}}\nabla U(x_0),\\
& x^* = x_0 + \sqrt{a(2-a)}\tilde{\zeta} = x_0 + \epsilon \tilde{\zeta},\\
& \zeta^* = \tilde{\zeta} - \sqrt{\frac{a}{2-a}}\nabla U(x^*) = \tilde{\zeta}- \frac{\epsilon}{1+\sqrt{1-\epsilon^2}}\nabla U(x^*),
\end{align*}
where the Metropolis ratio is $\rho=\exp(U(x_0) - U(x^*) + \frac{1}{2}\zeta^\T \zeta - \frac{1}{2}(\zeta^*)^\T \zeta^*)$.
Hence $u_0$ remains unchanged in HAMS-A, and is negated in HAMS-B, although the update of $u_0$ is irrelevant in this case. The update of $x_0$ to $x^*$
and acceptance-rejection coincide with modified pMALA in Section~\ref{sec:review}, which differs from ordinary pMALA because the step size $\epsilon^2/(1+\sqrt{1-\epsilon^2})$
is associated with $\nabla U(x_0)$ for updating $x_0$, instead of $\epsilon^2/2$.

The third case is when $ a = 0 $ (or equivalently $\epsilon=0$). This case is not interesting because $x$ remains constant.
Our discussion is for completeness. When $a = 0$, HAMS-B sets all variables constant:
$x^* = x_0, u^* = u_0$, and $\zeta^* = \zeta$. HAMS-A gives the updates \vspace{-.05in}
\begin{align*}
& \zeta \sim\mathcal N(\bo,I),\qquad x^* = x_0,\\
& u^* = (b-1) u_0 + \sqrt{b(2-b)}\zeta, \quad
\zeta^* = (1-b)\zeta + \sqrt{b(2-b)} u_0.
\end{align*}
In this case, the Metropolis ratio is always $1$. Hence HAMS-A can be viewed as an autoregressive process on $u$
while $x$ remains constant.\par

Finally, we note that our HAMS-A/B algorithms differ from UDL \citep{Bussi2007}, which uses two noise vectors per iteration,
although UDL also recovers leapfrog and pMALA in the extreme cases of $c=1$ and $c=0$ respectively.

%----------------------------------------------------------------------------------
\subsection{Preconditioning}
\label{subsec:precondition}

As commonly recognized in MCMC literatures, if there is information about the variance structure of the target
density, then the performance of MCMC samplers can be improved by applying a linear transformation, i.e., preconditioning.
Suppose that $\Sigma$ is an approximation to $\Var(x)$, or $M$ is an approximation to $(\Var(x))^{-1}$.
Then RWM and pMALA involve preconditioning using the approximate variance $\Sigma$ on $x$,
whereas HMC and UDL involve preconditioning using $M$ as the momentum variance.
These two approaches are conceptually equivalent, as discussed in the context of HMC by \cite{Neal2011},
although one can be more preferable than the other in computational implementations.

We use the first approach of preconditioning: applying a linear transformation to the original variable $x$ while keeping the momentum $u \sim \mathcal N(\bo,I)$.
Let $L$ be the lower triangular matrix obtained from
the Cholesky decomposition $M = LL^\T $. The transformed variable is $\tilde x = L^\T  x$.
If $x$ is approximately $\mathcal N(0,M^{-1})$, then $\tilde x$ is approximately $\mathcal N(\bo,I)$.
Application of HAMS-A/B in Algorithm~\ref{alg2} to the transformed variable $\tilde x$ leads to HAMS-A/B algorithms with preconditioning, which are shown in Algorithm~\ref{alg3}.
The gradient of the potential after the transformation, denoted as $\nabla U(\tilde x)$, is $L^{-1}\nabla U(x)$.

\begin{algorithm}[t] %[H]
\SetAlgoLined
Initialize $x_0,u_0,\tilde x_0 = L^\T x_0$ and $\nabla U(\tilde x_0) = L^{-1}\nabla U(x_0)$.\\
 \For{ $t = 0,1,2,..., N_{iter}$}{
   Sample $w \sim \text{Uniform} [0,1]$ and $\zeta \sim \mathcal N(\bo,I)$ \\
   $\xi = \sqrt{ab}u_t + \sqrt{a(2-a-b)}\zeta, \quad \tilde x^* = \tilde x_t - a \nabla U(\tilde x_t) + \xi$\\
   Propose $x^* = (L^\T )^{-1}\tilde x^*$\\
   $\nabla U(\tilde x^*) = L^{-1} \nabla U(x^*), \quad \tilde{\xi} = \nabla U(\tilde x^*) + \nabla U(\tilde x_t)$\\
   $\rho = \exp\left\{U(x_t) - U(x^*) + \frac{1}{2-a}(\tilde{\xi})^\T (\xi - \frac{a}{2}\tilde{\xi} )\right\}$ \\
  \eIf{$w < \min(1,\rho)$}{
    $x_{t+1} = x^*$,\quad $\tilde x_{t+1} = \tilde x^*$,\quad $\nabla U(\tilde x_{t+1}) =  \nabla U(\tilde x^*) $\qquad\# Accept \\
    \If{HAMS-A}
    {
      $u_{t+1} = \left(\frac{2b}{2-a} - 1\right)u_{t} + \frac{2\sqrt{b(2-a-b)}}{2-a}\zeta - \frac{\sqrt{ab}}{2-a}\tilde{\xi}$\\
    }
    \If{HAMS-B}
    {
      $u_{t+1} = u_t - \frac{\sqrt{ab}}{2-a}\tilde{\xi}$ \\
    }

   }{
    $x_{t+1} = x_t, u_{t+1} = -u_t, \tilde x_{t+1} = \tilde x_t,\nabla U(\tilde x_{t+1}) = \nabla U(\tilde x_{t}) $ \qquad\# Reject
  }
 }
 \caption{HAMS-A/HAMS-B (with preconditioning)}   \label{alg3}
\end{algorithm}

Our Algorithm~\ref{alg3} is carefully formulated, such that transforming $x$ and keeping $u\sim\mathcal N(\bo,I)$ improves computational efficiency, compared with
using the original variable $x$ and $u\sim\mathcal N(\bo, M)$.
See the Appendix Section \ref{subsec:simplify:alg3} for details of simplification.
Excluding the evaluation of $U(x)$ and $\nabla U(x)$, Algorithm \ref{alg3} involves
$2$ matrix-by-vector multiplications per iteration, $(L^\T )^{-1}\tilde x^*$ and $L^{-1}\nabla U(x^*)$. Moreover, computation of the Metropolis ratio $\rho$
is also optimized, requiring only $1$ inner product instead of $4$ as in Algorithm \ref{alg2}.
In contrast, UDL as described in Section \ref{sec:review} needs $5$ matrix-by-vector multiplications per iteration: $2$
for sampling from $\mathcal N(\bo, M)$, $1$ for computing $x^*$, and $2$ in the Metropolis ratio.
In the simulation studies, we implement UDL with reduced runtime in a similar way as Algorithm \ref{alg3}, in order to make fair comparisons with HAMS-A/B.

\section{Generalized Metropolis--Hastings sampling} \label{sec:extension}

Our development in Section~\ref{sec:proposed} presents a concrete class of generalized reversible algorithms, HAMS, using an augmented target density
originated from a Hamiltonian in physics.
In this section, we discuss a flexible framework of generalized Metropolis--Hastings sampling for a target distribution satisfying an invariance property.
This framework not only accommodates and sheds light on our construction of HAMS at a more abstract level,
but also facilitates possible further development of irreversible MCMC algorithms.

\vspace{.1in}
\textbf{Importance of rejection.}
Before describing our generalization, it is instructive to discuss a fictitious generalization of Metropolis--Hastings sampling,
which satisfies a reversibility-like condition upon acceptance of a proposal, but in general fails to leave a target density invariant
due to improperness incurred when a proposal is rejected.

Let $\pi(y)$ be a pre-specified probability density function on a space $\mathcal Y$.
By abuse of notation, we allow that $\pi(y)$ be directly a target density $\pi(x)$ in the context of Section~\ref{sec:intro}
or an augmented target density $\pi(x,u)$ with auxiliary variables $u$.
Consider an MCMC algorithm with the following transition kernel given a current value $y_0$.

\textit{A fictitious generalization of Metropolis--Hastings sampling}.\vspace{-.1in}
\begin{itemize}\addtolength{\itemsep}{-.1in}
\item Sample $y^*$ from a (forward) proposal density $Q( \cdot | y_0)$;
\item Set $y_1 = y^*$ with the acceptance probability
\begin{align*}
\tilde \rho (y^*|y_0) = \min\left(1 , \frac{\pi(y^*) Q_b (y_0 | y^*) }{ \pi(y_0) Q(y^* | y_0) } \right),
\end{align*}
or set $y_1 = y_0$ with the remaining probability, where $Q_b (\cdot | y^*)$ is a backward proposal density.
\end{itemize}

Let $\tilde K (y_1 | y_0)$ be the (forward) transition kernel from $y_0$ to $y_1$ for the sampling scheme above.
Then for any $y_1\not=y_0$ (i.e., a proposal is accepted, $y_1=y^*$), it can be easily shown that
$ \tilde K( y_1 | y_0) = Q(y_1| y_0) \tilde\rho(y_1 |y_0)$ and, by a symmetry argument,
\begin{align}
\pi(y_0) \tilde K(y_1 | y_0) = \pi(y_1) \tilde K_b( y_0 | y_1),   \label{eq:irrev}
\end{align}
where  $\tilde K_b( y_0 | y_1) = Q_b(y_0| y_1) \tilde\rho(y_0 |y_1)$. If (\ref{eq:irrev}) were satisfied for $y_1=y_0$ as well (i.e., a proposal is rejected),
then integrating (\ref{eq:irrev}) over $y_0$ would indicate $\int \pi(y_0) \tilde K(y_1 | y_0) \,\dif y_0 = \pi(y_1)$, that is,
the transition kernel $\tilde K$ leaves $\pi(\cdot)$ invariant.
Standard Metropolis--Hastings sampling corresponds to choosing $Q_b = Q$, in which case (\ref{eq:irrev}) holds trivially for $y_1=y_0$
as well as for $y_1\not= y_0$, Such a condition (\ref{eq:irrev}) with $\tilde K_b=\tilde K$ is known as detailed balance or reversibility.
For $Q_b \not= Q$, however, (\ref{eq:irrev}) may not hold for $y_1= y_0$, in spite of the fact that (\ref{eq:irrev}) is satisfied for $y_1 \not=y_0$.
Therefore, the preceding sampling scheme in general fails to leave $\pi(\cdot)$ invariant, for the complication caused by rejection of a proposal.

Our discussion above uses an heuristic interpretation of the transition kernel $\tilde K$ in the case of rejection of a proposal.
The issue is also reflected in the difficulty to obtain a more rigorous justification similar as in \cite{Tierney1994}.
See \cite{Ma2018}, Section 3.3, for a related discussion on a naive approach for constructing irreversible samplers.

\vspace{.1in}
\textbf{Generalized Metropolis--Hastings sampling.}
As motivated by our construction of HAMS algorithms, we propose generalized Metropolis--Hastings sampling
provided that a target density $\pi(y)$ is invariant under an orthogonal transformation.
Let $J$ be an orthogonal matrix $J$ such that $\pi(J^{-1} y) = \pi(y)$ for $y\in\mathcal Y$. By the change of variables with $|\det(J)|=1$,
this is equivalent to requiring that for any set $C \subset \mathcal Y$,
\begin{align}
\int_{J(C)} \pi(y) \,\dif y = \int_C \pi(y)\,\dif y . \label{eq:invariance}
\end{align}
where $J(C) = \{Jy: y \in C\} \subset \mathcal Y$.
Consider a sampling algorithm defined by the following transition kernel given a current value $y_0$.

\textit{Generalized Metropolis--Hastings sampling} (GMH).
\vspace{-.1in}
\begin{itemize}\addtolength{\itemsep}{-.1in}
\item Sample $y^*$ from a (forward) proposal density $Q( \cdot | y_0)$.
\item Set $y_1 = y^*$ with the acceptance probability
\begin{align}
\rho (y^*|y_0) = \min\left(1 , \frac{\pi(J^{-1} y^*) Q (Jy_0 | J^{-1} y^*) }{ \pi(y_0) Q(y^* | y_0) } \right), \label{eq:gMH-prob}
\end{align}
or set $y_1 = J y_0$ with the remaining probability.
\end{itemize}
Condition (\ref{eq:invariance}) is trivially satisfied for $J=I$ (the identity matrix), in which case the preceding algorithm reduces to standard Metropolis--Hastings sampling.

There are two notable differences compared with the fictitious generalization earlier.
First, the backward proposal density is explicitly defined as $Q (Jy_0 | J^{-1} y^*)$.
It is helpful to think of the proposal density $Q(y^* | y_0)$ as being induced by a stochastic mapping, $y^* = \mathcal M( y_0; Z)$ for a noise $Z$.
Then $Q (Jy_0 | J^{-1} y^*)$ corresponds to
the density of $J y_0$ given $J^{-1}y^*$ under the same mapping, $ J y_0= \mathcal M( J^{-1} y^*; Z^*)$, but with a new noise $Z^*$ considered to be identically distributed as $Z$.
See for example (\ref{eq:g2ms-f})--(\ref{eq:g2ms-b}) below.
Hence the forward and backward transitions of the proposals can be illustrated, similarly to (\ref{eq:transition-diagram}), as
\begin{align*}
y_0 \stackrel{Z}{\longrightarrow} y^*, \quad J^{-1} y^* \stackrel{Z^*}{\longrightarrow} J y_0, %\label{eq:transition-diagram2}
\end{align*}
where the two arrows denote the same mapping, depending on $Z$ or $Z^*$.
Second, the next variable $y_1$ is defined as $J y_0$ instead of $y_0$, in the case of rejection.
The generalization can be shown to be valid in leaving the target distribution $\pi(y)$ invariant.

\begin{prop} \label{prop:gMH}
Suppose that invariance (\ref{eq:invariance}) is satisfied. Let $K (y_1 | y_0)$ be the (forward) transition kernel from $y_0$ to $y_1$ for generalized Metropolis--Hastings sampling.
Then generalized detailed balance holds for any $y_1\not= J y_0$:
\begin{align}
\pi(y_0) K(y_1 | y_0) = \pi(J^{-1} y_1) K ( J y_0 | J^{-1} y_1),  \label{eq:gMH-gDB}
\end{align}
Moreover, the target density $\pi(y)$ is a stationary density of the Markov chain defined by the transition kernel $K(y_1| y_0)$.
\end{prop}

To connect with HAMS, generalized Metropolis--Hastings sampling is discussed above in terms of continuous variables.
However, our framework can be broadened to accommodate both continuous and discrete variables, by allowing $J y$ to be
an orthogonal-like mapping, for example, flipping a binary variable from one value to the other.
In the Supplement, we show that the irreversible jump sampler (I-Jump) in \cite{Ma2018} can be obtained as a special
case of generalized Metropolis--Hastings sampling with a symmetric, binary auxiliary variable.
Hence our HAMS algorithm differs from I-Jump in using momentum as an auxiliary variable,
and exploiting symmetry of mean-zero normal distributions.

\vspace{.1in}
\textbf{Generalized gradient-guided Metropolis sampling.}
The framework of generalized Metropolis--Hastings sampling allows a flexible specification of the proposal density $Q$.
Our HAMS algorithms use a proposal scheme which takes a gradient step and then adds Gaussian noises.
Using a similar update scheme, (\ref{eq:g2ms-f}) below, in generalized Metropolis--Hastings sampling leads to a class of gradient-guided sampling algorithms.
%provided that invariance (\ref{eq:invariance}) holds for an orthogonal matrix $J$.
Similarly as in Section~\ref{subsec:aux}, let $\bo \le A \le 2I$ be a symmetric matrix
in the order on positive semi-definite matrices. For a target $\pi(y)$, a potential function $U(y)$ is defined such that $\pi(y) \propto \exp\{-U(y)\}$.
This potential $U(y)$ can be the augmented potential $U(x) + u^\T u/2$ in Section~\ref{sec:proposed}.

\textit{Generalized gradient-guided Metropolis sampling} (G2MS).
\vspace{-.1in}
\begin{itemize}\addtolength{\itemsep}{-.1in}
\item Generate $y^*$ as \vspace{-.2in}
\begin{align}
y^* = y_0 - B \nabla U(y_0) + Z, \quad Z\sim \mathcal N(\bo, 2 A - A^2), \label{eq:g2ms-f}
\end{align}
where $B=I - (I-A)J $ and $2 A - A^2 = B+B^\T - B B^\T$. Compute $Z^*$ by \vspace{-.1in}
\begin{align}
J y_0 = J^{-1} y^* -B\nabla U(J^{-1}y^*)+ Z^*, \label{eq:g2ms-b}
\end{align}
obtained by replacing $(y_0,y^*)$ with $(J^{-1}y^*, Jy_0)$ and $Z$ with $Z^*$ in (\ref{eq:g2ms-f}).
\item Set $y_1 = y^*$ with the acceptance probability (\ref{eq:gMH-prob}), simplified as
\begin{align}
\rho (y^*|y_0) = \min\left(1 , \frac{\pi(y^*) \mathcal N(Z^* | \bo, 2A-A^2) }{ \pi(y_0) \mathcal N(Z| \bo, 2A-A^2) } \right), \label{eq:g2ms-prob}
\end{align}
or set $y_1 = J y_0$ with the remaining probability.
\end{itemize}

\begin{cor} \label{cor:g2ms}
Suppose that invariance (\ref{eq:invariance}) is satisfied. The conclusions of Proposition~\ref{prop:gMH} hold with transition kernel $K$ defined by generalized gradient-guided Metropolis sampling.
\end{cor}

In addition to exploiting gradient information,  the G2MS algorithm is carefully designed to achieve the rejection-free property when
the target density $\pi(y)$ is $\mathcal N(\bo, I)$, which satisfies invariance (\ref{eq:invariance}) for any orthogonal matrix $J$. In this case,
$U(y) = y^\T y/2$ with gradient $\nabla U(y) = y$, and hence the proposal scheme (\ref{eq:g2ms-f}) becomes
\begin{align}
    y^* = (I- A)J y_0 + Z, \quad Z\sim \mathcal N(\bo,2A - A^2 ) . \label{eq:g2ms-f2}
\end{align}
The update from $y_0$ to $y^*$ defines a VAR(1) process, which admits $\mathcal N(\bo,I)$ as a stationary distribution,
that is, if $y_0 \sim \mathcal N(\bo, I)$ then $y^* \sim \mathcal N(\bo,I)$, by similar calculation as in (\ref{eq:var-calc}).
However, stationarity of (\ref{eq:g2ms-f2}) with respect to $\mathcal N(\bo,I)$ does not automatically imply rejection-free.
In fact, because $(I-A)J$ may be asymmetric, the VAR(1) process in (\ref{eq:g2ms-f2}) is in general irreversible.
Standard Metropolis--Hastings sampling using the proposal scheme (\ref{eq:g2ms-f2}) is not rejection-free when $\pi(y)$ is $\mathcal N(\bo,I)$.
Otherwise, the resulting Markov chain is irreversible, which contradicts reversibility of standard Metropolis--Hastings sampling.
Nevertheless, the G2MS algorithm achieves rejection-free when $\pi(y)$ is $\mathcal N(\bo,I)$, due to the combination of the proposal scheme (\ref{eq:g2ms-f2}) with the generalized
acceptance probability (\ref{eq:g2ms-prob}).
In other words, the backward proposal density induced from (\ref{eq:g2ms-b})
agrees with the conditional density of $y_0$ given $y^*$ if $y_0\sim  \mathcal N(\bo, I)$ and $y^*$ is generated  by (\ref{eq:g2ms-f2}). See the proof for details.

\begin{cor} \label{cor:rejection-free2}
 Suppose that the target density $\pi(y)$ is  $\mathcal N(\bo,I)$. Then the generalized acceptance probability (\ref{eq:g2ms-prob})
 reduces to 1, and hence $y^*$ from the proposal scheme (\ref{eq:g2ms-f}) is always accepted under the G2MS algorithm.
\end{cor}

From the preceding discussion, the G2MS algorithm can be seen as being extended from a VAR(1) process in the form (\ref{eq:g2ms-f2}).
For completeness, we remark that the form of (\ref{eq:g2ms-f2}) depending on $A$ and $J$ is universal. In fact, consider a general VAR(1) process
\begin{align}
    y^* = (I- \tilde B) y_0 + Z, \quad Z\sim \mathcal N(\bo, \tilde B + \tilde B^\T - \tilde B \tilde B^\T ) , \label{eq:g2ms-f3}
\end{align}
where $\tilde B$ is a possibly asymmetric matrix such that $\tilde B + \tilde B^\T - \tilde B \tilde B^\T$ is (symmetric and) positive semi-definite.
Let $I-\tilde B = O_1 \Lambda O_2$ be a singular value decomposition, where $O_1$ and $O_2$ are orthogonal matrices, $\Lambda$ is a diagonal matrix containing the singular values of $I-\tilde B$.
Then $I - \tilde B$ can be written as
\begin{align*}
I - \tilde B = (O_1 \Lambda O_1^\T ) (O_1 O_2) = (I - \tilde A) \tilde J,
\end{align*}
where $\tilde A = I - O_1 \Lambda O_1^\T$ is symmetric and $\tilde J =O_1 O_2$ is orthogonal.
Moreover, the noise variance becomes $\tilde B + \tilde B^\T - \tilde B \tilde B^\T = I- (1-\tilde B)(I-\tilde B)^\T = I - (I-\tilde A)^2 = 2\tilde A - \tilde A^2$.
Therefore, the VAR(1) process (\ref{eq:g2ms-f3}) can be put in the form (\ref{eq:g2ms-f2}).

\vspace{.1in}
\textbf{Back to HAMS.}
The invariance (\ref{eq:invariance}) can be satisfied by an augmented target density defined with auxiliary variables.
In fact, our HAMS algorithms can be recovered as special cases of generalized Metropolis--Hastings sampling,
with $\pi(y)=\pi(x,u)$ in (\ref{eq:aug-density2}) and $J$ a block-diagonal matrix with $(I,-I)$ on the diagonal.
The invariance (\ref{eq:invariance}) is satisfied due to evenness of mean-zero normal distributions.
The HAMS algorithm studied in Proposition~\ref{prop1} is a special case of G2MS with the $A$ matrix in (\ref{eq:matrix-A}).
The HAMS algorithm in Proposition~\ref{prop1-b} is not contained in G2MS due to a modification with $\phi\not=0$,
but can still be treated in the framework of generalized Metropolis--Hastings sampling,
with the forward and backward proposal schemes discussed in Section~\ref{subsec:general}.
The general discussion here broadens our understanding of HAMS algorithms and opens doors for further development.

%%%%%%%%%%%%%%%%%%%%%%%%%%%%%%%%%%%%%%%%%%%%%%%%%%%%%%%%%%%%%%%%%%%%%%%%%%%%%%%%%%%%%%%%%%
%%%---------------------------------------------------------------------------------------
%%%%%%%%%%%%%%%%%%%%%%%%%%%%%%%%%%%%%%%%%%%%%%%%%%%%%%%%%%%%%%%%%%%%%%%%%%%%%%%%%%%%%%%%%%

\section{Simulation Studies}
\label{sec:sim}

We report simulation studies comparing HAMS-A/B with RWM, pMALA, pMALA*, HMC, UDL, and GMC
(see Section~\ref{sec:review}). We include RWM as a performance baseline.
The simulations include a multivariate normal distribution, a
stochastic volatility model and a log-Gaussian Cox model.
For space limitation, the normal experiment and results from pMALA* and GMC in the other two experiments are deferred to the Supplement.

For ease of comparison and tuning,  we use the $(\epsilon, c)$ parameterization for
HAMS-A and HAMS-B, equivalent to the $(a,b)$ parametrization by (\ref{eq:abepsc}). We fix the number of leapfrog steps for HMC  similarly as in \cite{Girolami2011}:
$nleap=50$ in sampling latent variables or $nleap=6$ in sampling parameters.
When preconditioning is applied, the $c$ values for HAMS-A/B as well as UDL and GMC
are determined in terms of $\epsilon$, by translating the default choices of $b$ given $a$ in (\ref{eq:b-choice}).
Without preconditioning, the $c$ values are specified by the following consideration.
Recall that the first momentum update of UDL is $u^+ = \sqrt{c} u_0 + \sqrt{1-c}Z_1$ in the form of an AR(1) process.
With a standard normal noise, the lag-$h$ auto-covariance for AR(1) is $\gamma(h) = c^{h/2}$.
To match resampling of momentum in HMC, we require $c= \gamma(h)^{2/h}$ with $h = nleap$ and a small value, $0.001$, for $\gamma(h)$.
Hence we set $c=0.76$ or $0.1$ corresponding to $nleap=50$ or $6$.

For tuning, we adjust step size $\epsilon$ during a burn-in period to achieve reasonable acceptance rates: around $30\%$ for RWM and $70\%$ for all other methods.
See the Supplement Section~\ref{subsec:tuning} for details. Samples are then collected after the burn-in.

To evaluate MCMC samples, a useful metric is the effective sample size,
$\mbox{ESS} = n /\{ 1+2\sum_{k=1}^{\infty}\rho(k)\}$,
where $n$ is the total number of draws and $\rho(k)$ is the lag-$k$ correlation.
To deal with irreversible Markov chains obtained by HAMS-A/B as well as UDL, we use the Bartlett window estimator of ESS similarly as in \cite{Ma2018}: \vspace{-.05in}
\begin{equation} \label{eq:ess}
\mbox{ESS} = \frac{n}{1+2\sum_{k=1}^K \left(1-\frac{k}{K}\right)\rho(k)},
\end{equation}
where the cutoff value $K$ is a large number (taken to be $3000$ in our results).
Moreover, ESS can be estimated from each coordinate for a multi-dimensional distribution.
As suggested in \cite{Girolami2011}, we report the minimum ESS over all coordinates, adjusted by runtime,
as a measure of computational efficiency.

%----------------------------------------------------------------------------------

\subsection{Stochastic volatility model}
\label{subsec:sim2}

Consider a stochastic volatility model \citep{Kim1998}, where latent volatilities are generated as
\begin{equation}
  \label{eq:sveq1}
x_t = \phi x_{t-1} + \eta_t, \quad \eta_t \sim\mathcal N(0,\sigma^2),\quad t= 2,3,...,T ,
\end{equation}
with $x_1 \sim\mathcal N(0,\sigma^2/(1-\phi^2))$, and the observations are generated as
\begin{equation}
  \label{eq:sveq2}
  y_t = z_t \beta \exp\{x_t/2\},\quad z_t\sim\mathcal N(0,1),\qquad t = 1,...,T.
\end{equation}
The parameters of interest are $\theta = (\beta,\sigma,\phi)^\T $. We simulate  $T = 1000$ observations from (\ref{eq:sveq1})--(\ref{eq:sveq2})
using parameter values $\beta = 0.65, \sigma = 0.15$ and $\phi = 0.98$. Let $\bx = (x_1,...,x_T)^\T $ and $ \by = (y_1,...,y_T)^\T $.
Two sets of experiments are conducted. First, we fix parameter values and sample latent variables
from $p(\bx|\by, \theta)$. Then we perform Bayesian analysis and sample  both the parameters and latent
variables from $p(\by|\bx,\theta)$. See Supplement Section  \ref{subsec:SVexpression} for expressions of gradients
and preconditioning matrices used.

For the first experiment, we fix parameters at their true values and perform sampling for latent variables only. The joint distribution of
$(x_1,\ldots,x_T)$ is $\mathcal N(\bo,C)$, with entries of the covariance matrix given by $C [i,j] = \phi^{|i-j|}\sigma^2/(1-\phi^2)$.
Its inverse $C^{-1}$ retains a simple tri-diagonal form. Following \cite{Girolami2011}, the
inverse variance $[\Var(\bx)]^{-1}$ can be approximated by $-\EE[\nabla^{2}\log p(\bx|\by, \theta)] = C^{-1} + \frac{1}{2}I$.
Hence for preconditioning, we set $M = C^{-1} + \frac{1}{2}I$ for HAMS-A/B, UDL and HMC,
and $\Sigma = M^{-1}$ for pMALA and RWM. As mentioned earlier, we use $nleap=50$ for HMC and choose $c$ given $\epsilon$ by (\ref{eq:abepsc}).
All algorithms are run for $5000$ burn-in
iterations, and then samples are collected from $5000$ iterations. The simulation process is repeated for $50$ times.

\begin{table}[t] % [H]
  \caption{Runtime and ESS comparison for sampling latent variables in the stochastic volatility model.
   Results are averaged over $50$ repetitions.  \label{tab:tabone}}
  \centering
  \begin{tabular}{|cccc|}
  \hline
  Method & Time (s) & \begin{tabular}[c]{@{}c@{}}ESS\\[-.1in] (min, median, max)\end{tabular} & $\frac{\mbox{minESS}}{\mbox{Time}}$ \\ \hline
  HAMS-A  & 98.7     & (2420, 3660, 6668)                                               & 24.51        \\
  HAMS-B  & 99.6     & (1915, 3404, 6229)                                            & 19.23      \\
  UDL    & 98.4     & (657, 1020, 1661)                                               & 6.68        \\
  HMC    & 1250.1   & (1125, 3698, 11240)                                                & 0.90         \\
  pMALA  & 120.5    & (374, 610, 990)                                                   & 3.11        \\
  RWM    & 51.7     & (7, 12, 20)                                                      & 0.14         \\ \hline
\end{tabular}
\end{table}

\begin{figure}[t] %[H]
  \begin{center}
  \includegraphics[width = 0.95\textwidth]{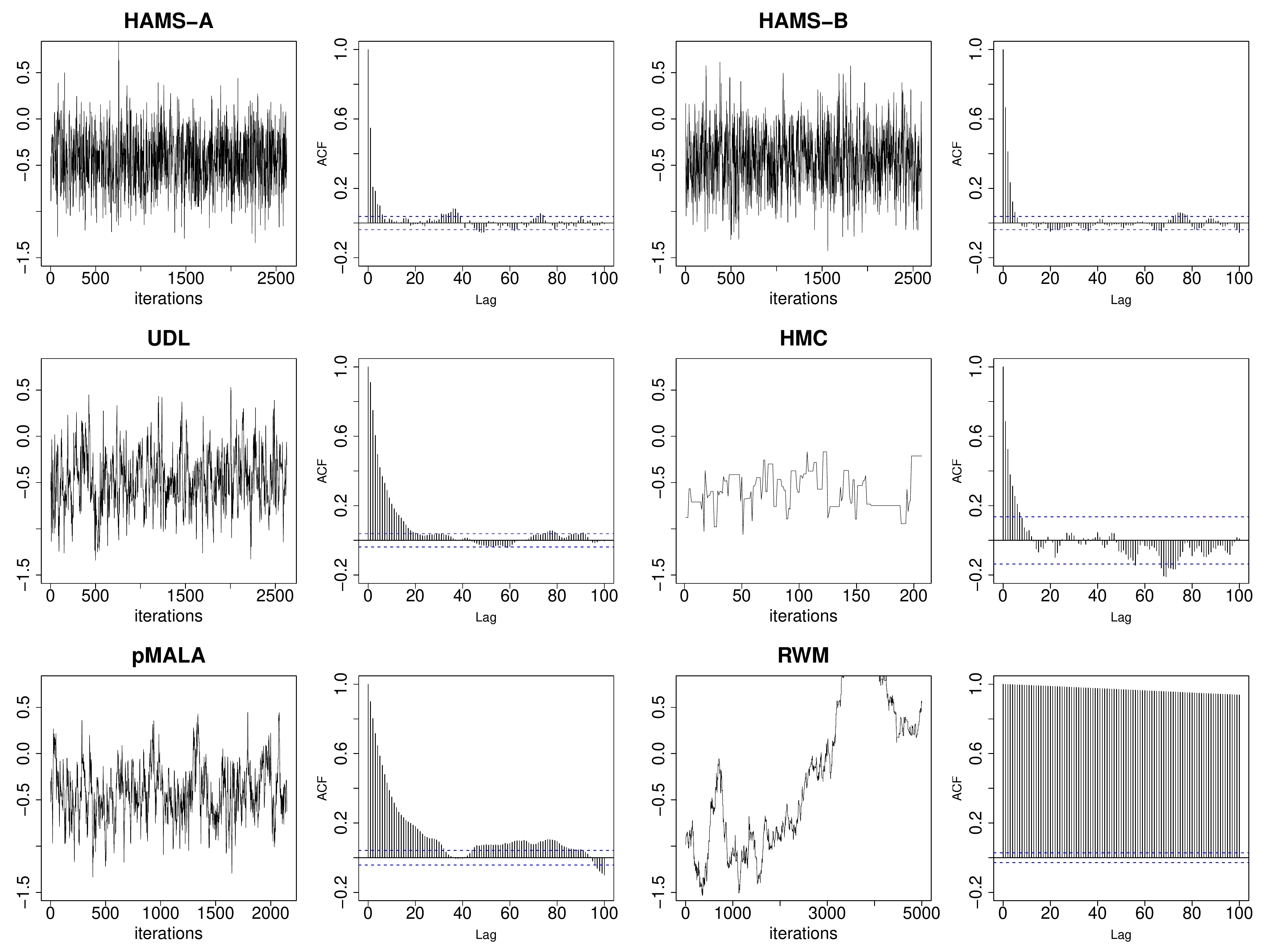}
  \end{center}
  \caption{Time-adjusted trace and ACF plots of one latent variable from an individual run for sampling latent variables in the stochastic volatility model. \vspace{-.1in}
  \label{fig:svtrace}}
\end{figure}

Table \ref{tab:tabone} shows the runtime and ESS comparison. Clearly, HAMS-A has the best performance in terms
of time-adjusted minimum ESS, followed by HAMS-B.
An interesting phenomenon about the ESSs from HAMS-A/B as well as HMC is that
an ESS value estimated by (\ref{eq:ess}) can exceed the actual number of draws collected, due to negative auto-correlations.
Figure \ref{fig:svtrace} shows trace plots of one latent variable
and corresponding autocorrelation function (ACF) plots from an individual run. The plots for each method are adjusted for runtime after burn-in:
we keep the number of draws inversely proportional to the runtime, with RWM keeping all $5000$
draws as the baseline. All time-adjusted plots are produced similarly in this and next sections. From the
trace plots, HAMS-A and HAMS-B appear to mix better than other methods. Moreover, the ACFs of HAMS-A and HAMS-B decay faster to 0 compared with other methods,
while exhibiting negative auto-correlations.

\begin{figure}[t] % [H]
  \begin{center}
  \includegraphics[width = 0.95\textwidth]{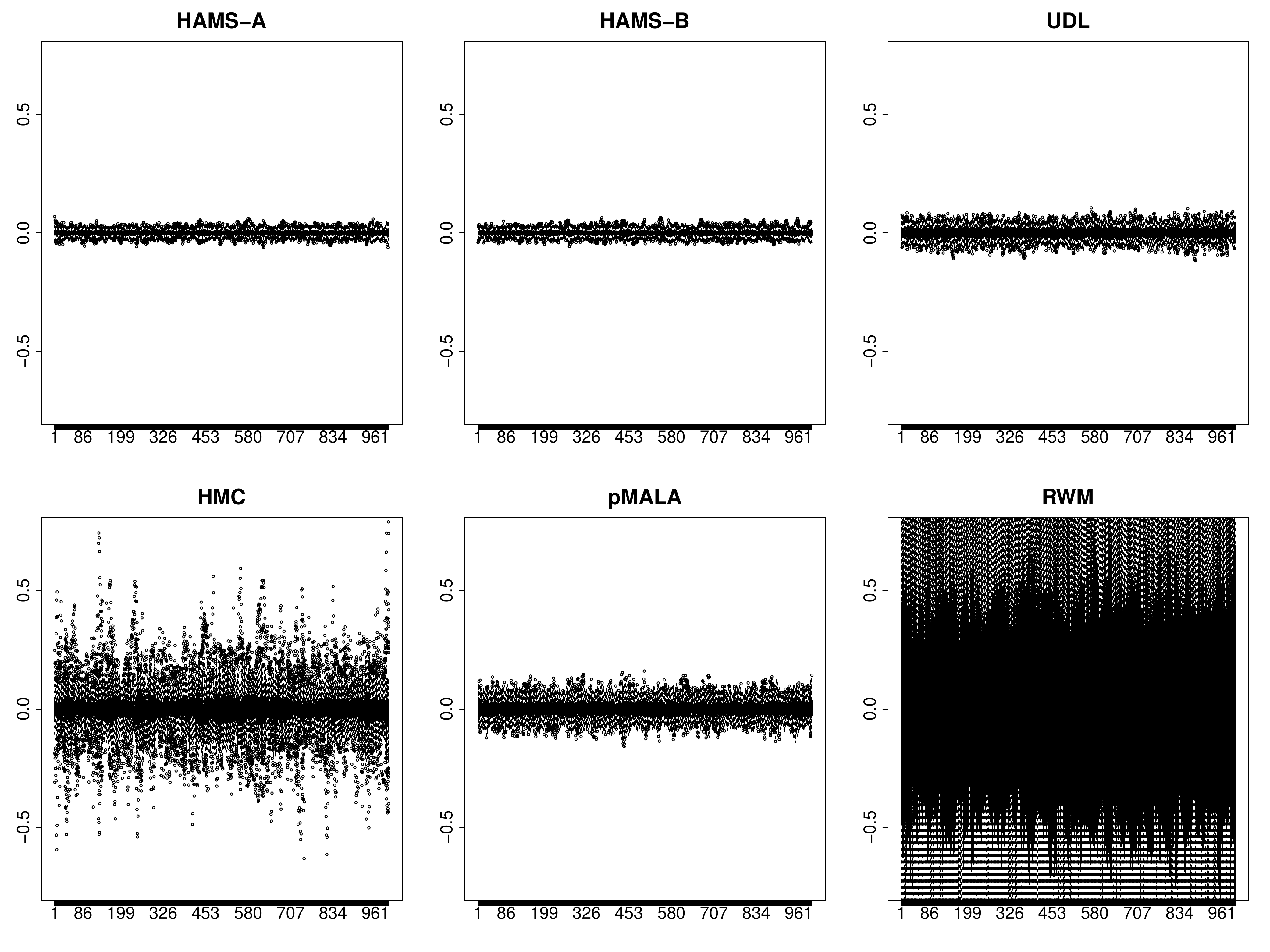}
  \end{center} \vspace{-.2in}
  \caption{Time-adjusted and centered boxplots of sample means of all latent variables over 50 repetitions for sampling latent variables in the stochastic volatility model.
  \label{fig:svbox}}
\end{figure}

Figure \ref{fig:svbox} shows the time-adjusted boxplots of the sample means of all latent variables for each method over $50$ repeated runs.
The boxplots are centered at the corresponding averages, and narrower boxplots indicate that a method is more consistent
across repeated simulations. Clearly, HAMS-A and HAMS-B are the most consistent, followed by UDL and pMALA.\
Much more variability is associated with HMC and RWM.

The superior performances of HAMS-A/B can be attributed to the fact that larger step sizes are used by HAMS-A/B than other methods, while
similar acceptance rates are obtained. See the Supplement Figure~\ref{fig:suppsv1}.
A possible explanation for the step size differences is that HAMS-A/B satisfies the rejection-free property and hence is more capable
of achieving reasonable acceptance rates with relatively large step sizes when the target density is not far from a normal density through preconditioning.

In the second experiment, we perform Bayesian analysis and sample both latent variables and parameters from the posterior $p(\bx,\theta|\by)$.
The priors are, independently, $\pi(\beta)\propto \beta^{-1}$, $\sigma^2 \sim \mbox{Inv-}\chi^2(10,0.05)$ and $(\phi+1)/2\sim \mbox{Beta}(20,1.5)$.
Moreover, we use the transformations $\sigma = \exp(\gamma)$ and $\phi = \tanh(\alpha)$ to ensure that $\sigma >0$ and $|\phi|<1$.
We employ a Gibbs-sampling scheme, alternating between $p(\bx|\by,\theta)$ and $p(\theta|\by,\bx)$, similarly as in \cite{Girolami2011}.
In the first experiment, the preconditioning matrix for latent variables needs to be computed only once because the parameters are fixed.
In the current experiment, to avoid re-evaluating the preconditioning matrix every Gibbs iteration, we first run each algorithm
without any preconditioning to obtain a crude estimate of the parameters, and then fix the preconditioning
matrix evaluated at this estimate. For HMC, the numbers of leapfrog steps are $50$ for latent variables and $6$ for parameters.
The initial values of parameters are dispersed over the following intervals $\beta\in[0.5,2]$, $\sigma\in[0.1,1]$, and $\phi\in[0,0.3]$.
For all methods, $10000$ draws are collected after a burn-in of
$10000$ iterations, which include two stages without preconditioning and one stage of tuning with preconditioning.
The simulation process is repeated for $20$ times.

Table \ref{tab:tabtwo} shows the results of posterior sampling. Except for RWM, the methods yield similar averages of sample means of the parameters.
However, HAMS-A and HAMS-B produce smaller standard deviations of sample means than the remaining methods, except that
pMALA gives a smaller standard deviation of sample means in $\sigma$, although substantially lower ESSs in all the parameters than HAMS-A/B.
In fact, HAMS-A and HAMS-B clearly outperform the other methods in terms of ESSs in all three parameters.

\begin{figure}[h]  %[H]
\centering
\begin{subfigure}{.49\linewidth}
    \centering
    \includegraphics[width = \textwidth]{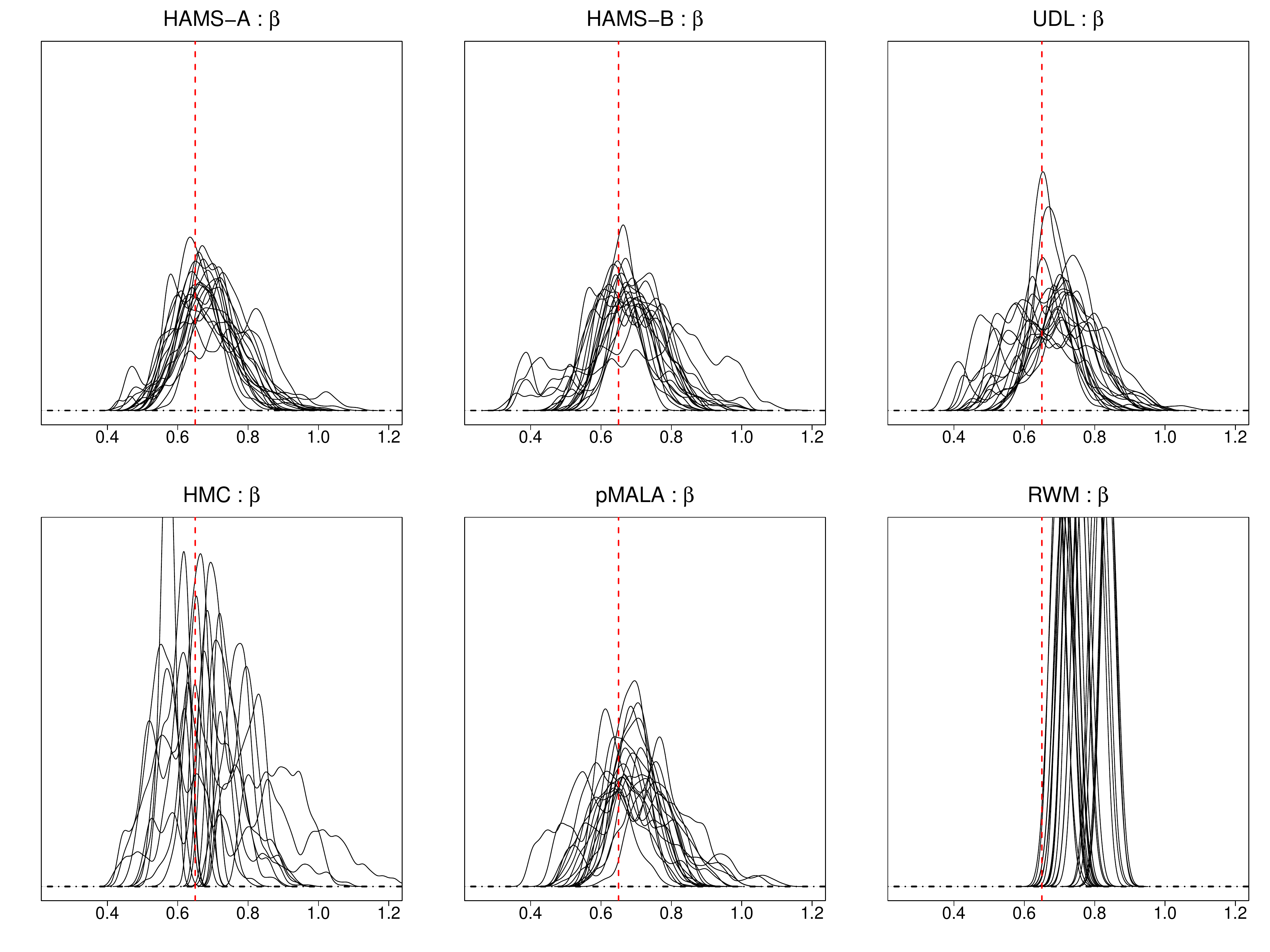}
    \caption{Densities of $\beta$}\label{fig:image1-SVD}
\end{subfigure}
    \hfill
\begin{subfigure}{.49\linewidth}
    \centering
    \includegraphics[width = \textwidth]{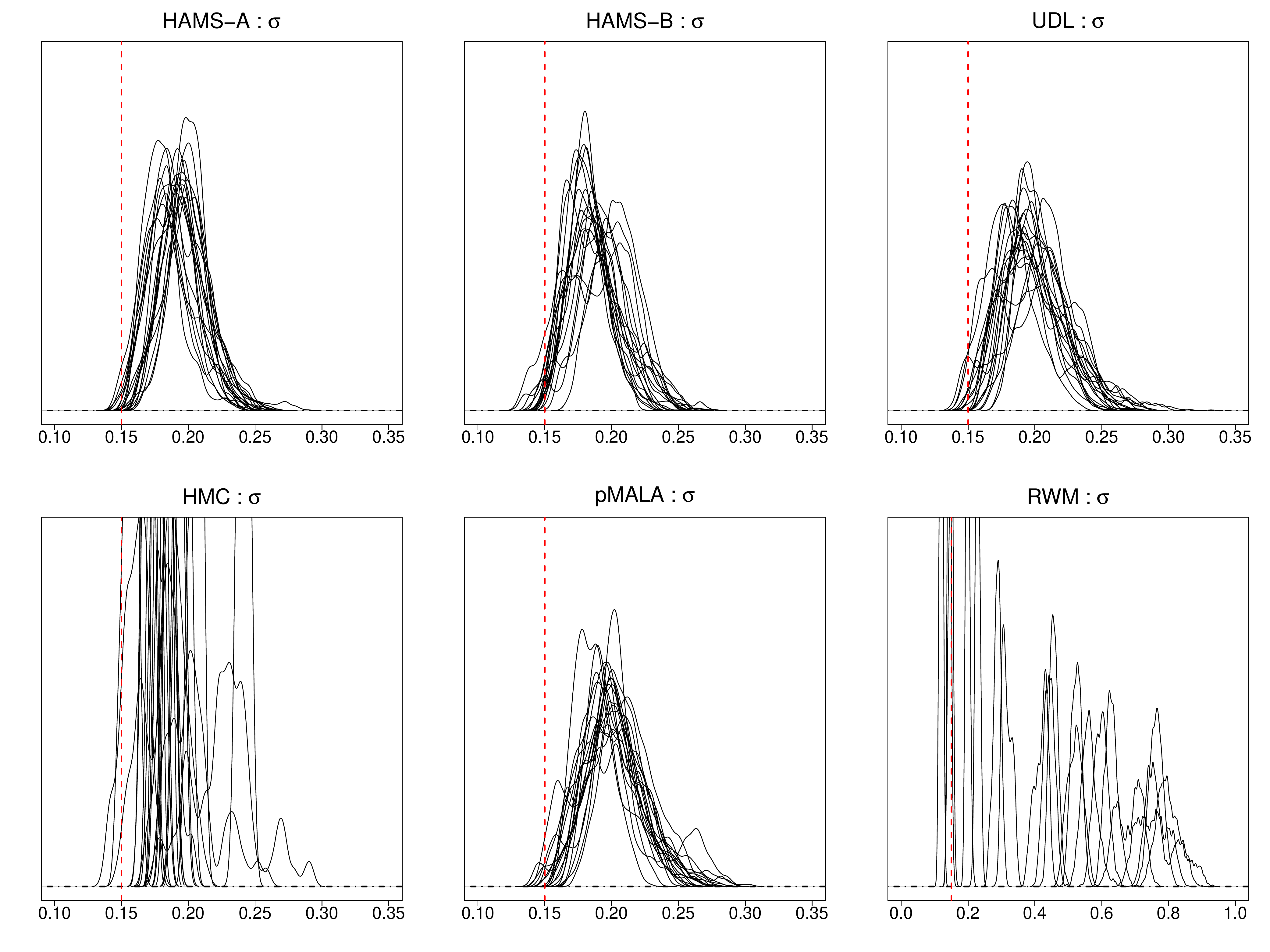}
    \caption{Densities of $\sigma$}\label{fig:image2-SVD}
\end{subfigure}

% \bigskip  %%%%<<<---
\begin{subfigure}{\linewidth}
  \centering
  \includegraphics[width = 0.49\textwidth]{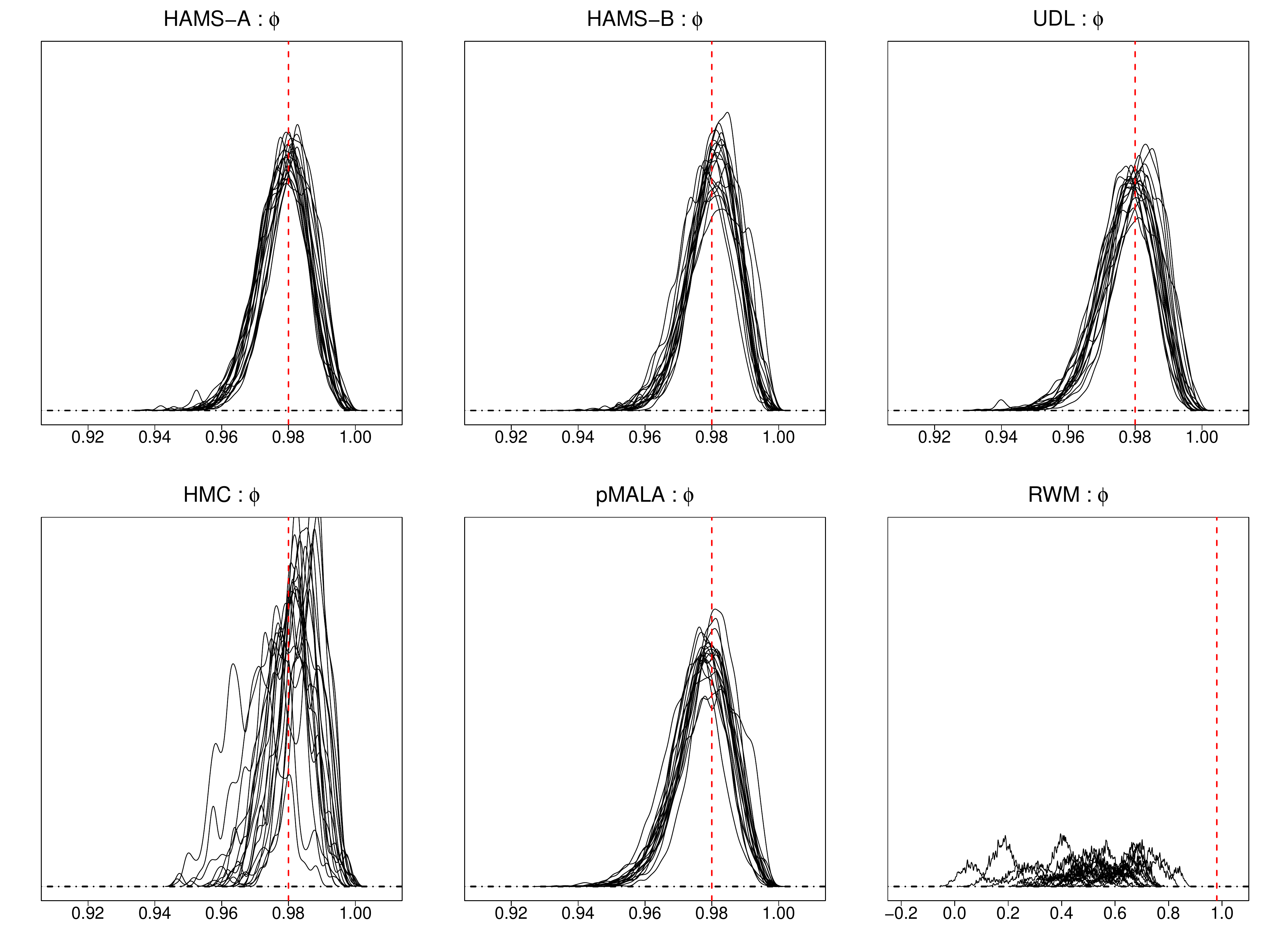}
  \caption{Densities of $\phi$}\label{fig:image3-SVD}
\end{subfigure}  \vspace{-.2in}  %%%%<<<---
\RawCaption{\caption{Time-adjusted posterior density plots of parameters ($20$ repetitions overlaid) in the stochastic volatility model. The true parameter
values are marked by vertical lines.}
\label{fig:svdensity}}
\end{figure}

Figure \ref{fig:svdensity} shows time-adjusted density plots for the parameters. Each plot
shows densities from $20$ repeated runs overlaid together.
Clearly, HAMS-A yields the most consistent density curves for all three parameters, followed by HAMS-B, UDL, and pMALA which sometimes produce outlying curves,
especially in $\beta$ and $\sigma$.

\begin{table}[t] %[H]
  \caption{Comparison of posterior sampling in the stochastic volatility model. Standard deviations of sample means are in parentheses.
  Results are averaged over $20$ repetitions.  \label{tab:tabtwo}} \vspace{-.2in}  %%%%<<<---
  \centering
  \begin{tabular}{|ccccccc|}
  \hline
  Method & \multicolumn{1}{c|}{Time (s)} & \begin{tabular}[c]{@{}c@{}}\ \\[-.1in] $\beta$ (sd)\end{tabular} & \begin{tabular}[c]{@{}c@{}}Sample Mean\\[-.1in] $\sigma$ (sd)\end{tabular} & \multicolumn{1}{c|}{\begin{tabular}[c]{@{}c@{}}\ \\[-.1in] $\phi$ (sd)\end{tabular}} & \begin{tabular}[c]{@{}c@{}}ESS\\[-.1in] ($\beta,\sigma,\phi$)\end{tabular} & $\frac{\mbox{minESS}}{\mbox{Time}}$ \\ \hline
  HAMS-A  & 1951.3                        & 0.68 (0.034)                                                          & 0.19 (0.006)                                                      & 0.98 (0.001)                                                                              & (30, 73, 220)                                                 & 0.015      \\
  HAMS-B  & 1942.3                        & 0.68 (0.037)                                                          & 0.19 (0.007)                                                      & 0.98 (0.001)                                                                              & (25, 59, 188)                                                 & 0.013      \\
  UDL    & 1945.8                        & 0.68 (0.039)                                                          & 0.20 (0.008)                                                      & 0.98 (0.002)                                                                              & (29, 37, 87)                                                 & 0.015      \\
  HMC    & 20920.2                       & 0.69 (0.050)                                                          & 0.19 (0.014)                                                      & 0.98 (0.003)                                                                              & (19, 12, 78)                                                 & 0.001      \\
  pMALA  & 2013.0                        & 0.68 (0.040)                                                          & 0.20 (0.005)                                                      & 0.98 (0.001)                                                                              & (15, 30, 76)                                                  & 0.008      \\
  RWM    & 1311.1                        & 0.76 (0.050)                                                          & 0.47 (0.229)                                                      & 0.51 (0.149)                                                                              & (89, 12, 7)                                                   & 0.006      \\ \hline
  \end{tabular} \vspace{-.2in}  %%%%<<<---
\end{table}

\clearpage

%----------------------------------------------------------------------------------
\subsection{Log-Gaussian Cox model}
\label{subsec:sim3}

\begin{table}[t] %[H]
  \caption{Runtime and ESS comparison for sampling latent variables in the log-Gaussian Cox model ($n=1024$).
   Results are averaged over $50$ repetitions.  \label{tab:tabthree}}
  \centering
  \begin{tabular}{|cccc|}
  \hline
  Method & Time (s) & \begin{tabular}[c]{@{}c@{}}ESS\\[-.1in] (min, median, max)\end{tabular} & $\frac{\mbox{minESS}}{\mbox{Time}}$ \\ \hline
  HAMS-A  & 81.0    & (803, 1655, 5461)                                                 & 9.91         \\
  HAMS-B  & 78.8    & (619, 1376, 4831)                                              & 7.86        \\
  UDL    & 78.8    & (322, 622, 1761)                                                & 4.08         \\
  HMC    & 1285.9   & (935, 1621, 4523)                                              & 0.73         \\
  pMALA  & 116.4    & (184, 340, 1002)                                                 & 1.58         \\
  RWM    & 51.1     & (8, 13, 22)                                                      & 0.16        \\ \hline
  \end{tabular}
\end{table}

Consider a log-Gaussian Cox model, where the latent variables $\bx = (x_{ij})_{i,j = 1,...,m}$ are associated with an $m\times m$ grid
\citep{Christensen2005,Girolami2011}.
Assume that $x_{ij}$'s are normal with means $0$ and a covariance function
$ C[(i,j),(i',j')] = \sigma^2\exp(-\sqrt{(i-i')^2 + (j - j')^2}/(m\beta))$. By abuse of notation, we denote
$\bx \sim \mathcal N(\bo, C)$, of dimension $n=m^2$. The observations $(y_{ij})_{i,j = 1,...,m}$
are independently Poisson, where the mean of $y_{ij}$ is
$\lambda_{ij} = n^{-1}\exp(x_{ij} + \mu)$, with $\mu$ treated as known. Hence the unknown parameters are
$\theta = (\sigma^2,\beta)^\T $. Given a prior $\pi(\theta)$, the posterior density is
\begin{equation}
  \label{eq:coxeq1}
  p(\bx,\theta|\by ) \propto \pi(\theta) |\det(C)|^{-1/2} \exp\left\{-\frac{1}{2}x^\T C^{-1}x\right\} \exp\left\{\sum_{i,j}(y_{ij}(x_{ij}+\mu) -\lambda_{ij} )\right\}.
\end{equation}
As in Section~\ref{subsec:sim2}, we conduct two sets of experiments: one is sampling latent variables with fixed parameters, and the other is sampling both parameters and
latent variables.

\begin{figure}[t] %[H]
  \begin{center}
  \includegraphics[width = 0.95\textwidth]{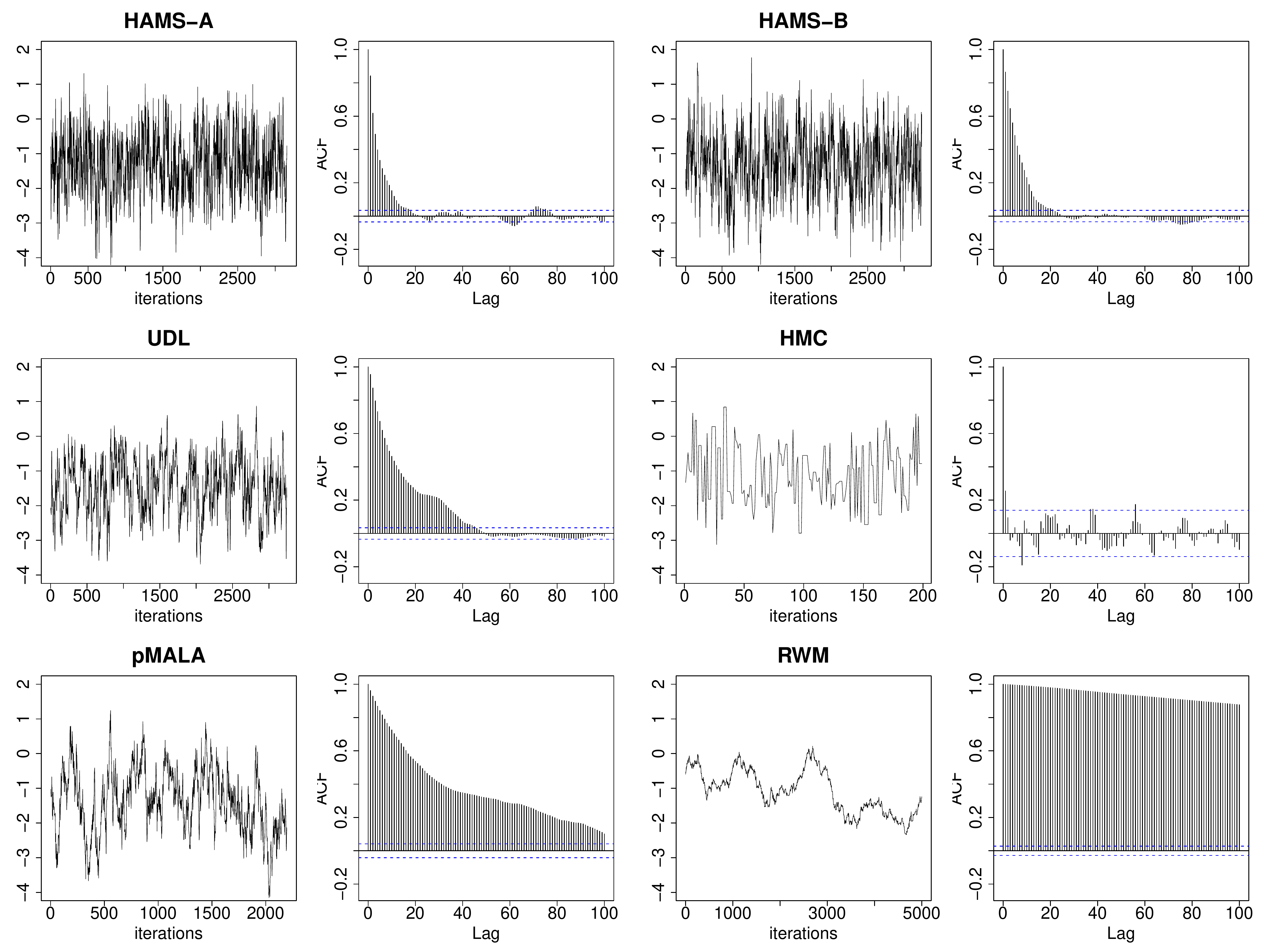}
  \end{center} \vspace{-.2in}
  \caption{Time-adjusted trace and ACF plots of one latent variable from an individual run for sampling latent variables in the log-Gaussian Cox model ($n=1024$).
  \label{fig:coxtrace}} \vspace{-.2in}
\end{figure}

\begin{figure}[t] %[H]
  \begin{center}
  \includegraphics[width = 0.95\textwidth]{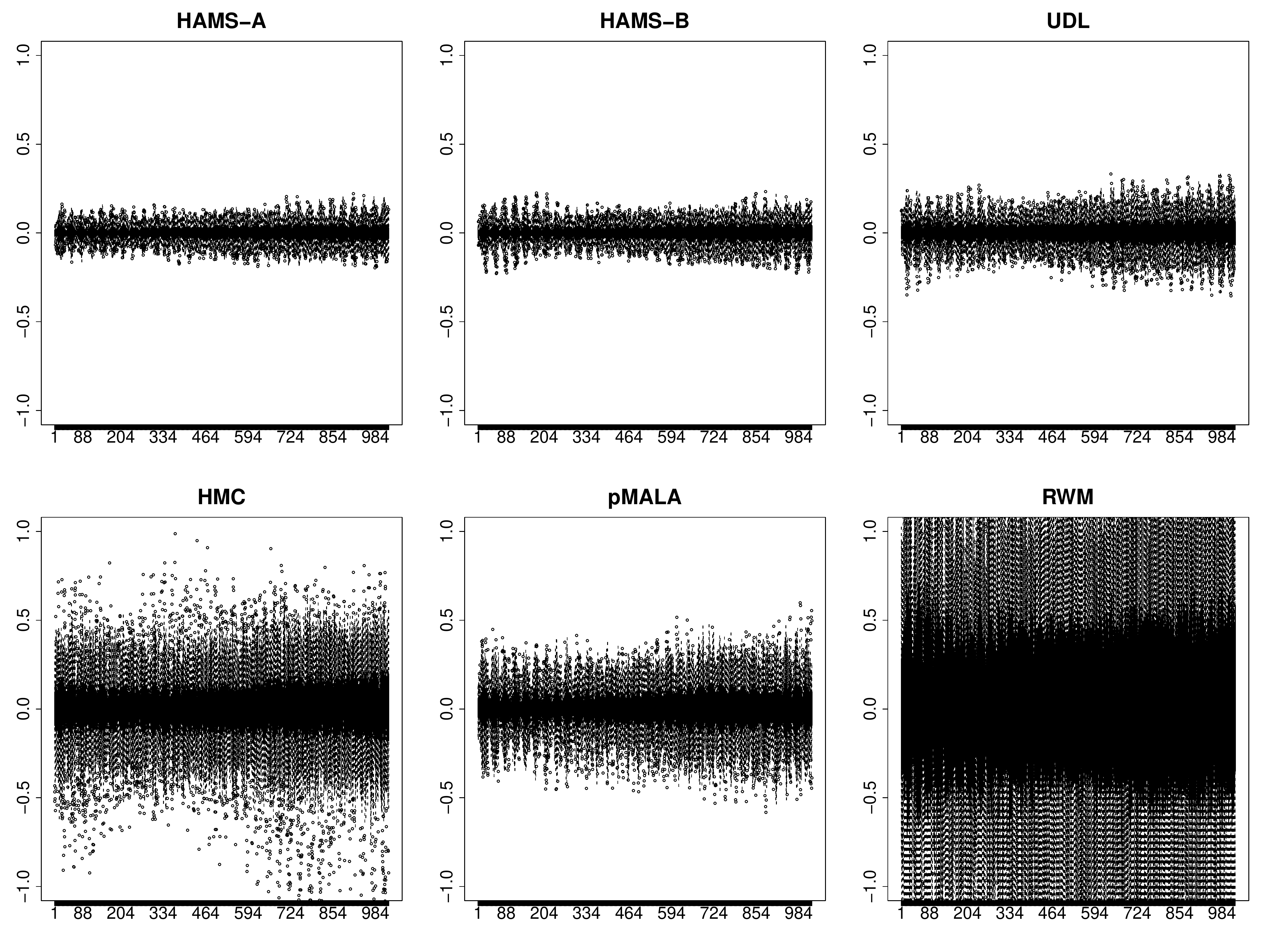}
  \end{center} \vspace{-.2in}
  \caption{Time-adjusted and centered boxplots of sample means of all latent variables for sampling latent variables in the log-Gaussian Cox model ($n=1024$).
  \label{fig:coxbox}}
\end{figure}

For latent variables sampling, we take $m = 32$ and generate $n = 32^2 = 1024$ observations using
the parameter values $\sigma^2 = 1.91, \beta = 0.3$ and $\mu = \log(126) - 0.5(1.91)$.
The example in \cite{Christensen2005} and \cite{Girolami2011} used $\beta = 1/33$. Here we increase $\beta$ to introduce more correlations in $\bx$
which makes the problem more challenging and leads to clearer comparison between different methods.
From (\ref{eq:coxeq1}),  the gradient of the negative log-likelihood is
$\nabla U(\bx) = n^{-1}\exp(\bx + \mu) + C^{-1}\bx - \by$. The expected Hessian is
$\EE[\nabla^{2}U(\bx)] = D + C^{-1}$, taken with respect to the prior of $\bx$, where $D$ is a
diagonal matrix with diagonal elements $n^{-1}\exp(\mu + \frac{1}{2}\sigma^2)$. Hence for
preconditioning, we set $\Sigma^{-1} = M = D + C^{-1}$.
The number of leapfrog steps is $50$ for HMC. For all methods,
$5000$ draws are collected after a burn-in of $5000$. The simulation process is repeated for $50$ times.

Table \ref{tab:tabthree} summarizes runtime and ESSs. Similarly as in Section~\ref{subsec:sim2}, HAMS-A has the best performance in terms of time-adjusted minimum ESS, followed by HAMS-B.
Notice that HMC has large raw ESSs than UDL and pMALA, but its performance is worse after adjusting for runtime.
Figure \ref{fig:coxtrace} shows time-adjusted trace plots of one latent variable and corresponding ACF plots taken from an
individual run. From both plots, HAMS-A and HAMS-B appear to mix better than the other methods.
Figure \ref{fig:coxbox} shows the time-adjusted and centered boxplots of sample means
for each method over $50$ repetitions. The spread of these boxplots corroborate the ESS results:
HAMS-A and HAMS-B are less variable than the remaining methods over repeated simulations.

Similarly as in Section~\ref{subsec:sim2}, the superior performances of HAMS-A/B are related to the rejection-free property of HAMS-A/B, which facilitates
use of relatively large step sizes
while reasonable acceptance rates are obtained. See  Supplement Figure~\ref{fig:suppcox1}.

Our final experiment is sampling both latent variables and parameters for Bayesian analysis of the log-Gaussian Cox model.
Unlike the stochastic volatility model where the inverse of covariance matrix of latent variables admits a
closed-form expression, the matrix $C$ here needs be inverted numerically whenever we evaluate the
density (\ref{eq:coxeq1}) or its gradient. For large $n$, sampling both parameters and latent variables is computationally demanding.
Hence we consider a reduced size $m=16$ and $n=256$. We still simulate observations $\by$ using
the ground truth $\sigma^2 = 1.91, \beta = 0.3$ and $\mu = \log(126) - 0.5(1.91)$. The priors are $\sigma^2,\beta \sim \mbox{Gamma}(2,0.5)$, independently.
Then we perform Gibbs sampling, alternating between $p(\bx|\by,\theta)$ and $p(\theta|\by,\bx)$, after the transformation
$\sigma^2 = \exp(\varphi_1)$ and $\beta = \exp(\varphi_2)$.
See Supplement Section \ref{subsec:Coxexpression} for details of associated calculations. HMC takes $50$ leapfrog steps for latent variables and $6$ for parameters.
For each method, $5000$ draws are collected after a burn-in period of $9000$, which include two stages without preconditioning and one stage of tuning with preconditioning.
The simulation process is repeated for $20$
times using dispersed starting values for the parameters $\sigma^2\in [0.25,4]$ and $\beta\in[0.05,1]$.

\begin{figure}[t] %[H]
  \centering
  \begin{subfigure}{.49\linewidth}
      \centering
      \includegraphics[width = \textwidth]{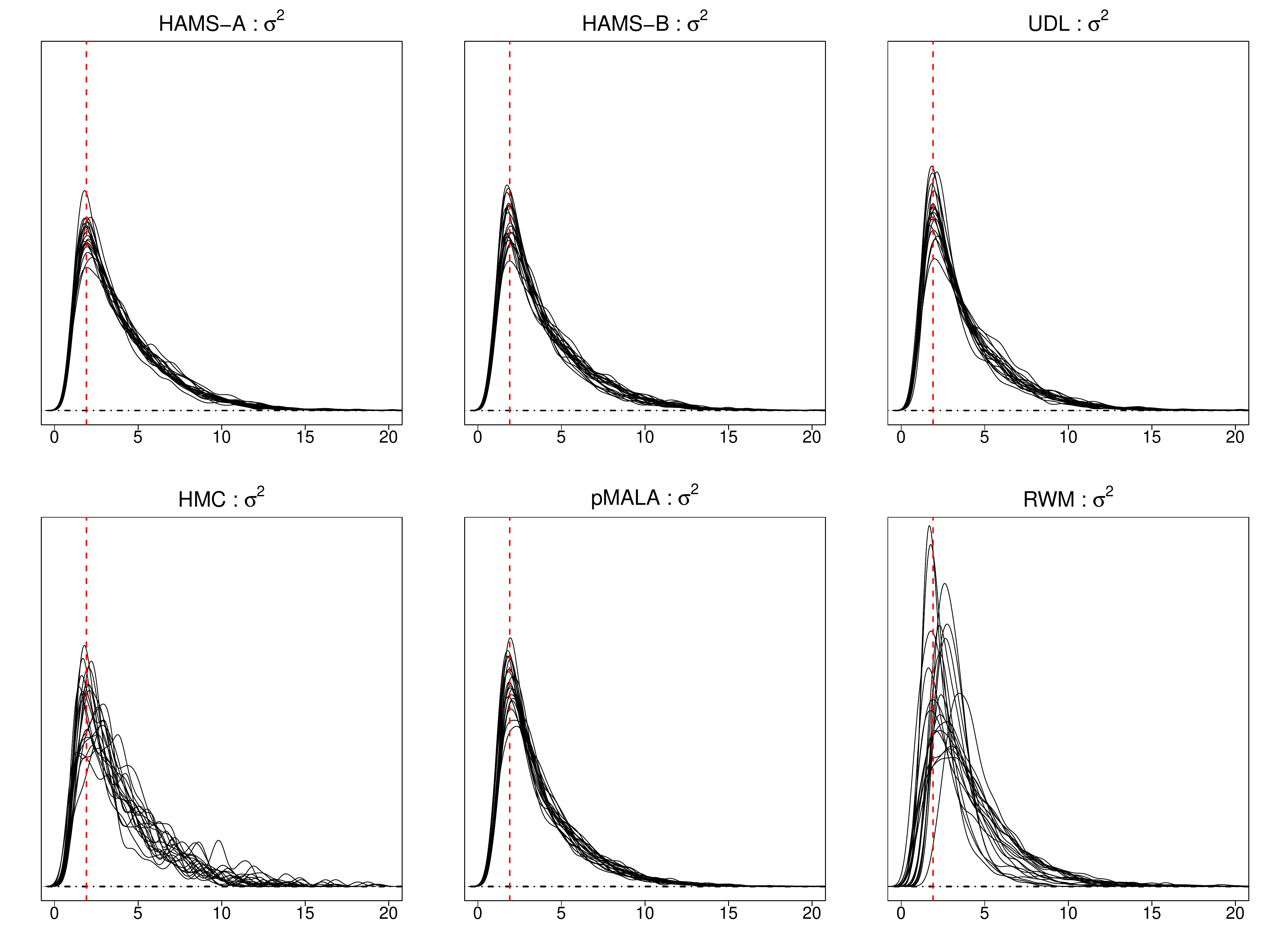}
      \caption{Densities of $\sigma^2$}\label{fig:image1-CoxD}
  \end{subfigure}
      \hfill
  \begin{subfigure}{.49\linewidth}
      \centering
      \includegraphics[width = \textwidth]{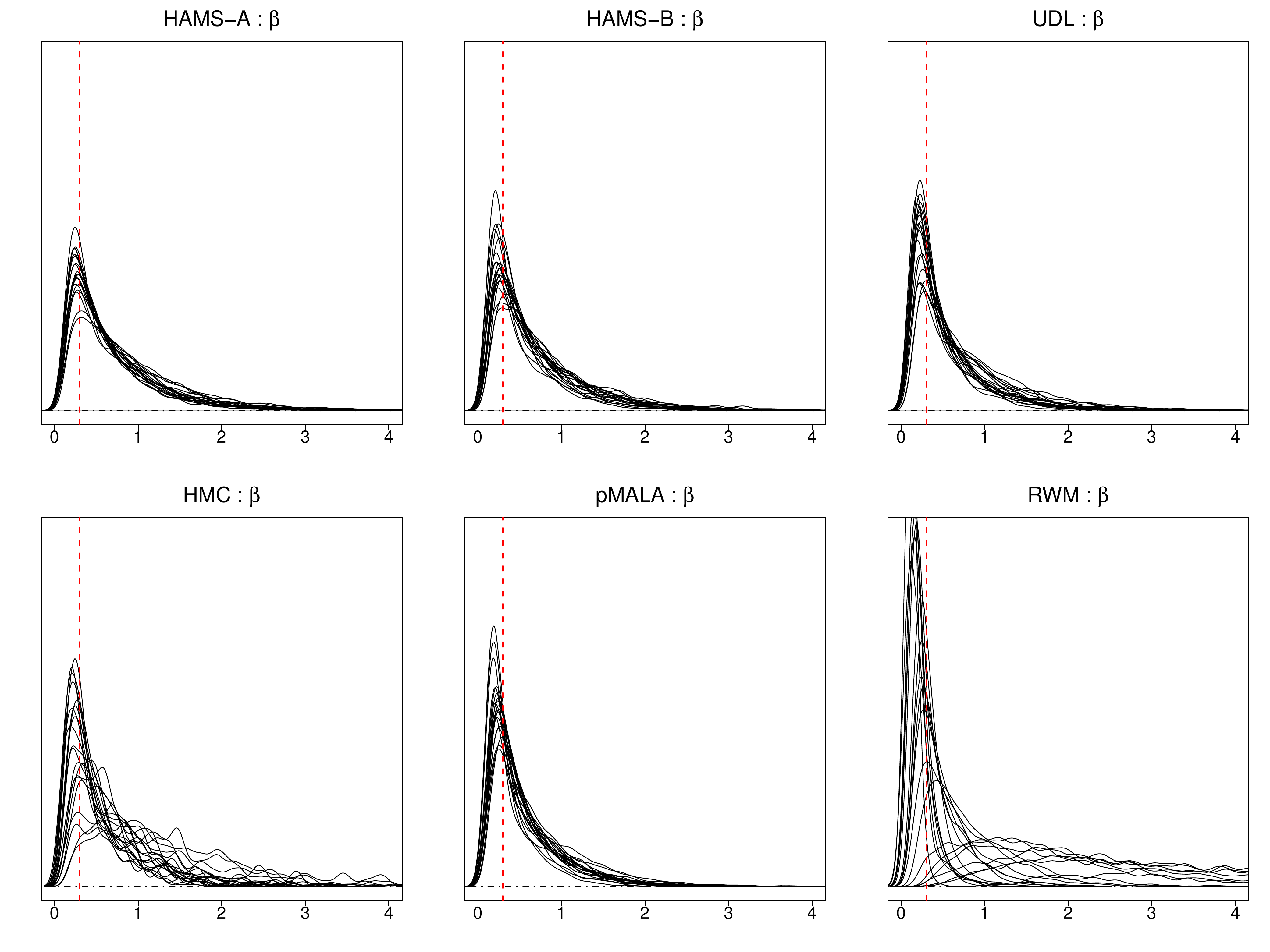}
      \caption{Densities of $\beta$}\label{fig:image12-CoxD}
  \end{subfigure}\vspace{-.2in}

  \RawCaption{\caption{Time-adjusted posterior density plots ($20$ repetitions overlaid) in log-Gaussian Cox model ($n=256$). The true parameter
values are marked by vertical lines.}
  \label{fig:coxdensity}}
\end{figure}

Table \ref{tab:tabfour} summarizes the results of posterior sampling. Figure \ref{fig:coxdensity} shows time-adjusted overlaid density plots for the parameters.
As shown in these plots, the posterior distributions of both $\sigma^2$ and $\beta$ are highly right-skewed. Accounting for this skewness, we consider the sample means
roughly aligned between different methods excluding RMW. From Table \ref{tab:tabfour}, while HMC has
the smallest standard deviation and largest ESS in $\sigma^2$, it shows poor performance in $\beta$ with the largest standard deviation and smallest ESS.
Among the remaining four methods, HAMS-A has the smallest standard deviation in both $\sigma^2$ and $\beta$, the largest ESS in $\sigma^2$ while HAMS-B has the
largest ESS in $\beta$.\par

\begin{table}[t] %[H]
  \caption{Comparison of posterior sampling in log-Gaussian Cox model ($n=256$). Standard deviations of sample means are in parentheses.
  Results are averaged over $20$ repetitions. \label{tab:tabfour}} \vspace{-.2in}
  \centering
  \begin{tabular}{|cccccc|}
  \hline
  Method & \multicolumn{1}{c|}{Time (s)} & \begin{tabular}[c]{@{}c@{}}Sample Mean\\[-.1in] $\sigma^2$ (sd)\end{tabular} & \multicolumn{1}{c|}{\begin{tabular}[c]{@{}c@{}}\ \\[-.1in] $\beta$ (sd)\end{tabular}} & \begin{tabular}[c]{@{}c@{}}ESS\\[-.1in] ($\sigma^2$,$\beta$)\end{tabular} &  $\frac{\mbox{minESS}}{\mbox{Time}}$ \\ \hline
  HAMS-A  & 2766.8                        & 3.90 (0.155)                                                          & 0.68 (0.073)                                                                                 & (978, 207)                                                     & 0.075         \\
  HAMS-B  & 2762.8                        & 3.93 (0.190)                                                          & 0.69 (0.106)                                                                                 & (838, 263)                                                     & 0.095         \\
  UDL    & 2759.1                        & 3.79 (0.171)                                                          & 0.59 (0.105)                                                                                 & (755, 246)                                                     & 0.089         \\
  HMC    & 25386.0                       & 3.88 (0.084)                                                          & 0.75 (0.113)                                                                                 & (2253, 139)                                                    & 0.005         \\
  pMALA  & 2755.3                        & 3.76 (0.189)                                                          & 0.57 (0.101)                                                                                 & (528, 178)                                                     & 0.065         \\
  RWM    & 1752.2                        & 3.70 (0.662)                                                          & 1.26 (1.434)                                                                                 & (226, 87)                                                      & 0.050         \\ \hline
  \end{tabular} \vspace{-.1in}
\end{table}

\section{Conclusion}

We propose a broad class of HAMS algorithms and develop two specific algorithms, HAMS-A/B, with convenient tuning and preconditioning strategies.
These algorithms achieve two distinctive properties: generalized reversibility and, for a normal target with a
pre-specified variance, rejection-free. Our numerical experiments demonstrate advantages of the proposed algorithms compared with existing ones.
Nevertheless, there are various topics of interest for further research.
In addition to HAMS-A/B, alternative algorithms can be derived by choosing a nonsingular noise variance $2 A - A^2$, which corresponds to two noise vectors per iteration.
These algorithms can be studied, together with HAMS-A/B and other algorithms related to underdamped Langevin dynamics.
In addition, it is desired to provide quantitative analysis of performances of sampling algorithms with or without the rejection-free property.
Finally, our framework of generalized Metropolis--Hastings can be exploited to develop other possible irreversible sampling algorithms.

%\bibliographystyle{plainnat}
%\bibliography{mcmcrefs-old}

\spacingset{1.1}

\bibliographystyle{apalike}
%\bibliography{mcmcrefsnew.bib}

\bibliography{mcmcrefsnew}

\spacingset{1.5}

%%%%%%%%%%%%%%%%%%%%%%%%%%%%%%%%%%%%%%%%%%%%%%%%%%%%%%%%%%%%%%%%%%%%%%%%%%%%%%%%%%%%%%%%%%
%%%---------------------------------------------------------------------------------------
%%%%%%%%%%%%%%%%%%%%%%%%%%%%%%%%%%%%%%%%%%%%%%%%%%%%%%%%%%%%%%%%%%%%%%%%%%%%%%%%%%%%%%%%%%

\clearpage

\setcounter{page}{1}

\setcounter{section}{0}
\setcounter{equation}{0}

\setcounter{figure}{0}
\setcounter{table}{0}

\renewcommand{\theequation}{S\arabic{equation}}
\renewcommand{\thesection}{\Roman{section}}

\renewcommand\thefigure{S\arabic{figure}}
\renewcommand\thetable{S\arabic{table}}

\setcounter{lem}{0}
\renewcommand{\thelem}{S\arabic{lem}}

\begin{center}
{\Large Supplementary Material for}

{\Large ``Hamiltonian Assisted Metropolis Sampling"}

\vspace{.1in} {\large Zexi Song and Zhiqiang Tan}
\end{center}

\section{Auxiliary variable derivation of proposal schemes} \label{sec:auxiliary}

We show that the proposal scheme (\ref{eq:proposal}) can also be derived through an auxiliary variable argument related to \cite{Titsias2018}, combined
with an over-relaxation technique as in \cite{Adler1981} and \cite{Neal1998}.
Compared with \cite{Titsias2018},  our derivation deals with the augmented density of $(x,u)$, instead of $x$ alone.
More importantly, our derivation incorporates an over-relaxation technique to accommodate all possible proposal schemes (\ref{eq:proposal}).
Finally, our derivation invokes a different normal approximation to the target distribution and, when applied without the momentum variable,
would lead to the modified pMALA algorithm as discussed in Section~\ref{sec:review}.

The starting point of our derivation is to introduce auxiliary variables $(y,v)$
and further augment the target density as $\pi(x,u,y,v) = \pi(x,u)\pi(y,v|x,u)$.
The conditional density $\pi(y,v|x,u)$ can be defined from a random walk update,
\begin{equation} \label{eq:cond-proposal}
  (y,v) | (x, u)
  \sim \mathcal N ( (x,u) , S ),
\end{equation}
where $S$ is a $(2k)\times (2k)$ variance matrix independent from $(x,u)$.
Given $(x_0,u_0)$, consider the following steps to sample from the new target:
\begin{itemize}
\item sample $(y,v)|(x_0,u_0)\sim \pi(y,v|x_0,u_0)$ directly according to (\ref{eq:cond-proposal}),
\item sample $(x_1,u_1)|(y,v)\sim \pi(x_1,u_1|y,v)$ by drawing $(x^*,u^*)$ from a conditional proposal density $q(x^*,u^*|y,v,x_0,u_0)$ and accepting $(x_1,u_1)=(x^*,u^*)$
with the usual Metropolis--Hastings probability or otherwise setting $(x_1,u_1)=(x_0,u_0)$.
\end{itemize}
The two steps can be identified as Gibbs sampling and Metropolis--Hastings within Gibbs sampling respectively.
Next, the proposal density $q(x^*,u^*|y,v,x_0,u_0)$ can be defined as an approximation to $\pi( x^*, u^* | y,v)$,
based on an approximation to $\pi(x)$ by a normal density with an identity variance anchored at $x_0$:
\begin{align}
  \tilde\pi (x; x_0) & \propto \exp\left\{-U(x_0) - (x - x_0)^\T \nabla U(x_0) -\frac{1}{2}(x - x_0)^\T (x - x_0) \right\} \nonumber \\
         & \propto \mathcal N (x|x_0 - \nabla U(x_0), I).  \label{eq:normal-approx}
\end{align}
Specifically, $\tilde\pi (x; x_0)$ is determined such that the gradient of $-\log \tilde\pi (x; x_0)$ at $x_0$ coincides with $ \nabla U(x_0)$, the gradient of $U(x)= -\log \pi(x)$ at $x_0$.
We take $q(x^*,u^*|y,v,x_0,u_0)=\tilde\pi(x^*,u^*|y,v; x_0 )$, the induced conditional density by (\ref{eq:cond-proposal2}) in Lemma~\ref{lem:normal-approx}.
This result can be shown by similar calculation as in Gelman et al.~(2014, Section 3.5).  % Bayesian data analysis book

\begin{lem} \label{lem:normal-approx}
Define $\tilde\pi(x,u; x_0 ) \propto \tilde\pi(x; x_0) \exp(-u^\T u/2)$. Then the joint density defined by $\tilde\pi (x,u; x_0) \pi( y,v | x,u)$ induces the conditional density
\begin{align}
\tilde\pi(x,u|y,v; x_0 ) = \mathcal N (x,u | \mu_{x_0}, A),  \label{eq:cond-proposal2}
\end{align}
where $\pi( y,v | x,u )$ is as in (\ref{eq:cond-proposal}), and
\begin{align*}
  A = (I + S^{-1})^{-1},
  \quad
  \mu_{x_0} = A \left[ \begin{pmatrix}
      x_0 - \nabla U(x_0)\\
      \bo
    \end{pmatrix} + S^{-1}\begin{pmatrix}
      y \\
      v
    \end{pmatrix}\right].
\end{align*}
\end{lem}

\vspace{.1in}
Similarly as in \cite{Titsias2018}, the auxiliary variables $(y,v)$ can be integrated out to obtain a marginal scheme from $(x_0,u_0)$ to $(x^*,u^*)$ as
\begin{align}
  & q(x^*,u^*| x_0, u_0 ) =  \int \tilde\pi(x^*,u^*|y,v; x_0 )\pi(y,v|x_0,u_0)\,d(y,v) \nonumber \\
  & =
  \mathcal N\left(
    \begin{pmatrix}
      x^*\\
      u^*
    \end{pmatrix}\middle\vert
    \begin{pmatrix}
      x_0\\
      u_0
    \end{pmatrix} - A\begin{pmatrix}
      \nabla U(x_0)\\
      u
    \end{pmatrix},
    AS^{-1}A + A
  \right), \label{eq:proposal3}
\end{align}
where $AS^{-1}A + A = 2A - A^2$ for  $A = (I + S^{-1})^{-1}$. Hence the proposal scheme (\ref{eq:proposal3}) from the auxiliary variable argument
retains the same form as (\ref{eq:proposal}). This discussion also confirms the previous observation that
when the target density $\pi(x)$ is $\mathcal N(\bo,I)$, the proposal $(x^*,u^*)$ in (\ref{eq:proposal2}) is always accepted, because
the normal approximation $\tilde\pi(x;x_0)$ becomes exact and hence $(x^*,u^*)$ is obtained from just two-block Gibbs sampling.

There is, however, a caveat in the link between (\ref{eq:proposal}) and (\ref{eq:proposal3}). Using the auxiliary variables leads the proposal (\ref{eq:proposal3}),
with the relation $A = (I + S^{-1})^{-1}$. Because $S$ is positive semi-definite as a variance matrix, this relation
imposes the constraint that $A\leq I$. For the proposal scheme (\ref{eq:proposal}), it is only required that $\bo\leq A\leq 2I$.
When $I<A\leq 2I$, the scheme (\ref{eq:proposal}) remains valid, but cannot be deduced from (\ref{eq:proposal3}).
Hence (\ref{eq:proposal}) encapsulates a broader class of proposal distributions
than directly derived via auxiliary variables.

Next we show that the
over-relaxation technique \citep{Adler1981, Neal1998} can be exploited to define an auxiliary proposal density $q(x^*,u^*|y,v,x_0,u_0)$
more flexible than above, so that the entire class of proposal distributions (\ref{eq:proposal}) can be recovered.
By over-relaxation based on normal distributions, consider the proposal density
\begin{align*}
  q_\alpha (x^*,u^*|y,v,x_0,u_0) = \mathcal N\left(x^*, u^* | \mu_{x_0} + \alpha ( (x_0^\T,u_0^\T)^\T - \mu_{x_0}), (1-\alpha^2) A \right),
\end{align*}
where $\mu_{x_0}$ and $A$ are defined as in Lemma~\ref{lem:normal-approx}, and $-1\leq \alpha \leq 1$ controls the degree of over-relaxation.
Setting $\alpha=0$ gives the previous choice $q(x^*,u^*|y,v,x_0,u_0)=\tilde\pi(x^*,u^*|y,v; x_0 )$ and leads to the marginal proposal density (\ref{eq:proposal3}).

\begin{lem} \label{lem:overrelax}
Let $A_\alpha = (1-\alpha)A$. The marginal proposal density obtained by integrating out $(y,v)$ from $q_\alpha (x^*,u^*|y,v,x_0,u_0)$ is
\begin{align}
  & q_\alpha (x^*,u^*| x_0, u_0 ) =  \int  q_\alpha (x^*,u^*|y,v,x_0,u_0) \pi(y,v|x_0,u_0)\,d(y,v) \nonumber \\
  & =
  \mathcal N\left(
    \begin{pmatrix}
      x^*\\
      u^*
    \end{pmatrix}\middle\vert
    \begin{pmatrix}
      x_0\\
      u_0
    \end{pmatrix} - A_\alpha\begin{pmatrix}
      \nabla U(x_0)\\
      u
    \end{pmatrix},
    2A_\alpha - A_\alpha^2
  \right). \label{eq:proposal3-overrelax}
\end{align}
%Moreover, $A_\alpha S^{-1}A_\alpha + (1+\alpha)A_\alpha = 2A_\alpha - A_\alpha^2$.
\end{lem}

\vspace{.1in} By the preceding result, the marginal scheme (\ref{eq:proposal3-overrelax}) is still of the form (\ref{eq:proposal}), with $A$ replaced by $A_\alpha$.
The matrix $A_\alpha$ is determined from $\alpha$ and $S$ as $A_\alpha = (1-\alpha)(I+S^{-1})^{-1}$.
The constraints $-1\le \alpha\le 1$ and $S \ge \bo$ imply that  $\bo \le A_\alpha \le 2 I$. Conversely,
any matrix $\bo \le A \le 2I$ can be obtained as $A_\alpha$ for some $-1\le \alpha \le 1$ and $S \ge \bo$.
The choice $A = 2I$ corresponds to the limit case $\alpha=-1$ and $S \to \infty$.
In this sense, the proposal scheme (\ref{eq:proposal}) with any choice $\bo \le A \le 2I$ can be identified
as a marginal scheme from the auxiliary variable argument while incorporating over-relaxation.

Finally, as might by noted by readers, the foregoing development (including the over-relaxation) remains valid when the momentum variable $u$ is dropped.
In this case, the proposal density in (\ref{eq:proposal3}) reduces to
\begin{align*}
q ( x^* |  x_0) = \mathcal N( x^* | x_0 - A \nabla U(x_0), 2 A -A^2),
\end{align*}
where $A$ is a $k \times k$ symmetric matrix satisfying $\bo \le A \le I$ before over-relaxation.
Taking $A = \frac{\epsilon^2}{1+\sqrt{1-\epsilon^2}} I= (1- \sqrt{1-\epsilon^2}) I$ leads to the proposal scheme
\begin{align}
x^* = x_0 - \frac{\epsilon^2}{1+\sqrt{1-\epsilon^2}}  \nabla U(x_0) + Z, \quad Z \sim \mathcal N(0, \epsilon^2 I), \label{eq:proposal-MALA}
\end{align}
which is precisely the proposal scheme in modified pMALA with $\Sigma=I$ (or modified MALA).
In general, our auxiliary variable argument can be applied with an arbitrary choice of variance matrix $\Sigma$,
to obtain a proposal scheme in the form
\begin{align}
x^* = x_0 - A  \nabla U(x_0) + Z, \quad Z \sim \mathcal N(0, 2 A -A \Sigma^{-1} A), \label{eq:proposal-Sigma}
\end{align}
where $A$ is a $k \times k$ symmetric matrix satisfying $A^{-1} \ge \Sigma^{-1}$.
Taking $A = \frac{\epsilon^2}{1+\sqrt{1-\epsilon^2}} \Sigma= (1- \sqrt{1-\epsilon^2}) \Sigma$ leads to modified pMALA described in Section~\ref{sec:review}.

It is informative to compare our schemes with \cite{Titsias2018} in the Bayesian setting with $\pi(x)\propto \exp\{-U(x)\} \propto \exp\{\ell(x)\}\mathcal N(x|\bo,C)$,
where $\ell(x)$ is the log-likelihood and $C$ a prior variance.
As discussed in Section~\ref{sec:review}, the proposal scheme (\ref{eq:proposal-TP}) in \cite{Titsias2018}
differs from that in modified pMALA with general $\Sigma$, except
for equivalence in the special case $C=I$, where both proposal schemes reduce to (\ref{eq:proposal-MALA}).
This difference can be understood as follows.
Given the current value $x_0$, the normal approximation of $\pi(x)$ used in \cite{Titsias2018} is
\begin{align}
  \tilde\pi_{\mbox{\scriptsize TP}} (x; x_0) & \propto \exp\left\{ \ell (x_0) + (x - x_0)^\T \nabla \ell(x_0) -\frac{1}{2} x^\T C^{-1} x \right\} \nonumber \\
         & \propto \exp\left\{ -(x - x_0)^\T \nabla U (x_0) -\frac{1}{2} (x-x_0)^\T C^{-1} (x-x_0) \right\} , \label{eq:normal-TP}
\end{align}
where $ \nabla \ell(x_0) = -\nabla U (x_0) + C^{-1} x_0$. Apparently, the normal density (\ref{eq:normal-TP}) in general differs from (\ref{eq:normal-approx}) used in
our derivation unless $C=I$.

For the second-order algorithm in the Supplement of \cite{Titsias2018},
the proposal scheme, after correcting a typo to match the first-order scheme (\ref{eq:proposal-TP}) when $G=\bo$,
can be written as
\begin{align}
x^*  = \frac{2}{\delta}C^\dag x_0 + C^\dag (\nabla \ell (x_0) - G x_0)  + Z = x_0 -C^\dag \nabla U(x_0)  + Z,
\quad Z \sim \mathcal N(\bo,  \frac{2}{\delta} C^{\dag 2} + C^\dag), \label{eq:proposal-TP2}
\end{align}
where $C^\dag  = (\frac{2}{\delta}I +C^{-1} -G)^{-1}$,
and $G$ is the Hessian $\nabla^2 \ell(x_0)$ or an approximation.
%The associated normal approximation of $\pi(x)$ is
%\begin{align}
%  \tilde\pi_{\mbox{\scriptsize TP2}} (x; x_0) & \propto \exp\left\{ \ell (x_0) + (x - x_0)^\T \nabla \ell(x_0) + \frac{1}{2}(x-x_0)^\T G (x-x_0)-\frac{1}{2} x^\T C^{-1} x \right\} \nonumber \\
%         & \propto \exp\left\{ -(x - x_0)^\T \nabla U (x_0) -\frac{1}{2} (x-x_0)^\T (C^{-1}-G) (x-x_0) \right\} .\label{eq:normal-TP2}
%\end{align}
For simplicity, assume that $G$ is independent of $x_0$. The corresponding
approximation to the variance of the target $\pi(x)$ is then $\Sigma= (C^{-1} - G)^{-1}$.
Moreover, the proposal scheme (\ref{eq:proposal-TP2}), by direct calculation, can be expressed as (\ref{eq:proposal-Sigma}) with
$\Sigma= (C^{-1} - G)^{-1}$ and
$A = C^\dag = (\Sigma^{-1} + \frac{2}{\delta} I)^{-1}$.
Therefore, the second-order algorithm of \cite{Titsias2018} and modified pMALA use proposal schemes both in the class (\ref{eq:proposal-Sigma}),
but with different choices of $A$ matrix, after the approximate variance $\Sigma$ is matched.

\subsection{Proof of Lemma~\ref{lem:overrelax}}

Given the current variables $(x_0,u_0)$, the variables $(y,v)$ are generated as
\begin{align}
  (y,v) | (x_0,u_0) \, \sim \mathcal N ( (x_0, u_0),\, S).\label{eq:lemmaS2pf1}
\end{align}
The variables $(x^*,u^*)$ are then generated from $q_\alpha$ as
\begin{align}
  & (x^*,u^*)| (y,v,x_0,u_0) \, \sim  \mathcal N\left((1-\alpha) \mu_{x_0} +
   \alpha  \begin{pmatrix}
    x_0\\
    u_0
  \end{pmatrix},\, (1-\alpha^2)A\right),  \label{eq:lemmaS2pf2-a}
\end{align}
where $-1\leq \alpha \leq 1$, and
\begin{align}
  & A = (I+ S^{-1})^{-1},\,\mu_{x_0} = A\left(\begin{pmatrix}
    x_0 - \nabla U(x_0)\\
    \bo
  \end{pmatrix}
   + S^{-1}
   \begin{pmatrix}
    y\\
    v
  \end{pmatrix}
  \right). \label{eq:lemmaS2pf2-b}
\end{align}
Then $(x^*,u^*)$ and $(y,v)$ are jointly normal given $(x_0,u_0)$ and hence $(x^*,u^*)|(x_0,u_0)$ is also normally distributed. It suffices to determine its mean and variance.

First, we compute $\EE( x^*, u^*| x_0, u_0)$. By (\ref{eq:lemmaS2pf1}) and (\ref{eq:lemmaS2pf2-b}),
\begin{align}
  & \EE[\mu_{x_0}|x_0,u_0]
  = A \left(
    \begin{pmatrix}
      x_0 - \nabla U(x_0)\\
      \bo
    \end{pmatrix}
    + S^{-1}
   \begin{pmatrix}
    x_0\\
    u_0
  \end{pmatrix}
  \right) \nonumber \\
  & = A \left(
      (I+ S^{-1})
      \begin{pmatrix}
       x_0\\
       u_0
     \end{pmatrix}
     -
     \begin{pmatrix}
      \nabla U(x_0)\\
      u_0
    \end{pmatrix}
  \right)
   =   \begin{pmatrix}
    x_0\\
    u_0
  \end{pmatrix}   - A
  \begin{pmatrix}
    \nabla U(x_0)\\
    u_0
  \end{pmatrix}. \label{eq:lemmaS2pf3}
\end{align}
Therefore, by (\ref{eq:lemmaS2pf2-a}) and (\ref{eq:lemmaS2pf3}),
\begin{align*}
& \EE( x^*, u^*| x_0, u_0)
 =  \EE[\,\EE (x, u^* |y,v, x_0, u_0 )\,| x_0, u_0] \\
& = \EE\left[(1-\alpha) \mu_{x_0} +
\alpha  \begin{pmatrix}
 x_0\\
 u_0
\end{pmatrix} \middle|\, x_0,u_0 \right]
= \begin{pmatrix}
  x_0\\
  u_0
\end{pmatrix} - A_{\alpha}
\begin{pmatrix}
  \nabla U(x_0)\\
  u_0
\end{pmatrix}, %\label{eq:lemmaS2pf4}
\end{align*}
where $A_{\alpha} = (1-\alpha) A$.

Next, we compute $\Var( x^*, u^*| x_0,u_0)$. By (\ref{eq:lemmaS2pf2-a})--(\ref{eq:lemmaS2pf2-b}),
\begin{align}
& \Var[\,\EE ( x^*, u^* |y,v, x_0, u_0 )\,| x_0,u_0]
= \Var\left[ (1-\alpha)\mu_{x_0} + \alpha
\begin{pmatrix}
  x_0\\
  u_0
\end{pmatrix}\middle | x_0,u_0 \right] \nonumber \\
& = (1-\alpha)^2 \Var[\mu_{x_0}|x_0,u_0]
= (1-\alpha)^2AS^{-1}A = A_{\alpha}S^{-1}A_{\alpha}, \label{eq:lemmaS2pf5}\\
 & \EE[\,\Var( x^*, u^* |y,v, x_0, u_0)\, | x_0, u_0] = \EE[(1-\alpha^2)A | x_0,u_0] \nonumber \\
& = (1-\alpha^2) A = (1+\alpha) A_{\alpha}. \label{eq:lemmaS2pf6}
\end{align}
Combining (\ref{eq:lemmaS2pf5}) and (\ref{eq:lemmaS2pf6}) yields
\begin{align*}
  & \Var( x^*, u^*| x_0,u_0) \\
  & = \EE[\,\Var( x^*, u^* |y,v,x_0,u_0 (\,|x_0,u_0] + \Var[\,\EE( x^*, u^* |y,v,x_0,u_0)\,|x_0,u_0] \nonumber \\
  & = A_{\alpha} S^{-1} A_{\alpha} + (1+\alpha) A_{\alpha} . %\label{eq:lemmaS2pf7}.
\end{align*}

Finally, we show that $A_{\alpha}S^{-1}A_{\alpha} + (1+\alpha) A_{\alpha} = 2A_{\alpha} - A_{\alpha}^2$.
Because $A = (I + S^{-1})^{-1}$, we have $A(I + S^{-1})  = I$ and hence $A^2 + AS^{-1}A = A$. Then
\begin{align*}
& (1-\alpha)^2 AS^{-1}A = (1-\alpha)^2 A - (1-\alpha)^2 A^2 \\
& \Rightarrow A_{\alpha}S^{-1}A_{\alpha} = (1-\alpha) A_{\alpha} - A_{\alpha}^2 \\
& \Rightarrow A_{\alpha}S^{-1}A_{\alpha} + (1+\alpha) A_{\alpha} = 2A_{\alpha} - A_{\alpha}^2.
\end{align*}
This completes the proof of Lemma S2.

\section{Demonstration of validity of UDL}
\label{sec:UDLMR}

We demonstrate that UDL is valid in leaving the augmented target $\pi(x,u)$ invariant.
Similarly as HAMS, by Proposition~\ref{prop:gMH}, it suffices to verify that the acceptance probability stated for UDL in Section~\ref{sec:review}
can be written in the form of generalized Metropolis--Hastings probability (\ref{mr1}) for the associated (forward) proposal density $Q$.

First, we calculate the generalized Metropolis--Hastings probability (\ref{mr1}) with the (forward) proposal density $Q$ from UDL.
The proposal scheme in UDL is defined as
\begin{align*}
& \text{Sample } Z_1, Z_2 \sim \mathcal N(\bo, M) \text{ independently, } \\ %\label{eq:udlapp1} \\
& u^+ = \sqrt{c} u_0 + \sqrt{1- c} Z_1,\\
& \tilde{u} = u^+ - \frac{\epsilon}{2}\nabla U(x_0), \quad
 x^* = x_0 + \epsilon M^{-1}\tilde{u}, \quad
 u^- = \tilde{u} - \frac{\epsilon}{2} \nabla U(x^*),\\
& u^* = \sqrt{c} u^- + \sqrt{1 - c} Z_2.  %\label{eq:udlapp2}
\end{align*}
The noises $(Z_1,Z_2)$ can be expressed as
\begin{align}
  & Z_1 = \left(\frac{M}{\epsilon} (x^* - x_0) + \frac{\epsilon}{2}\nabla U(x_0) - \sqrt{c}u_0 \right)(1-c)^{-1/2}, \label{eq:udlapp5} \\
  & Z_2 = \left(\frac{\sqrt{c}M}{\epsilon} (x_0 - x^*) + \frac{\epsilon\sqrt{c}}{2}\nabla U(x^*) + \sqrt{c}u^* \right)(1-c)^{-1/2}.\label{eq:udlapp6}
\end{align}
Suppose that the mapping above from $(x_0,u_0)$ to $(x^*,u^*)$ is applied from $(x^*,-u^*)$ to $(x_0,-u_0)$, but using new noises $(Z_3,Z_4)$.
By exchanging $(x_0,u_0)$ and $(x^*,-u^*)$, the new noises $(Z_3,Z_4)$ can be calculated as
\begin{align}
  & Z_3 = \left(\frac{M}{\epsilon} (x_0 - x^*) + \frac{\epsilon}{2}\nabla U(x^*) + \sqrt{c}u^* \right)(1-c)^{-1/2}, \label{eq:udlapp7} \\
  & Z_4 = \left(\frac{\sqrt{c}M}{\epsilon} (x^* - x_0) + \frac{\epsilon\sqrt{c}}{2}\nabla U(x_0) - \sqrt{c}u_0 \right)(1-c)^{-1/2}.\label{eq:udlapp8}
\end{align}
Then the forward and backward transitions of the proposals for UDL can
be illustrated in a similar manner to (\ref{eq:transition-diagram}) as
\begin{align}
  \begin{pmatrix}
    x_0 \\
    u_0
  \end{pmatrix}
  \stackrel{(Z_1,Z_2)}{\longrightarrow}
  \begin{pmatrix}
    x^* \\
    u^*
  \end{pmatrix},
  \qquad
  \begin{pmatrix}
    x^* \\
    -u^*
  \end{pmatrix}
  \stackrel{(Z_3,Z_4)}{\longrightarrow}
  \begin{pmatrix}
    x_0 \\
    -u_0
  \end{pmatrix}, \label{eq:transition-diagram-udl}
\end{align}
where the arrows denote the {\it same} mapping, depending on $(Z_1,Z_2)$ or $(Z_3,Z_4)$.

Because $(Z1,Z2)$ are the only sources of randomness, the (forward) proposal density from $(x_0,u_0)$ to $(x^*,u^*)$ is
\begin{align}
  Q(x^*,u^*|x_0,u_0) &= \mathcal N(Z_1|\bo, M )\mathcal N(Z_2|\bo, M ) \nonumber \\
  & \propto \exp\left(-\frac{1}{2}Z_1^\T M^{-1} Z_1 - \frac{1}{2}Z_2^\T M^{-1} Z_2 \right).\label{eq:udlapp3}
\end{align}
Evaluation of the {\it same} proposal density from  $(x^*, -u^*)$ to $(x_0, -u_0)$ gives
\begin{align}
 Q(x_0, -u_0 |x^*, -u^*) & = \mathcal N(Z_3|\bo, M )\mathcal N(Z_4|\bo, M ) \nonumber \\
 & \propto \exp\left(-\frac{1}{2}Z_3^\T M^{-1} Z_3 - \frac{1}{2}Z_4^\T M^{-1} Z_4 \right).\label{eq:udlapp4}
\end{align}
Using (\ref{eq:udlapp5}) to (\ref{eq:udlapp4}), the log ratio of proposal densities is
\begin{align}
  &\log\left(\frac{Q(x_0,-u_0|x^*,-u^*) }{ Q(x^*,u^*|x_0,u_0)}\right) =\frac{1}{2}(x^* - x_0)^\T (\nabla U(x^*) + \nabla U(x_0) ) \nonumber \\
  & - \frac{\epsilon^2}{8}\left([\nabla U(x^*)]^\T M^{-1}\nabla U(x^*) - [\nabla U(x_0)]^\T M^{-1}\nabla U(x_0) \right)
  -\frac{1}{2}\left(u_0^\T M^{-1}u_0 - (u^*)^\T M^{-1}u^* \right).  \label{eq:udlapp9}
\end{align}
Furthermore, the log ratio of target densities at $(x_0,u_0)$ and $(x^*, -u^*)$ is
\begin{align}
  \log\left(\frac{\pi(x^*,-u^*)}{\pi(x_0,u_0)}\right) = U(x_0) - U(x^*) + \frac{1}{2}\left(u_0^\T M^{-1}u_0 - (u^*)^\T M^{-1}u^*\right).
  \label{eq:udlapp10}
\end{align}
From (\ref{eq:udlapp9}) and (\ref{eq:udlapp10}), the generalized Metropolis--Hastings probability (\ref{mr1})
is
\begin{align}
    \min\left(1, \exp\left\{  U(x_0) - U(x^*) +\frac{(x^* - x_0)^\T }{2}(\nabla U(x_0) + \nabla U(x^*))\right.\right. \nonumber\\
    \qquad\left.\left. - \frac{\epsilon^2}{8}\left([\nabla U(x^*)]^\T M^{-1}\nabla U(x^*) - [\nabla U(x_0)]^\T M^{-1}\nabla U(x_0) \right) \right\}\right).
    \label{eq:udlapp11}
\end{align}

Second, we show that generalized Metropolis--Hastings probability (\ref{eq:udlapp11}) reduces to the acceptance probability
stated in Section \ref{sec:review}:
\begin{equation}
    \min\left(1,\exp(H(x_0,u^+) - H(x^*, u^-))\right). \label{eq:udlapp12}
\end{equation}
In fact, direct calculation using $u^- = u^+ - \frac{\epsilon}{2}\left(\nabla U(x^*) + \nabla U(x_0)\right)$ yields
\begin{align*}
 (u^-)^\T M^{-1}u^- %= (u^+ - \frac{\epsilon}{2}\left(\nabla U(x^*) + \nabla U(x_0)\right))^\T  M^{-1} (u^+ - \frac{\epsilon}{2}\left(\nabla U(x^*) + \nabla U(x_0)\right)) \\
 & = (u^+)^\T M^{-1} u^+ + \frac{\epsilon^2}{4}(\nabla U(x_0) + \nabla U(x^*))^\T  M^{-1}(\nabla U(x_0) + \nabla U(x^*)) \\
 & \quad - \epsilon (u^+)^\T  M^{-1}(\nabla U(x_0) + \nabla U(x^*)),
\end{align*}
and hence
\begin{align}
& \frac{1}{2}(u^+)^\T M^{-1} u^+ - \frac{1}{2}(u^-)^\T M^{-1} u^- \nonumber \\
& = \frac{\epsilon}{2}(u^+)^\T  M^{-1}(\nabla U(x_0) + \nabla U(x^*)
- \frac{\epsilon^2}{8}(\nabla U(x_0) + \nabla U(x^*))^\T  M^{-1}(\nabla U(x_0) + \nabla U(x^*)) \nonumber \\
%& = \frac{\epsilon}{2}\left(u^+ - \frac{\epsilon}{2}\nabla U(x_0)\right)^\T  M^{-1} (\nabla U(x_0) + \nabla U(x^*))
%- \frac{\epsilon^2}{8}\left( [\nabla U(x^*) ]^\T  M^{-1}\nabla U(x^*) - [\nabla U(x_0) ]^\T  M^{-1}\nabla U(x_0)  \right)\nonumber \\
%& = \frac{ (\epsilon M^{-1} \tilde{u})^\T  }{2} (\nabla U(x_0) + \nabla U(x^*))
%- \frac{\epsilon^2}{8}\left[\nabla U(x^*) ]^\T  M^{-1}\nabla U(x^*) - [\nabla U(x_0) ]^\T  M^{-1}\nabla U(x_0)  \right)\nonumber \\
& = \frac{1}{2} (x^* - x_0)^\T (\nabla U(x_0) + \nabla U(x^*))  - \frac{\epsilon^2}{8}\left( [\nabla U(x^*) ]^\T  M^{-1}\nabla U(x^*) - [\nabla U(x_0) ]^\T  M^{-1}\nabla U(x_0)  \right). \label{eq:udlapp13}
\end{align}
By the definition of the Hamiltonian, we have
\begin{align*}
H(x_0,u^+) - H(x^*, u^-) = U(x_0) - U(x^*) + \frac{1}{2}(u^+)^\T M^{-1} u^+ & - \frac{1}{2}(u^-)^\T M^{-1} u^-.
\end{align*}
Substituting (\ref{eq:udlapp13}) into the above, we see that (\ref{eq:udlapp12}) equals (\ref{eq:udlapp11}).

\section{Generalized Metropolis--Hastings sampling} \label{sec:gMH}

We give a broader definition of generalized Metropolis--Hastings sampling in Section~\ref{sec:extension},
to accommodate both continuous and discrete variables.

Let $\pi(y)$ be a pre-specified probability density function on $\mathcal Y$, with respect to possibly a product of Lebesgue and counting measures.
Assume that $J: \mathcal Y \to \mathcal Y$ is an invertible mapping, such that for any set $C\subset \mathcal Y$ and integrable function $h$,
\begin{align}
&\int_{J(C)} \pi(y) \,\dif y = \int_C \pi(y)\, \dif y, \label{eq:invariance2-a}\\
&\int_{J(C)} h( J^{-1} y) \,\dif y = \int_C h(y)\, \dif y, \label{eq:invariance2-b}
\end{align}
where $J^{-1}$ denote the inverse mapping of $J$, and $J(C) = \{Jy: y \in C\}$.
While (\ref{eq:invariance2-a}) is restated from (\ref{eq:invariance}), condition (\ref{eq:invariance2-b}) is analogous to saying that
the Jacobian determinant of $J$ is $\pm 1$ in the case where $\mathcal Y$ is Euclidean endowed with the Lebesgue measure.
With this interpretation of $Jy$, generalized Metropolis--Hastings sampling is still defined as in Section~\ref{sec:extension}.
More importantly, Proposition~\ref{prop:gMH} can be seen to remain valid, by substituting (\ref{eq:invariance2-b}) for all the change-of-variables calculation in the proof.

Next we show that the irreversible jump sampler (I-Jump) in \cite{Ma2018} can be obtained as a special case of generalized Metropolis-Hastings sampling,
when a binary auxiliary variable $s \in \{1,-1\}$ is introduced for sampling from an original target density $\pi(x)$ on $\mathcal X$.
Given current variables $(x_0,s_0)$, an iteration of I-Jump can be described as follows, where
$f(\cdot|x_0)$ and $g(\cdot|x_0)$ are two possibly different proposal densities.

\textit{Irreversible jump sampler} (I-Jump).
\vspace{-.1in}
\begin{itemize}\addtolength{\itemsep}{-.1in}
\item Sample $w \sim \text{Uniform} [0,1]$.
\item If $s_0=1$, sample $x^* \sim f(\cdot|x_0)$ and compute
\begin{align*}
\rho (x^*|x_0) = \min\left(1 , \frac{\pi(x^*) g(x_0 |x^*)  }{ \pi(x_0) f(x^*|x_0)} \right);
\end{align*}
else sample $x^* \sim g(\cdot|x_0)$ and compute
\begin{align*}
\rho (x^*|x_0) = \min\left(1 , \frac{\pi(x^*) f(x_0 |x^*)  }{ \pi(x_0) g(x^*|x_0)} \right).
\end{align*}

\item If $w < \rho(x^*|x_0)$, then set $(x_1,s_1) = (x^*, s_0)$; else set $(x_1,s_1) = (x_0, -s_0)$.
\end{itemize}

To recast I-Jump, consider the augmented target density $\pi(x,s) = \pi(x) /2$ on the product space $\mathcal Y=\mathcal X \times \{1,-1\}$,
that is, $x$ and $s$ are independent and $s$ takes value 1 or $-1$ with equal probabilities.
The mapping defined by $J(x,s) = (x,-s)$ satisfies conditions (\ref{eq:invariance2-a})--(\ref{eq:invariance2-b}).
Define the proposal density $Q$ as
\begin{align*}
Q( x^*, s^* | x_0, s_0) = \left\{\begin{array}{ll}
f(x^*|x_0), & \mbox{if } s^*=s_0=1,\\
g(x^* | x_0), & \mbox{if } s^*=s_0 = -1,\\
0 , & \mbox{if } s^* \not = s_0.
\end{array} \right.
\end{align*}
Then the acceptance probability in I-Jump can be expressed as
\begin{align*}
\rho (x^*, s^* |x_0,s_0) = \min\left(1 , \frac{\pi(x^*,-s^*) Q(x_0, -s_0 |x^*, -s^*)  }{ \pi(x_0,s_0) Q( x^*, s^*|x_0, s_0)} \right).
\end{align*}
by noticing that $s^*=s_0$ and $\pi(x^*,-s^*)/\pi(x_0,s_0)=\pi(x^*)/\pi(x_0)$. Therefore, the I-Jump algorithm can be seen as
generalized Metropolis--Hastings sampling.

As a concrete example of I-Jump, \cite{Ma2018} proposed an irreversible MALA (I-MALA) algorithm. The proposal schemes $f(\cdot | x_0)$ and $g(\cdot|x_0)$
are defined as discretizations of irreversible continuous Markov processes. Each proposal scheme can be related to (\ref{eq:g2ms-f}) in our G2MS algorithm with $y_0$ replaced by $x_0$:
\begin{align*}
x^* = x_0 - B \nabla U(x_0) + Z , \quad Z \sim \mathcal N(\bo, B+B^\T - B B^\T).
\end{align*}
For $B= \epsilon^2 B_0$ with $\epsilon\approx 0$, the preceding scheme is approximately
\begin{align}
x^* &=  x_0 - \epsilon^2 (D_0+C_0) \nabla U(x_0) + Z , \quad Z \sim \mathcal N(\bo, 2\epsilon^2 D_0 ), \label{eq:Ma-proposal}
\end{align}
where $D_0 = (B_0+B_0^\T)/2$ is symmetric and $C_0 = (B_0 - B_0^\T)/2$ is skew-symmetric.
It is interesting that the form of (\ref{eq:Ma-proposal}) matches the proposal schemes derived by discretizing general Markov procsses in \cite{Ma2018}.

Although both HAMS and I-MALA can be subsumed by generalized Metropolis--Hastings sampling, there remain important differences.
The HAMS algorithm uses momentum as an auxiliary variable
and hence is able to exploit symmetry in the momentum distribution, whereas I-MALA relies on lifting with a binary variable \citep{Gustafson1998,Vucelja2016}
and needs to split the original variable $x$ to specify symmetric and skew-symmetric matrices $D_0$ and $C_0$ when
defining proposal schemes based on irreversible Markov processes in $x$.
Further research is desired to compare and connect these algorithms.

\section{Proofs} \label{sec:proofs}

\subsection{Proof of Propositions~\ref{prop1} and \ref{prop1-b}}

The results follow from Proposition~\ref{prop:gMH}, by the discussion at the end of Section~\ref{sec:extension}.

\subsection{Proof of Proposition~\ref{prop:gMH}}

First, the transition kernel $K(y_1 | y_0)$ can be expressed as
\begin{align}
K(y_1 | y_0) \,\dif y_1 = Q (y_1 |y_0) \rho(y_1 | y_0) \dif y_1 + (1- r(y_0)) \delta_{J y_0} (\dif y_1), \label{prf:gMH-1}
\end{align}
where $r(y_0) = \int Q (y_1 |y_0) \rho(y_1 | y_0) \dif y_1$ and $\delta_y$ denotes point mass at $y$.
Then for $y_1 \not= J y_0$,
\begin{align*}
\pi(y_0) K(y_1 | y_0) = \pi(y_0) Q (y_1 |y_0) \rho(y_1 | y_0).
\end{align*}
Replacing $(y_0, y_1)$ with $(J^{-1} y_1, J y_0)$ above shows that for $J y_0 \not= y_1$,
\begin{align*}
\pi(J^{-1} y_1) K( Jy_0 | J^{-1} y_1) = \pi( J^{-1} y_1) Q ( J y_0 | J^{-1} y_1) \rho( J y_0 | J^{-1} y_1 ),
\end{align*}
where $ \pi( J^{-1} y_1)  = \pi (y_1)$ by the invariance property (\ref{eq:invariance}) and
$$
\rho( J y_0 | J^{-1} y_1 ) = \min\left(1 , \frac{\pi(y_0) Q ( y_1 | y_0) }{ \pi(J^{-1} y_1) Q( Jy_0 | J^{-1}y_1) } \right).
$$
Hence (\ref{eq:gMH-gDB}) holds for $J y_0 \not= y_1$, because
\begin{align}
 &\pi(y_0) Q (y_1 |y_0) \rho(y_1 | y_0) =\pi( J^{-1} y_1) Q ( J y_0 | J^{-1} y_1) \rho( J y_0 | J^{-1} y_1 ) \nonumber \\
 &=\min\left( \pi(y_0) Q ( y_1 | y_0), \pi(J^{-1} y_1) Q( Jy_0 | J^{-1}y_1) \right), \label{prf:gMH-2}
\end{align}
which holds whether $J y_0 = y_1$ or not.

The proof that $\pi(y)$ is a stationary distribution is a generalization of \cite{Tierney1994}.
It suffices to show that for any set $C \subset \mathcal Y$,
\begin{align}
\int_C \left(\int \pi(y_0) K(y_1 | y_0)\,\dif y_0 \right) \,\dif y_1 = \int_C \pi(y_1) \,\dif y_1. \label{prf:gMH-3}
\end{align}
By (\ref{prf:gMH-1}), the left-hand side of (\ref{prf:gMH-3}) can be calculated as
\begin{align*}
& \int_C \left(\int \pi(y_0) Q (y_1 |y_0) \rho(y_1 | y_0) \,\dif y_0 \right) \,\dif y_1 + \int_{J^{-1}(C)} (1-r(y_0)) \pi(y_0)\,\dif y_0 \\
%& = \int_C \left(\int\pi( J^{-1} y_1) Q ( J y_0 | J^{-1} y_1) \rho( J y_0 | J^{-1} y_1 ) \,\dif y_0 \right) \,\dif y_1 + \int_{J^{-1}(C)} (1-r(y_0)) \pi(y_0)\,\dif y_0 \\
& = \int_C \left(\int Q ( J y_0 | J^{-1} y_1) \rho( J y_0 | J^{-1} y_1 ) \,\dif y_0 \right) \pi( J^{-1} y_1) \,\dif y_1 + \int_{J^{-1}(C)} (1-r(y_0)) \pi(y_0)\,\dif y_0 \\
& = \int_C   r(J^{-1} y_1 ) \pi( J^{-1} y_1) |\det(J^{-1})| \,\dif y_1 + \int_{J^{-1}(C)} (1-r(y_0)) \pi(y_0)\,\dif y_0 \\
& = \int_C   r(J^{-1} y_1 ) \pi( J^{-1} y_1) |\det(J^{-1})|  \,\dif y_1 + \int_C (1-r(J^{-1} y_0)) \pi(J^{-1} y_0) |\det(J^{-1})| \,\dif y_0 \\
& = \int_C   \pi( J^{-1} y_1) |\det(J^{-1})| \,\dif y_1 = \int_{J(C)}   \pi(y_1)\,\dif y_1,
\end{align*}
which yields the right-hand side of (\ref{prf:gMH-3}) by the invariance property (\ref{eq:invariance}).
The first equality follows from (\ref{prf:gMH-2}), the second from the definition of $r(\cdot)$ and the change of variables, and the third and fifth both from the change of variables.

\subsection{Proof of Corollary~\ref{cor:rejection-free}}

The result follows from Corollary~\ref{cor:rejection-free2}, by the discussion at the end of Section~\ref{sec:extension}.

\subsection{Proof of Corollary~\ref{cor:rejection-free2}}

The backward proposal scheme (\ref{eq:g2ms-b}) becomes $J y_0 = (I- A) y^* + Z^*$.
%\begin{align}
%    J y_0 = (I- A) y^* + Z^* . \label{eq:g2ms-b2}
%\end{align}
The new noise $Z^*$ can be directly calculated using (\ref{eq:g2ms-f2}) as
\begin{align*}
Z^* = J y_0 -(I- A) y^* = (2A - A^2) J y_0 - (I-A)Z,
\end{align*}
which is distributed as $\mathcal N(\bo,2A - A^2 )$, if $y_0 \sim \mathcal N(\bo, I)$, independently of $Z \sim \mathcal N(\bo, 2 A - A^2)$.
Hence if $y_0 \sim \mathcal N(\bo, I)$ and $y^*$ is generated by (\ref{eq:g2ms-f2}) with $Z$ independent of $y_0$, then
the conditional density of $y^*$ given $y_0$ is $p(y^*| y_0) = \mathcal N(Z| \bo, 2A-A^2)$ and
the conditional density of $J y_0$ given $y^*$ is $p(J y_0 | y^*) = \mathcal N(Z^* | \bo, 2A-A^2)$.
By the change of variables, the conditional density of $y_0$ given $y^*$ is also $p(y_0 |y^*) = \mathcal N(Z^* | \bo, 2A-A^2)$ because $| \det(J) |=1$.
Therefore, the acceptance probability (\ref{eq:g2ms-prob}) reduces to 1, because $\pi(y_0) p(y^*| y_0) = \pi(y^*) p(y_0 | y^*)$:
both $\pi(y_0) p(y^*| y_0)$ and $\pi(y^*) p(y_0 | y^*)$ give the joint density of $(y_0, y^*)$.

\subsection{Proof of Lemma \ref{lem:hamsa}}

\label{subsec:proof:leam:hamsa}

The HAMS-A proposal described in Section \ref{subsec:hams-ab} is
\begin{align*}
    & \tilde Z = Z - a\nabla U(x_0) + \sqrt{ab} u_0,\quad  Z\sim \mathcal N(\bo,a(2-a-b)I),\\
    & x^* = x_0 + \tilde Z,\quad u^* = -u_0 + \sqrt{\frac{b}{a}}\tilde Z + \phi (\tilde Z + \nabla U(x_0) - \nabla U(x^*)), \\
    & Z^* = \tilde Z - a\nabla U(x^*) - \sqrt{ab} u^*.
\end{align*}
We express $x^*,u^*$ and $Z^*$ in terms of $x_0,u_0,Z$ and $\nabla U(x^*)$:
\begin{align}
    x^* & = x_0 - \nabla U(x_0) + \sqrt{ab}u_0 + Z, \label{eq:hamsamrapp1} \\
    u^* & = [\phi - \phi a - \sqrt{ab}]\nabla U(x_0) - \phi \nabla U(x^*) + [\phi \sqrt{ab} + b - 1] u_0
    + \left[\phi + \sqrt{\frac{b}{a}}  \right] Z,  \\
    Z^* & = [ab + \phi a\sqrt{ab} - \phi \sqrt{ab} - a] \nabla U(x_0) + (\sqrt{ab}\phi - a)\nabla U(x^*) \nonumber \\
    & \quad   +[2\sqrt{ab} - \phi ab - b\sqrt{ab} ] u_0 + \left[1-\phi\sqrt{ab} -b\right] Z. \label{eq:hamsamrapp2}
\end{align}

Suppose that the target density $\pi(x)$ is $\mathcal N(\bo, \gamma^{-1}I)$. Then $x^*,u^*$ and $Z^*$ from (\ref{eq:hamsamrapp1})--(\ref{eq:hamsamrapp2})
can be expressed in terms of only  $x_0,u_0$ and $Z$ as
\begin{align}
  x^* & = (-a\gamma + 1)x_0 + \sqrt{ab}u_0 + Z, \label{eq:hamsamrapp3} \\
  u^* & = [ \underbrace{ a\phi \gamma^2}_{(i)} - (a\phi + \sqrt{ab})\gamma ]x_0 + [\underbrace{-\phi\sqrt{ab}\gamma}_{(iii)} + \phi \sqrt{ab} + b - 1]u_0
          + \left[\underbrace{-\phi\gamma}_{(v)} + \phi + \sqrt{\frac{b}{a}}\right]Z,\\
  Z^* & = [\underbrace{ (a^2 - \phi a\sqrt{ab})\gamma^2}_{(ii)} + (ab - 2a + \phi a\sqrt{ab})\gamma  ] x_0 \nonumber \\
      & + [\underbrace{(\phi ab - a\sqrt{ab})\gamma}_{(iv)} + 2\sqrt{ab} - b \sqrt{ab} -\phi ab]u_0 \nonumber \\
      & + [\underbrace{(\phi\sqrt{ab} - a)\gamma}_{(vi)} + 1 - b -\phi\sqrt{ab}] Z.\label{eq:hamsamrapp4}
\end{align}
The quantity inside the exponential in (\ref{mr2}) is
\begin{align}
  & H(x_0,u_0) - H(x^*,u^*) + \frac{Z^\T Z-(Z^*)^\T Z^*}{2a(2-a-b)}   \nonumber \\
  & = \frac{\gamma}{2}x_0^\T x_0 - \frac{\gamma}{2}(x^*)^\T x^* + \frac{1}{2}u_0^\T u_0 - \frac{1}{2}(u^*)^\T u^* + \frac{Z^\T Z}{2a(2-a-b)} - \frac{(Z^*)^\T Z^*}{2a(2-a-b)}. \label{eq:hamsamrapp5}
\end{align}
Substituting (\ref{eq:hamsamrapp3})--(\ref{eq:hamsamrapp4}) into the above shows that
(\ref{eq:hamsamrapp5}) can be expressed as a quadratic form in $x_0,u_0$ and $Z$:
\begin{align*}
  (x_0^\T, u_0^\T, Z^\T) G(\gamma) (x_0^\T, u_0^\T, Z^\T)^\T,
\end{align*}
where $G(\gamma)$ is a $3 \times 3$ block matrix. For $i,j=1,2,3$, the $(i,j)$th block of $G(\gamma)$ is of the form $g_{ij}(\gamma) I$,
where $g_{ij}(\gamma)$ is a scalar, polynomial of $\gamma$, with coefficients depending on $(a,b,\phi)$.

Now we compute the coefficients of the
leading terms (terms corresponding to highest power of $\gamma$) of $g_{11}(\gamma)$, $g_{22}(\gamma)$ and $g_{33}(\gamma)$.
Because we focus on only the leading terms,
it is sufficient to examine (\ref{eq:hamsamrapp3})--(\ref{eq:hamsamrapp4}) and account for the coefficients of $x_0,u_0,Z$, labeled as $(i),.., (v)$,
which lead to the highest power of $\gamma$ in $g_{11}(\gamma)$, $g_{22}(\gamma)$ and $g_{33}(\gamma)$.
The coefficient of the leading term of $g_{11}(\gamma)$ associated with $x_0^\T x_0$ is
\begin{align}
  & -\frac{(i)^2}{2} - \frac{(ii)^2}{2a(2-a-b)} = -\frac{1}{2}(a\phi)^2\gamma^4 - \frac{(a^2 - \phi a\sqrt{ab})^2 \gamma^4 }{2a(2-a-b)}\nonumber \\
  & = \frac{\gamma^4}{2(2-a-b)}(-a^2\phi^2(2-a-b) + 2\phi a^2\sqrt{ab} - \phi^2a^2b - a^3)\nonumber \\
  & = \frac{\gamma^4 a^2}{2(2-a-b)} (\phi^2(a-2) + \phi 2\sqrt{ab} - a). \label{eq:hamsamrapp6}
\end{align}
The coefficient of the leading term of $g_{22}(\gamma)$  associated with $u_0^\T u_0$ is
\begin{align}
  & -\frac{(iii)^2}{2} - \frac{(iv)^2}{2a(2-a-b)} = -\frac{1}{2}(\phi \sqrt{ab})^2 \gamma^2 - \frac{(\phi ab - a \sqrt{ab})^2 \gamma^2}{2a(2-a-b)} \nonumber \\
  & = \frac{\gamma^2}{2(2 - a -b)}(2\phi ab\sqrt{ab} - a^2b - \phi^2ab^2 - 2\phi^2ab + \phi^2 a^2b + \phi^2ab^2) \nonumber \\
  & = \frac{\gamma^2 ab}{2(2-a-b)}(\phi^2(a-2) + \phi 2\sqrt{ab} - a). \label{eq:hamsamrapp7}
\end{align}
The coefficient of the leading term of $g_{33}(\gamma)$ associated with $Z^\T Z$ is
\begin{align}
  &-\frac{(v)^2}{2} - \frac{(vi)^2}{2a(2-a-b)} = - \frac{1}{2}\phi^2\gamma^2 - \frac{(\phi\sqrt{ab} - a)^2\gamma^2}{2a(2-a-b)}\nonumber \\
  & = \frac{\gamma^2}{2(2-a-b)} (2\phi\sqrt{ab} - a - \phi^2b - 2\phi^2 + a\phi^2 + b\phi^2)\nonumber \\
  & = \frac{\gamma^2}{2(2-a-b)}(\phi^2(a-2) + \phi 2\sqrt{ab} - a). \label{eq:hamsamrapp8}
\end{align}
Notice that (\ref{eq:hamsamrapp6})--(\ref{eq:hamsamrapp8}) involve $\phi$ only through the same quadratic function of $\phi$:
\begin{align*}   %\label{eq:hamsamrapp9}
    h(\phi) = \phi^2(a-2) + \phi 2\sqrt{ab} - a.
\end{align*}
For $a>0,b\geq 0 $ and $a + b \leq 2$, we have $h(\phi)\leq 0$, because $(2\sqrt{ab})^2 +4a(a-2) = 4a(a+b-2)\leq 0$. Hence $|h(\phi)|$ is minimized at $\phi = -2\sqrt{ab}/(2(a-2)) = \sqrt{ab}/(2-a)$.

%----------------------------------------------------------------------------------
\subsection{Proof of Lemma \ref{lem:hamsb}}
\label{subsec:proof:lem:hamsb}
We use the following choice of $A$ in (\ref{eq:matrix-A}): $a_1 = 2 - a, a_2 = \sqrt{ab}, a_3 = 2-b$ with the constraints on $a,b$ that $a>0,b\geq 0$ and $a+b\leq 2$.
The noise terms are proportional: $Z_2 = -\sqrt{b/a} Z_1$.
The new noises $Z_1^*$ and $Z_2^*$, defined by (\ref{eq:forward-Z}), (\ref{eq:forward-xu2}), and (\ref{eq:backward-Zdef}), can be expressed in terms of $u_0,\nabla U(x_0), \nabla U(x^*)$ and $Z_1$ as
\begin{align*}
  Z_1^* &= \underbrace{\sqrt{ab}(b-\phi\sqrt{ab})}_{\theta_1}u_0 + \underbrace{(a+ab-2 - \phi(a-1)\sqrt{ab})}_{\theta_2} \nabla U(x_0)  \\
  & + \underbrace{(a-2 + \phi \sqrt{ab})}_{\theta_3}\nabla U(x^*) + \underbrace{(b+1-\phi\sqrt{ab})}_{\theta_4} Z_1, \\ %\label{eq:hamsbapp1} \\
  Z_2^* &= \underbrace{(2-b)(b-\phi\sqrt{ab})}_{\psi_1}u_0 + \underbrace{(\sqrt{ab}(1-b) - \phi(a-1)(2-b))}_{\psi_2} \nabla U(x_0) \\
  & + \underbrace{(-\sqrt{ab} + \phi (2-b) )}_{\psi_3}\nabla U(x^*) + \underbrace{(\sqrt{b/a}(1-b) - \phi(2-b) )}_{\psi_4} Z_1. %\label{eq:hamsbapp2}
\end{align*}
Suppose there exists $r\in\mathbb{R}$ such that $Z_2^* = r Z_1^*$ for arbitrary values of $x_0, u_0$ and $Z_1$. Then
the coefficients, denoted as $\theta_1,\ldots,\theta_4, \psi_1,\ldots,\psi_4$, satisfy
\begin{align}
  r\theta_1 = \psi_1, \quad r\theta_2 = \psi_2, \quad r\theta_3 =\psi_3, \quad r\theta_4 =\psi_4 .\label{eq:hamsbapp4}
\end{align}
We study the following possibilities.

First, suppose that $\theta_1\neq 0$. Then $r = \frac{\psi_1}{\theta_1} = \frac{2-b}{\sqrt{ab}}$  by (\ref{eq:hamsbapp4}).
Substituting this into $r\theta_4 = \psi_4$ in (\ref{eq:hamsbapp4}) yields
\begin{align*}
  & r\theta_4 = \psi_4 \Rightarrow \frac{2-b}{\sqrt{ab}}(b+1 - \phi\sqrt{ab}) =  \sqrt{b/a}(1-b) - \phi(2-b) \\
  & \Rightarrow \frac{(2-b)(b+1)}{\sqrt{ab}} = \sqrt{b/a}(1-b)
  \Rightarrow  (2-b)(b+1)  = b(1-b) \\
  & \Rightarrow b - b^2 +2 = b - b^2 \Rightarrow 0 = 2,
\end{align*}
which is a contradiction. Hence $\theta_1 = \psi_1 = 0$, which gives two possibilities: either $b = 0$ or $ \phi = \sqrt{b/a}$.

Next suppose that $b = 0$. Then $\theta_4 = 1$ and $\psi_4 = -2\phi$, and hence  $r = \psi_4/\theta_4 = -2\phi$ by (\ref{eq:hamsbapp4}).
Moreover, $\theta_2 = a-2$ and $\psi_2 = -2\phi(a - 1)$, and
\begin{align*}
r \theta_2 = \psi_2 \Rightarrow -2\phi(a-2) = -2\phi(a-1),
\end{align*}
which implies that $\phi = 0$. Thus if $b = 0$, then $\phi = 0$ as well. This gives the
trivial case that $r = 0$ and $Z_2^* \equiv 0$.

Finally suppose that $\phi = \sqrt{b/a}$. Then  $Z_2^* = rZ_1^*$ is satisfied with $r = -\sqrt{b/a}$ by the following calculation:
\begin{align*}
  & \theta_1 = \psi_1 = 0, \\
  & \theta_2 = a + ab - 2 - b(a - 1) = a + b - 2, \\
  & \psi_2 = \sqrt{ab}(1-b) - \sqrt{\frac{b}{a}}(a-b)(2-b) = -\sqrt{\frac{b}{a}}(a + b - 2 ) = r\theta_2, \\
  & \theta_3 = a - 2 + b,  \quad
  \psi_3 = -\sqrt{ab} + \sqrt{\frac{b}{a}}(2-b) = - \sqrt{\frac{b}{a}}(b - 2 + a) = r\theta_3, \\
  & \theta_4 = b + 1 - b = 1, \quad
  \psi_4 = \sqrt{\frac{b}{a}}(1-b) - \sqrt{\frac{b}{a}}(2-b)  = -\sqrt{\frac{b}{a}} = r\theta_4.
\end{align*}
Therefore $Z_2^* = r Z_1^*$ if and only if $r = - \sqrt{b/a}$ and $\phi = \sqrt{b/a}$, which also includes the trivial case, $r = \phi = b = 0$.

%----------------------------------------------------------------------------------
\subsection{Proof of Lemma~\ref{lem:b-choice}}

By the rejection-free property, $(x_1, u_1) = (x^*, u^*)$ when the target density $\pi(x)$ is $\mathcal N(\bo,I)$.
We give a proof for HAMS-A and HAMS-B separately.

For HAMS-A, the lag-1 auto-covariance matrix is
\[C_A =
 \Cov((x^*,u^*), (x_0,u_0))= \begin{pmatrix}
   (1-a) I & \sqrt{ab} I \\
   -\sqrt{ab} I & (b-1) I
 \end{pmatrix}.
\]
The eigenvalues of $C_A$ are the eigenvalues of $C_A$ with $I=1$, each with multiplicities $k$.
Henceforth we assume $I=1$. The two eigenvalues of $C_A$ are
\[
\lambda_1=  \frac{1}{2} ( b-a+ \sqrt{\Delta}), \quad \lambda_2 =  \frac{1}{2} (b-a- \sqrt{\Delta}),
\]
where
$$
\Delta = (a+b-2)^2-4ab = \{ 2 - (\sqrt{a}-\sqrt{b})^2 \} \{ 2- (\sqrt{a} + \sqrt{b})^2 \}.
$$
Given $a \in(0,2)$, we show that the choice of $b \in (0,2-a)$ which minimizes $\max( |\lambda_1| , |\lambda_2|)$ is
$b^* = (\sqrt{2} -\sqrt{a})^2$, where $|\cdot|$ denotes the modulus. For this choice $b^*$, $\Delta =0$ and the two eigenvalues are identical, $\lambda_1^*=\lambda_2^* = 1 -\sqrt{2a}$.
We distinguish three cases.
\begin{itemize}
\item[(i)] Suppose $(\sqrt{a}+\sqrt{b})^2 > 2$. Then $\lambda_1$ and $\lambda_2$ are complex, and
\begin{align*}
& |\lambda_1 |^2 = |\lambda_2|^2 = \lambda_1 \lambda_2 = b+a -1 \\
& > (\sqrt{2} -\sqrt{a})^2 + a -1 = (\sqrt{2a}-1)^2 = \lambda_1^{* 2}.
\end{align*}

\item[(ii)] Suppose $(\sqrt{a}+\sqrt{b})^2 < 2$ and $b \ge a$. Then $\lambda_1 (>0)$ and $\lambda_2$ are real, and $\max(|\lambda_1|, |\lambda_2|) = \lambda_1$.
For fixed $a$, the derivative of $\lambda_1$ with respect to $b$ is
\begin{align*}
\frac{\dif \lambda_1}{\dif b} = \frac{1}{2} \left( 1 + \frac{b-a-2}{\sqrt{\Delta}} \right) \le \frac{1}{2} \frac{(2-a-b) + (b-a-2)}{\sqrt{\Delta}} = \frac{-a}{\sqrt{\Delta}} <0,
\end{align*}
where the first inequality uses $\sqrt{\Delta} \le 2-a-b$. Then $\lambda_1$ is decreasing in $b$, which is upper-bounded by $b^*=(\sqrt{2}-\sqrt{a})^2$. Hence $\lambda_1>\lambda_1^*$.

\item[(iii)] Suppose $(\sqrt{a}+\sqrt{b})^2 < 2$ and $b \le  a$. Then $\lambda_1$ and $\lambda_2 (<0)$ are real, and $\max(|\lambda_1|, |\lambda_2|) = -\lambda_2$.
For fixed $a$, the derivative of $\lambda_2$ with respect to $b$ is
\begin{align*}
\frac{\dif \lambda_2}{\dif b} = \frac{1}{2} \left( 1 - \frac{b-a-2}{\sqrt{\Delta}} \right) = \frac{1}{2} \left( 1+\frac{2+a-b}{\sqrt{\Delta}} \right) >0.
\end{align*}
Then $\lambda_2$ is increasing in $b$, which is upper-bounded by $\min(a, b^*)$.
If $b^* \le a$, then $|\lambda_2| = -\lambda_2 > -\lambda_2^* = |\lambda_2^*|$.
If $a < b^*$, then $|\lambda_2| = -\lambda_2$ is greater than the value of $-\lambda_2$ corresponding $b=a$,
which is identical to the value of $\lambda_1$ (due to $b=a$) and still greater than $|\lambda_1^*|$ by the conclusion from (iii).
\end{itemize}
Combining the three cases shows that $\max( |\lambda_1| , |\lambda_2|) \ge |\lambda_1^*| = |\lambda_2^*|$.

For HAMS-B, we work with the equations (\ref{eq:update-x2})--(\ref{eq:update-u2}) with $a_1 = 2-a,a_3 = 2-b$ and
$a_2 = \sqrt{ab}$, that is, before the reparametrization $ab = \tilde a \tilde b$ and $a(2-a-b) = \tilde a (2 - \tilde a - \tilde b)$.
Then the lag-1 auto-covariance matrix is
\[
  C_B =  \Cov((x^*,u^*), (x_0,u_0))= \begin{pmatrix}
    (-1+a) I & \sqrt{ab}I \\
    -\sqrt{ab}I & (1-b)I
  \end{pmatrix}.
\]
The eigenvalues of $C_B$ are the same as those of $C_A$.
Hence the maximum modulus of eigenvalues is also minimized by the choice $b = (\sqrt{2} - \sqrt{a})^2$.
By the reparametrization $ab = \tilde a \tilde b$ and $a(2-a-b) = \tilde a (2 - \tilde a - \tilde b)$,
the resulting choice is $\tilde b = \frac{\tilde a(2-\tilde a)}{(\sqrt{2} + \sqrt{2-\tilde a})^2}$,
which gives the desired expression with $(\tilde a, \tilde b)$ relabeled as $(a,b)$.

\subsection{Simplification of preconditioning for Algorithm \ref{alg3} }
\label{subsec:simplify:alg3}

\begin{algorithm}[t] %[H]
  \SetAlgoLined
  Initialize $x_0,u_0$\\
   \For{ $t = 0,1,2,..., N_{iter}$}{
     Sample $w \sim \text{Uniform} [0,1]$ and $\zeta \sim \mathcal N(\bo,I)$ \\
     Transform $\tilde x_t = L^\T x_t$ \\
     $\tilde x^* = \tilde x_t - a L^{-1}\nabla U(x_t) + \sqrt{ab} u_t + \sqrt{a(2-a-b)}\zeta$\\
     Propose $x^* = (L^\T)^{-1}\tilde x^*$\\
     \If{HAMS-A}
     {
        Propose $u^* = \left(\frac{2b}{2-a} - 1\right) u_t - \frac{\sqrt{ab}}{2-a}L^{-1}(\nabla U(x_t) + \nabla U(x^*)) + \frac{2\sqrt{b(2-a-b)}}{2-a}\zeta$\\
        $\zeta^* = \left(1-\frac{2b}{2-a}\right) \zeta - \frac{\sqrt{a(2-a-b)}}{2-a}L^{-1}(\nabla U(x_t) + \nabla U(x^*)) + \frac{2\sqrt{b(2-a-b)}}{2-a}u_t$
     }
     \If{HAMS-B}
     {  Propose $u^* = u_t - \frac{\sqrt{ab}}{2-a}L^{-1}(\nabla U(x_t) + \nabla U(x^*))$\\
        $\zeta^* = \zeta - \frac{\sqrt{a(2-a-b)}}{2-a}L^{-1}(\nabla U(x_t) + \nabla U(x^*))$
     }
     $\rho = \exp\left\{H(x_t,u_t) - H(x^*,u^*) + \frac{1}{2}\zeta^\T \zeta - \frac{1}{2}(\zeta^*)^\T \zeta^*\right\}$\\
    \eIf{$w < \min(1,\rho)$}{ $(x_{t+1},u_{t+1}) = (x^*,u^*)$ \qquad\# Accept
     }{
      $(x_{t+1},u_{t+1}) = (x_t,-u_t)$ \qquad\# Reject
    }
   }
   \caption{HAMS-A/HAMS-B (with preconditioning non-simplified)}  \label{alg2star}
\end{algorithm}

As discussed in Section \ref{subsec:precondition} for preconditioning, we apply the linear transformations $\tilde{x} = L^\T x$ and $\nabla U(\tilde x) = L^{-1}\nabla U(x)$ to HAMS-A/B in
Algorithm \ref{alg2}. We show the the resulting algorithm, stated as Algorithm \ref{alg2star} here, can be rearranged in an equivalent but computationally more efficient form as Algorithm \ref{alg3}.

Suppose that the equivalence holds for $(x_t,u_t)$. By the relation $\nabla U(\tilde x_t) = L^{-1} \nabla U(x_t)$ and the definition of $\xi$ in Algorithm~\ref{alg3}, we have
\begin{align*}
  \tilde x^* & = \tilde x_t - a\nabla U(\tilde x_t) + \xi \\
             & = \tilde x_t - a L^{-1} \nabla U(x_t) + \sqrt{ab} u_t + \sqrt{a(2-a-b)}\zeta.
\end{align*}
Hence, when the proposal is accepted, $x_{t+1}=x^* = (L^\T)^{-1} \tilde x^*$ in both algorithms.
By the relation $\tilde{\xi} = \nabla U(\tilde x) + L^{-1}\nabla U(x^*) = L^{-1}(\nabla U(x_t) +\nabla U(x^*) )$, we see that
when the proposal is accepted, the expressions of $u_{t+1}$ are the same in both algorithms.
When the proposal is rejected, $(x_{t+1},u_{t+1})= (x_t,-u_t)$ is also the same in the two algorithms.

To show  the equivalence holds for $(x_{t+1},u_{t+1})$, it remains to check that the acceptance probabilities are equal in the two algorithms. We need to show
\begin{align*}
 U(x_t) - U(x^*) & + \frac{1}{2-a}(\tilde{\xi})^\T (\xi - \frac{a}{2}\tilde{\xi} )  = H(x_t,u_t) - H(x^*,u^*) + \frac{1}{2}\zeta^\T \zeta - \frac{1}{2}(\zeta^*)^\T \zeta^*,
\end{align*}
which is equivalent to
\begin{align*}
\frac{2}{2-a}(\tilde{\xi})^\T (\xi - \frac{a}{2}\tilde{\xi} ) =  u_t^\T u_t -  (u^*)^\T u^* +\zeta^\T \zeta - (\zeta^*)^\T \zeta^*,
\end{align*}
because $H(x_t,u_t) - H(x^*,u^*) =U(x_t) - U(x^*) + \frac{1}{2} u_t^\T u_t - \frac{1}{2} (u^*)^\T u^*$.

Consider the algorithm HAMS-B. We use the following fact
\begin{align}
  u_t^\T u_t -  (u^*)^\T u^* = (u_t - u^*)^\T(u_t + u^*),\quad \zeta^\T \zeta - (\zeta^*)^\T \zeta^*  = (\zeta_t - \zeta^*)^\T(\zeta_t + \zeta^*). \label{prcapp0}
\end{align}
By direct calculation, we have
\begin{align}
  & u_t - u^* = \frac{\sqrt{ab}}{2-a} L^{-1}(\nabla U(x_t) + \nabla U(x^*) ) = \frac{\sqrt{ab}}{2-a} \tilde{\xi} ,  \label{prcapp1} \\
  %& u_t + u^* = 2u_t + \frac{\sqrt{ab}}{2-a} L^{-1}(\nabla U(x_t) + \nabla U(x^*) ) = 2u_t - \frac{\sqrt{ab}}{2-a}\tilde{\xi}, \nonumber \\
  &  (u_t - u^*)^\T(u_t + u^*)  = \frac{\sqrt{ab}}{2-a} (\tilde{\xi})^\T \left(2u_t - \frac{\sqrt{ab}}{2-a}\tilde{\xi}\right), \label{prcapp2}
\end{align}
and
\begin{align}
  & \zeta - \zeta^{*} = \frac{\sqrt{a(2-a-b)}}{2-a}L^{-1}(\nabla U(x_t) + \nabla U(x^*) ) = \frac{\sqrt{a(2-a-b)}}{2-a}\tilde{\xi}, \label{prcapp3} \\
  %& \zeta + \zeta^{*} = 2\zeta - \frac{\sqrt{a(2-a-b)}}{2-a}L^{-1}(\nabla U(x_t) + \nabla U(x^*) ) = 2\zeta -  \frac{\sqrt{a(2-a-b)}}{2-a}\tilde{\xi}, \nonumber \\
  &  (\zeta - \zeta^{*})^\T (\zeta + \zeta^{*} ) = \frac{\sqrt{a(2-a-b)}}{2-a} (\tilde{\xi})^\T \left(2\zeta -  \frac{\sqrt{a(2-a-b)}}{2-a}\tilde{\xi}\right). \label{prcapp4}
\end{align}
Combining (\ref{prcapp0})--(\ref{prcapp3}) yields
\begin{align}
  &  u_t^\T u_t -  (u^*)^\T u^* +\zeta^\T \zeta - (\zeta^*)^\T \zeta^*  \nonumber \\ % = (u_t - u^*)^\T(u_t + u^*) + (\zeta - \zeta^{*})^\T (\zeta + \zeta^{*} ) \\
  & =(\tilde{\xi})^\T \left(\frac{2\sqrt{ab}}{2-a}u_t + \frac{2\sqrt{a(2-a-b)}}{2-a}\zeta - \left( \frac{ab}{(2-a)^2} + \frac{a(2-a-b)}{(2-a)^2}\right)\tilde{\xi}  \right) \nonumber  \\
  & = \frac{2}{2-a}(\tilde{\xi})^\T\left(\sqrt{ab} u_t + \sqrt{a(2-a-b)}\zeta - \frac{a}{2}\tilde{\xi} \right) \label{prcapp5} \\
  & = \frac{2}{2-a}(\tilde{\xi})^\T\left(\xi - \frac{a}{2}\tilde{\xi} \right).  \nonumber
\end{align}
Hence the acceptance probabilities match for HAMS-B in Algorithms~\ref{alg3} and \ref{alg2star}.

Finally consider the algorithm HAMS-A. Define intermediate variables
\begin{align*}
  & u^{\dagger} = \left(\frac{2b}{2-a} - 1\right)u_t + \frac{2\sqrt{b(2-a-b)}}{2-a} \zeta, \\
  & \zeta^{\dagger} = \left(1 - \frac{2b}{2-a} \right)\zeta + \frac{2\sqrt{b(2-a-b)}}{2-a} u_t.
\end{align*}
Then the following identities hold:
\begin{align}
  &(u^{\dagger})^\T u^{\dagger} + ( \zeta^{\dagger} )^\T  \zeta^{\dagger}  = u_t^\T u_t + \zeta^\T \zeta, \label{prcapp6} \\
  & \sqrt{ab} u^\dag + \sqrt{a(2-a-b)}\zeta^\dag  = \sqrt{ab} u_t + \sqrt{a(2-a-b)}\zeta \,(=\xi). \label{prcapp7}
\end{align}
Identity (\ref{prcapp6}) follows, because after expanding the inner products on the left hand side, the cross terms cancel out and the squared terms have coefficients
\[
  \left(\frac{2b}{2-a} - 1\right)^2 + \left(\frac{2\sqrt{b(2-a-b)}}{2-a} \right)^2 = 1.
\]
Identity (\ref{prcapp7}) follows because by direct calculation
\begin{align*}
& u^\dag -u_t =  \frac{2\sqrt{2-a-b}}{2-a}  (\sqrt{2-a-b} \,u_t + \sqrt{b} \zeta ),\\
& \zeta^\dag - \zeta = \frac{2\sqrt{b}}{2-a} ( - \sqrt{b} \zeta + \sqrt{2-a-b}\,u_t) .
\end{align*}
Moreover, it can be verified by definition that
\[
  u^{\dagger} - u^* = \frac{\sqrt{ab}}{2-a}\tilde{\xi}, \quad \zeta^{\dagger} - \zeta^* =    \frac{\sqrt{a(2-a-b)}}{2-a}\tilde{\xi}.
\]
Then (\ref{prcapp0})--(\ref{prcapp5}) remain valid with $u_t$ and $\zeta$ replaced by $u^{\dagger}$ and $\zeta^{\dagger}$.
From these equations together with the identities (\ref{prcapp6})--(\ref{prcapp7}), we find
\begin{align*}
  & u_t^\T u_t  -  (u^*)^\T u^* +\zeta^\T \zeta - (\zeta^*)^\T \zeta^* \\
  & = (u^{\dagger})^\T u^{\dagger}  -  (u^*)^\T u^* +(\zeta^{\dagger})^\T \zeta^{\dagger} - (\zeta^*)^\T \zeta^*  \\
  & = \frac{2}{2-a}(\tilde{\xi})^\T\left(\sqrt{ab} u^\dag + \sqrt{a(2-a-b)}\zeta^\dag - \frac{a}{2}\tilde{\xi} \right) \label{prcapp5} \\
  & = \frac{2}{2-a}(\tilde{\xi})^\T\left(\xi - \frac{a}{2}\tilde{\xi} \right).
\end{align*}
Hence the acceptance probabilities match for HAMS-A in Algorithms~\ref{alg3} and \ref{alg2star}.

%%%%%%%%%%%%%%%%%%%%%%%%%%%%%%%%%%%%%%%%%%%%%%%%%%%%%%%%%%%%%%%%%%%%%%%%%%%%%%%%%%%%%%%%%%
%%%---------------------------------------------------------------------------------------
%%%%%%%%%%%%%%%%%%%%%%%%%%%%%%%%%%%%%%%%%%%%%%%%%%%%%%%%%%%%%%%%%%%%%%%%%%%%%%%%%%%%%%%%%%
\section{Details for simulation studies} \label{sec:tech-simulation}

\subsection{Expressions for stochastic volatility model}
\label{subsec:SVexpression}
The stochastic volatility model is defined as
\begin{align*}
  x_t & = \phi x_{t-1} + \eta_t, \quad t = 2, ... , T, \quad x_1\sim\mathcal N\left(0,\frac{\sigma^2}{1-\phi^2}\right),\\
  y_t & = z_t \beta \exp(x_t/2), \quad z_t\stackrel{iid}{\sim} \mathcal N(0,1),\quad \eta_t\stackrel{iid}{\sim}\mathcal N(0,\sigma^2), \quad t = 1, ... ,T.
\end{align*}
Denote $\bx = (x_1,...,x_T)^\T,\by = (y_1,...,y_T)^\T,\bz =(z_1,...,z_T)^\T$ and $\theta = (\beta,\sigma,\phi)^\T$.  The joint density of $(\bx,\by,\theta)$ is
\begin{align*}
  p(\bx,\by,\theta) & = \pi(\theta)\cdot \underbrace{ p(x_1)\prod_{t=2}^T p(x_t|x_{t-1},\phi,\sigma) }_{\mathcal N(\bx|\bo,C)} \cdot
  \overbrace{\prod_{t = 1}^T p(y_t| x_t,\beta)}^{\mathcal N(\by| \bo, \beta^2 \exp(\bx))} \\
  & \propto \pi(\theta) |\det (C)|^{-1/2} \exp\left\{-\frac{1}{2}\bx^\T C^{-1}\bx\right\} \beta^{-T} \exp\left\{-\frac{1}{2}\sum_{t=1}^T(x_t + \beta^{-2}y_t^2\exp(-x_t)) \right\}.
\end{align*}
The matrix $C$ and its inverse are given by
\[ C =
  \frac{\sigma^2}{1-\phi^2}
  \begin{pmatrix}

    1 & \phi & \phi^2 & \cdots & \phi^{T-2}  & \phi^{T-1} \\
    \phi & 1 & \phi &  \cdots& \phi^{T-3} & \phi^{T-2} \\
    \phi^2 & \phi & 1 & \cdots & \phi^{T-4} & \phi^{T-3}\\
    \vdots & \vdots & \vdots &\ddots  &\vdots  & \vdots \\
    \phi^{T-2} & \phi^{T-3} & \phi^{T-4} & \cdots &  1& \phi\\
    \phi^{T-1} & \phi^{T-2} & \phi^{T-3} & \cdots & \phi & 1

  \end{pmatrix}
\]

\[
    \Longleftrightarrow
    C^{-1} =
    \frac{1}{\sigma^2}
  \begin{pmatrix}
      1 & -\phi & 0 & \cdots & 0 &0 \\
      -\phi & 1+\phi^2 & -\phi  & \cdots  & 0 &0 \\
      0 &  -\phi& 1+\phi^2 & \cdots & 0 &0 \\
      \vdots & \vdots & \vdots & \ddots &  \vdots& \vdots\\
      0 & 0 & 0 & \cdots & 1+\phi^2 & -\phi\\
      0 & 0 & 0 & \cdots & -\phi & 1
  \end{pmatrix}.
\]

The conditional posterior of the latent variables is
\[
  p(\bx|\by,\theta) \propto \exp\left\{-\frac{1}{2}\bx^\T C^{-1}\bx\right\}\exp\left\{ - \frac{1}{2}\sum_{t=1}^T(x_t + \beta^{-2}y_t^2\exp(-x_t))\right\}.
\]
Then the negative log-density (or potential function) is
\[
  U(\bx) = \frac{1}{2} \bx^\T C^{-1}\bx + \frac{1}{2}\sum_{t=1}^T(x_t + \beta^{-2}y_t^2\exp(-x_t)),
\]
where dependency on $(\by,\theta)$ is suppressed in the notation. The gradient is
\[
  \nabla U(\bx) = C^{-1}\bx  - \frac{1}{2}\beta^{-2}\by\exp(-\bx) + \frac{1}{2}\mathbf{1},
\]
where $\mathbf{1}$ is a vector of all $1$'s. The hessian is
\begin{align*}
\nabla^2 U(\bx) & = C^{-1} + \frac{1}{2} \diag[ \beta^{-2}\by^2\exp(-\bx) ].
\end{align*}
The square $\by^2$ is taken component-wise. Using the relation between $\by$ and $\bx$,
the diagonal elements in the second term can be expressed as
\[
\beta^{-2}\by^2\exp(-\bx)  = \beta^{-2}\exp(-\bx) \bz^2 \beta^2 \exp(\bx) = \bz^2.
\]
Hence
\[
\EE [ \nabla^2 U(\bx)] = C^{-1} + \frac{1}{2}I,
\]
which leads to the preconditioning in Section \ref{subsec:sim2}. The expectation above is taken over the marginal
distribution of $\bz$. \par

For the parameters, the priors are
\[
  \pi(\beta) \propto \beta^{-1},\quad \sigma^2 \sim \mbox{Inv-}\chi^2(10,0.05), \quad \frac{\phi+1}{2}\sim \mbox{Beta}(20,1.5).
\]
Then $\sigma$ and $\phi$ are also transformed by $\sigma = \exp(\gamma)$ and $\phi = \tanh(\alpha)$. The resulting potential
for the transformed parameters is
\[
  U(\beta,\alpha,\gamma) = (T+1)\log \beta - 20.5\log(1+\tanh \alpha ) - 2\log (1-\tanh \alpha ) \frac{1}{2} \bx^\T C^{-1}\bx + \frac{1}{2}\sum_{t=1}^T\beta^{-2}y_t^2\exp(-x_t),
\]
where dependency on $(\by,\bx)$ is suppressed in the notation. The gradient is
\begin{align*}
\frac{\partial U(\beta,\alpha,\gamma)}{\partial \beta} & = \frac{T+1}{\beta} - \frac{\sum_{t=1}^T y_t^2\exp(-x_t)}{\beta^3} ,\\
\frac{\partial U(\beta,\alpha,\gamma)}{\partial \alpha} & = 22.5 \tanh \alpha - 18.5 - \exp(-2\gamma) x_1^2 \tanh \alpha (1-\tanh^2\alpha),\\
&\quad - \exp(-2\gamma )\sum_{t=2}^T(x_t - \tanh \alpha x_{t-1}) x_{t-1}(1-\tanh^2\alpha),\\
\frac{\partial U(\beta,\alpha,\gamma)}{\partial \gamma} & = - \bx^\T C^{-1} \bx -\frac{1}{2}\exp(-2\gamma) + 10 +T.
\end{align*}
Finally the expected hessian computed with respect to the marginals of $\bx$ and $\bz$ is
\[
  \EE[\nabla^2 U(\beta,\alpha,\gamma)] = \begin{pmatrix}
    (2T-1)/\beta & 0  & 0 \\
    0 & \exp(-2\gamma) + 2T  & 2\tanh \alpha \\
    0 & 2\tanh \alpha & 21.5 - 19.5\tanh^2\alpha + (T-1)(1-\tanh^2\alpha)
    \end{pmatrix}.
\]
When sampling the parameters, we use $M = \Sigma^{-1} = \EE[\nabla^2 U(\beta,\alpha,\gamma)]$ for preconditioning.

%%%%%%%%%%%%%%%%%%%%%%%%%%%%%%%%%%%%%%%%%%%%%%%%%%%%%%%%%%%%%%%%%%%%%%%%%%%%%%%%%%%%%%%%%%
%%%---------------------------------------------------------------------------------------
%%%%%%%%%%%%%%%%%%%%%%%%%%%%%%%%%%%%%%%%%%%%%%%%%%%%%%%%%%%%%%%%%%%%%%%%%%%%%%%%%%%%%%%%%%

\subsection{Expressions for log-Gaussian Cox model}
\label{subsec:Coxexpression}

Denote $\bx = (x_{ij}), \by = (y_{ij}), i,j = 1,...,m$ and let $C$ be the matrix corresponding to the covariance function as described in
Section \ref{subsec:sim3}.
The joint posterior density is
\begin{align*}
 & p(\bx,\sigma^2,\beta|\by) \propto \\
 & \pi(\sigma^2)\pi(\beta)(det|C|)^{-1/2} \exp\left\{-\frac{1}{2}x^\T C^{-1}x\right\} \exp\left\{\sum_{i,j}(y_{ij}(x_{ij}+\mu) - n^{-1}\exp(x_{ij}+\mu) )\right\}.
\end{align*}

The potential function from the conditional posterior of the latent variables given $(\by,\sigma^2,\beta)$ is
\[
  U(\bx) = \frac{1}{2}x^\T C^{-1}x - \sum_{i,j}(y_{ij}x_{ij} - n^{-1}\exp(x_{ij}+\mu)),
\]
where dependency on $(\by,\sigma^2,\beta)$ is suppressed in the notation.
The gradient is
\[
  \nabla U(\bx) = C^{-1}\bx - \by + n^{-1}\exp(\bx + \mu).
\]
The hessian is
\[
  \nabla^2 U(\bx) = C^{-1} + n^{-1}\diag[\bx + \mu].
\]
Because marginally $\bx \sim \mathcal N(0,C)$, we take the expectation
\[
  \EE[\nabla^2 U(\bx)] = C^{-1} + n^{-1}\diag[\sigma^2/2 + \mu],
\]
which is used for preconditioning in Section \ref{subsec:sim3}.

For the parameters, we use the priors $\sigma^2\sim \mbox{Gamma}(2,0.5)$ and $\beta \sim \mbox{Gamma}(2,0.5)$ and the transformations
$\sigma^2 = \exp(\varphi_1),\beta = \exp(\varphi_2)$. Then the potential function from the conditional posterior of transformed parameters given $(\by,\bx)$ is
\[
  U(\varphi_1,\varphi_2) = \frac{1}{2}(\exp(\varphi_1)+ \exp(\varphi_2)) - 2(\varphi_1 + \varphi_2) + \frac{1}{2}\bx^\T C^{-1}\bx + \frac{1}{2}\log \det (C),
\]
where dependency on $(\by,\bx)$ is suppressed in the notation.
The gradient is
\[
  \frac{\partial U(\varphi_1,\varphi_2)}{\partial \varphi_1} = \frac{\exp(\varphi_1)}{2} - 2 + \frac{n}{2} - \frac{1}{2}\bx^\T C^{-1}\bx,
\]

\[
  \frac{\partial U(\varphi_1,\varphi_2)}{\partial \varphi_2} = \frac{\exp(\varphi_2)}{2} - 2 + \frac{1}{2} \tr \left(\frac{\partial C}{\partial \varphi_2}\right) - \frac{1}{2}\bx^ \T C^{-1} \frac{\partial C}{\partial \varphi_2} C^{-1}\bx,
\]
where
\begin{align*}
  & \frac{\partial C}{\partial \varphi_2} [(i,j),(i',j')]  = \\
  & m^{-1}\exp(\varphi_1) \exp(-\varphi_2) \sqrt{(i-i')^2 + (j-j')^2} \exp (-\sqrt{(i-i')^2 + (j-j')^2}/(m\exp(\varphi_2))).
\end{align*}
The marginal expected hessian is
\[
  \EE[\nabla^2 U(\varphi_1,\varphi_2)] = \begin{pmatrix}
    \frac{1}{2}(\exp(\varphi_1) + n) & \frac{1}{2} \tr(C^{-1} \frac{\partial C}{\partial \varphi_2} )  \\
    \frac{1}{2} \tr(C^{-1} \frac{\partial C}{\partial \varphi_2} ) & \frac{1}{2}(\exp(\varphi_1) + \tr( C^{-1} \frac{\partial C}{\partial \varphi_2}C^{-1} \frac{\partial C}{\partial \varphi_2} ))
    \end{pmatrix}.
\]
When sampling the parameters, we use $M = \Sigma^{-1} = \EE[\nabla^2 U(\varphi_1,\varphi_2)]$ for preconditioning.\par

\subsection{Step size tuning} \label{subsec:tuning}

\begin{figure}[H]
  \begin{center}
  \includegraphics[height = 0.3\textheight]{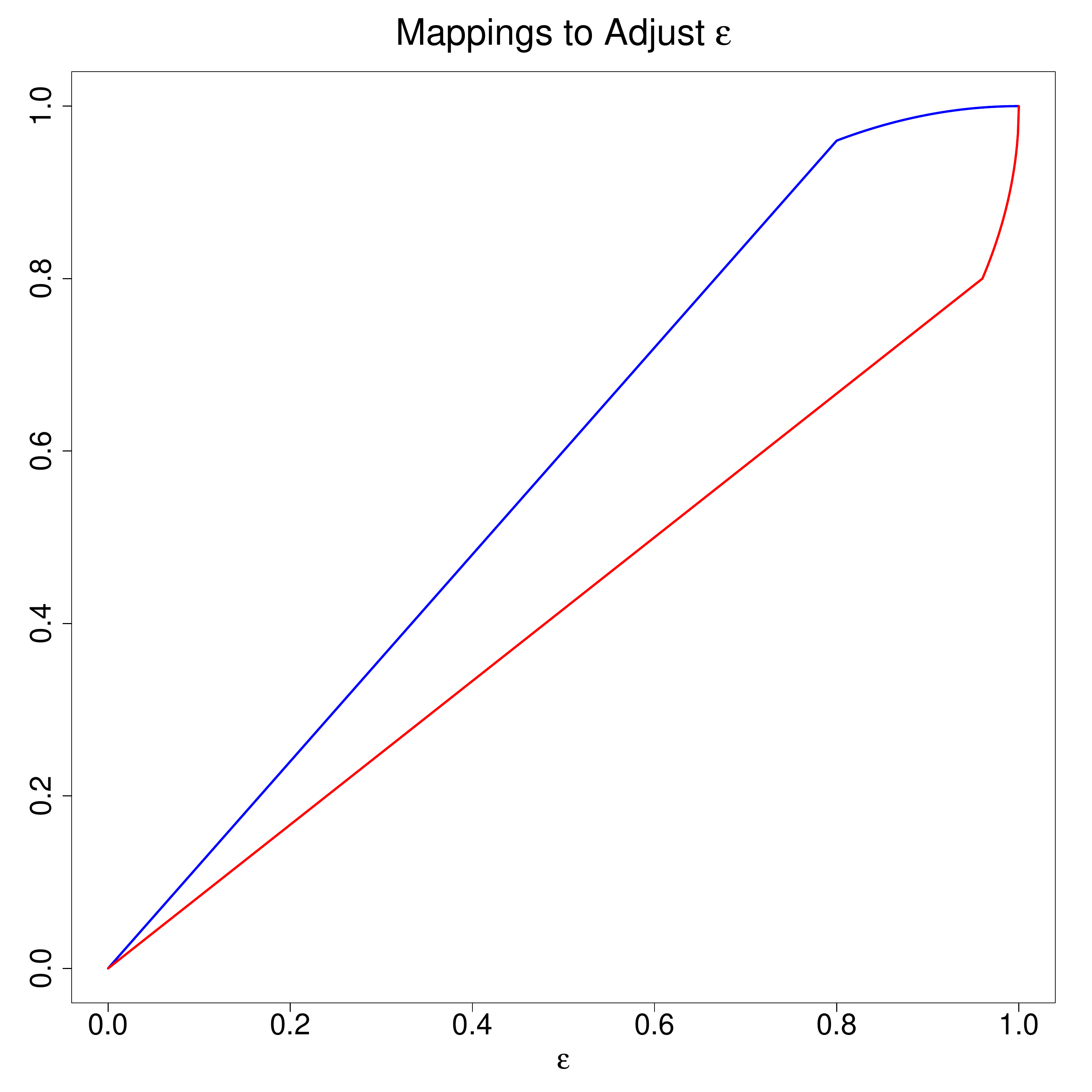}
  \end{center} \vspace{-.3in}
  \caption{ Tuning of step size $\epsilon$ with $\delta = 0.2$. Blue curve is mapping used to increase $\epsilon$. Red curve is mapping used
  to decrease $\epsilon$ \label{fig:tuning}}
\end{figure}

As mentioned in Section \ref{sec:sim}, we periodically adjust step size $\epsilon$ based on the acceptance rate during the burn-in period. When acceptance
is too low (smaller than a lower threshold), we decrease $\epsilon$ by the mapping $\epsilon \gets \max(1-\sqrt{1-\epsilon}, \frac{\epsilon}{1+\delta})$;
when acceptance is too high (larger than a upper threshold), we increase $\epsilon$ by
the mapping $\epsilon \gets \epsilon + \epsilon\cdot \min (1-\epsilon, \delta)$, where $\delta$ is an adjustment value taken to be $\delta = 0.2$ in all our simulations.
The increase and decrease mappings are, by design, inverse of each other, as illustrated in Figure \ref{fig:tuning}. The two mappings are mostly linear, but are curved when $\epsilon$ is close to $1$
to ensure that $\epsilon$ is always between $0$ and $1$ after the update.

\section{Additional simulation results}

We present an experiment with a multivariate normal distribution, and
additional simulation results including pMALA* and GMC from the experiments with the stochastic volatility model
and log-Gaussian Cox model.

%----------------------------------------------------------------------------------
\subsection{Multivariate normal distribution}

Consider the problem of sampling from a 100 dimensional normal distribution
with high correlations: $\pi (x) = \mathcal N(\bo, C)$ where the entries of $C$ are
\[
  C[i,j] = 0.9^{|i - j|},\quad i,j = 1,...,100.
\]
We do not employ any preconditioning here, although we still refer to pMALA and
pMALA* as such. This experiment is used to compare different algorithms
when the variance of the target distribution may not be readily approximated.
Hence potential advantages associated with the rejection-free property are removed from HAMS-A/B.

In terms of tuning, we set $\epsilon = 0.19$ for HAMS-A, HAMS-B, UDL, GMC,
pMALA and pMALA* to maintain acceptance rates around $70\%$. Through empirical trials
we find that HAMS-A, UDL and GMC have good performance using a large carryover ($c$ value), while HAMS-B favors a relatively
small carryover. Hence we set $c = 0.95$ for HAMS-A, UDL and GMC, $c = 0.25$ for HAMS-B. For HMC, we set $nleap = 50$ and
$\epsilon = 0.17$ which also yields a $70\%$ acceptance rate. For RWM, we set $\epsilon = 0.06$ and the resulting
acceptance is around $40\%$. To account for the additional computation cost due to leapfrog steps,
HMC is run for $200$ iterations and all other methods are run for $200\times 50 = 10000$ iterations. The simulation process is repeated for
$100$ times with a fixed starting value of $\bo$.\par

\begin{figure}[H]
  \begin{center}
  \includegraphics[width = 0.95\textwidth]{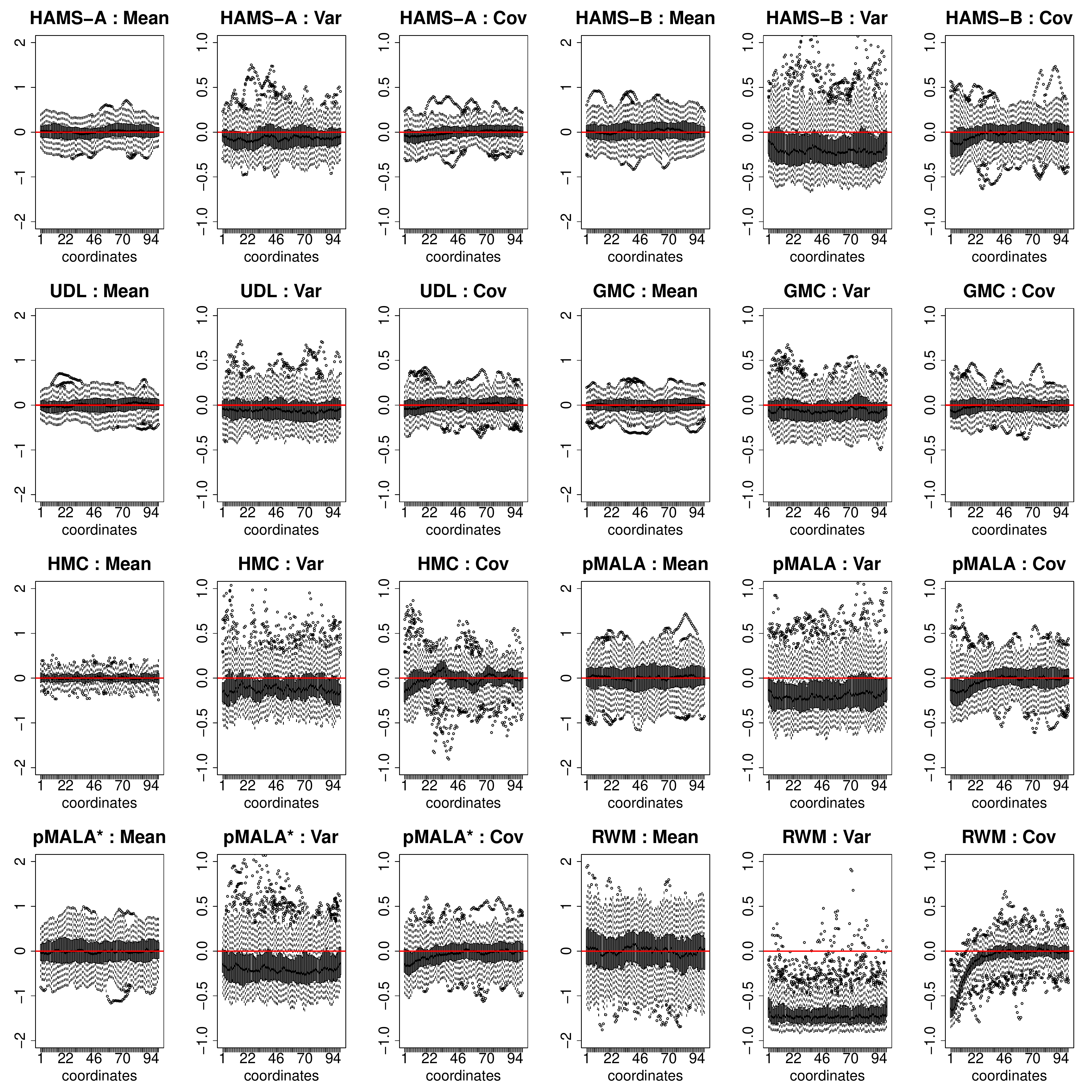}
  \end{center}
  \caption{Time-adjusted and centered boxplots of sample means, variances, and covariances of 100 coordinates
over 100 repetitions for sampling from the multivariate normal distribution. Red lines indicate zero.
\label{fig:suppnormal1}}
  \end{figure}

Figure \ref{fig:suppnormal1} shows boxplots of sample means and variances of 100 coordinates and sample covariances
of 100 coordinates with the first coordinate after centered about the true values. Hence deviations from $0$ (marked by red lines) show
divergence from the truth. From the boxplots, we see that HAMS-A, UDL and GMC are comparable to each other.
They are mostly accurate in the means and covariances while slightly underestimate the variances.
Sample means of HAMS-B are correctly centered but exhibit more variation. HAMS-B underestimates the variances more than HAMS-A, UDL, and GMC,
and also the covariances associated with the first several coordinates.
Compared to HAMS-B, pMALA shows similar underestimation of variances and covariances, but has an even wider spread in sample means. For pMALA*,
because $\epsilon = 0.18$ is  small, its performance is similar to that of the unmodified pMALA.
While HMC is good in terms of sample means, it underestimates variances and is inaccurate in
covariances with a considerable number of outliers. RWM performs poorly to capture neither
variance nor covariance.

\begin{figure}[H]
    \begin{center}
    \includegraphics[width = 0.95\textwidth]{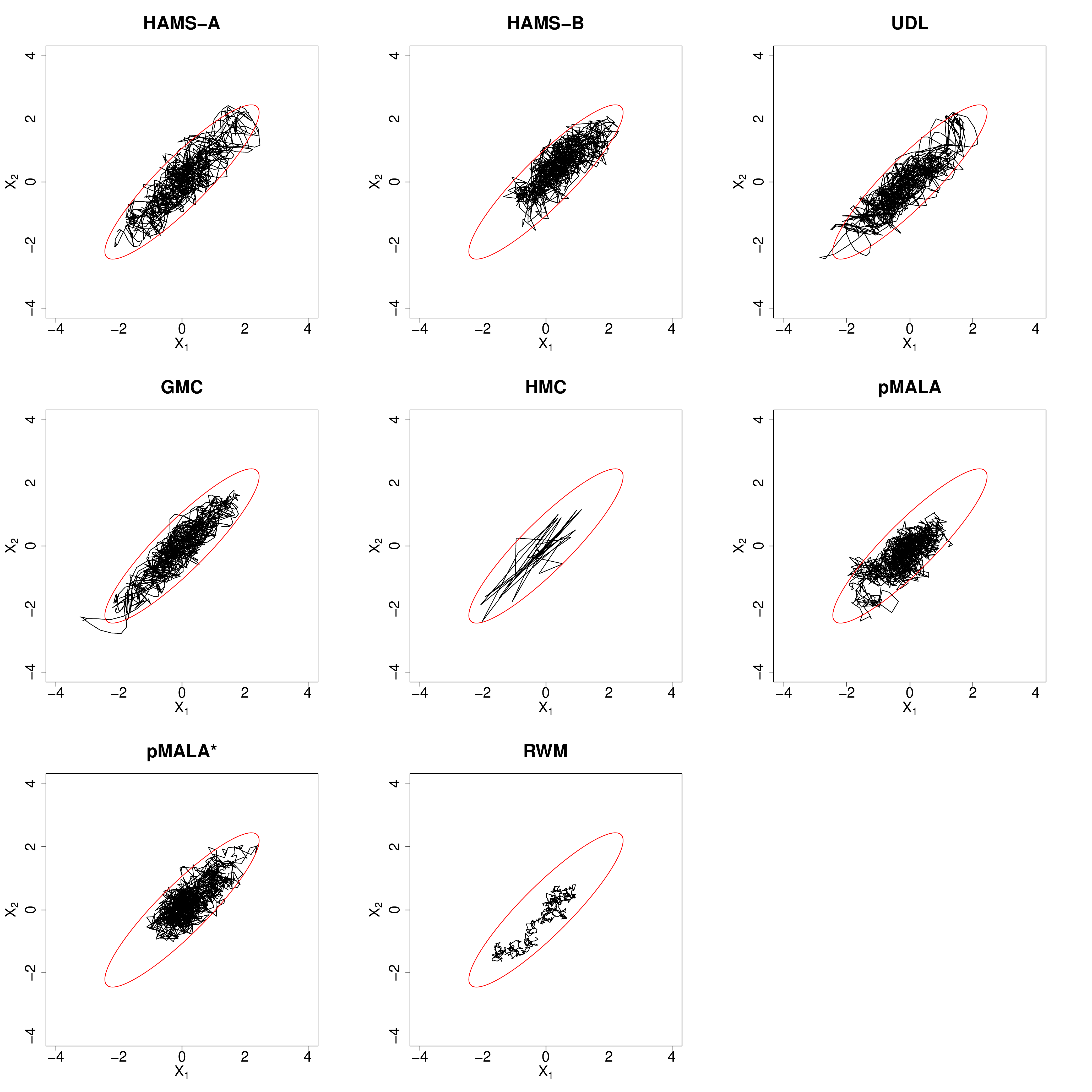}
    \end{center}
    \caption{Time-adjusted trace plots of the first two coordinates from first $2000$ iterations (first $40$ iterations for HMC)  for sampling from the multivariate normal distribution.
    Red ellipses indicate $95\%$ probability regions.
    \label{fig:suppnormal2}}
  \end{figure}

Figure \ref{fig:suppnormal2} shows trace plots of first $2000$ iterations (first $40$ iterations for HMC) from
an individual run. The first two coordinates are plotted and red ellipses mark regions containing $95\%$
probability of the marginal target density. HAMS-A best fills up the area. UDL and GMC
are also reasonable but leave a small part in the upper right blank.
HAMS-B, pMALA and pMALA* all cover smaller areas with parts of the corners missing.
The HMC trace misses the top right quadrant and its movement is only aligned to the long axis of the ellipse.
RWM performs poorly and covers the least amount of the area.

\clearpage
%%%---------------------------------------------------------------------------------------
\subsection{Stochastic volatility model}

\begin{figure}[H]
  \begin{center}
  \includegraphics[height = 0.35\textheight]{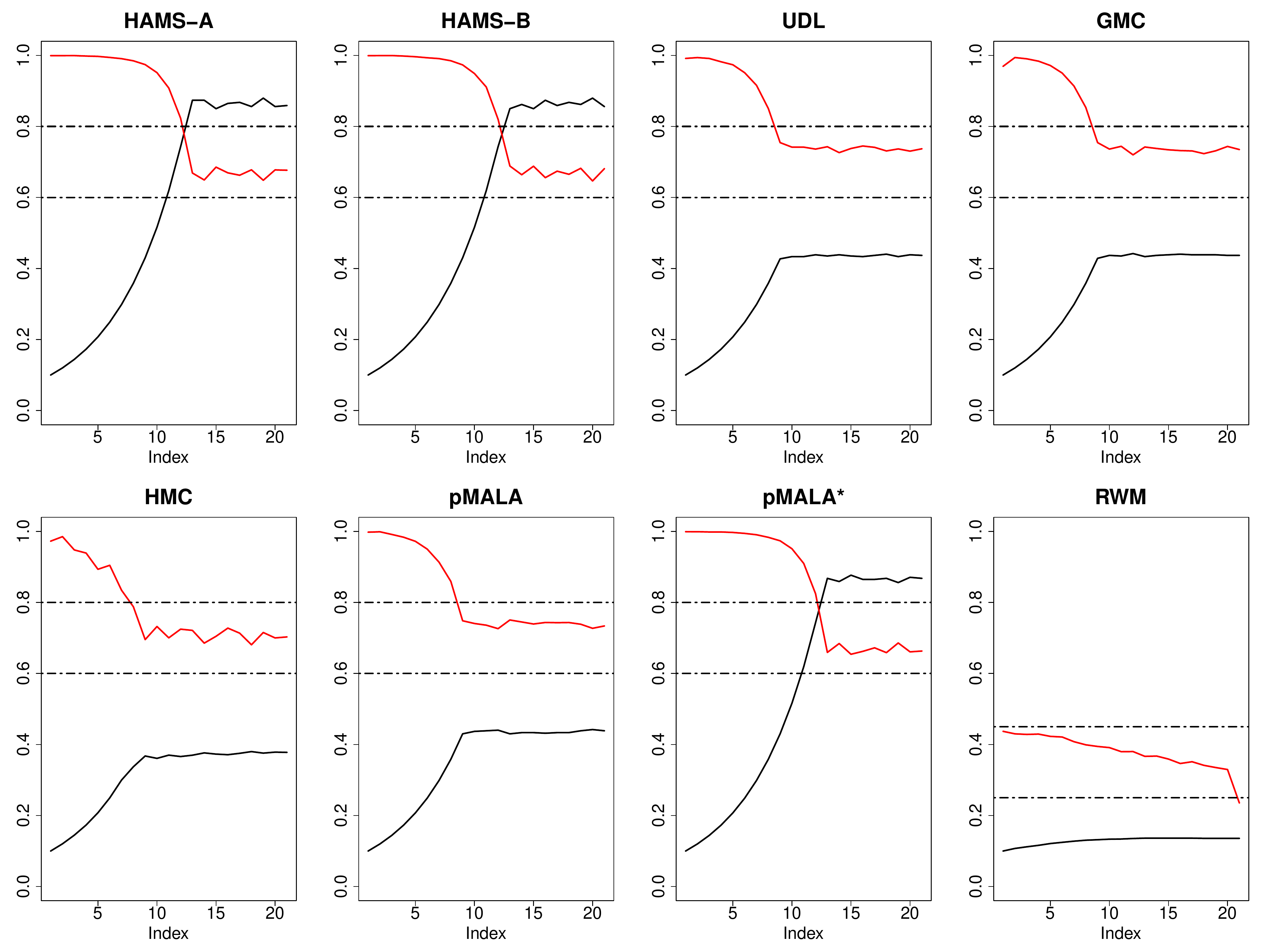}
  \end{center}
  \caption{Average step sizes (black) and acceptance rates (red) for sampling latent variables in the stochastic volatility model.
  For every 250 iterations, acceptance rates are calculated and step sizes adjusted.
  Results are averaged over $50$ repetitions.  \label{fig:suppsv1}}
\end{figure}

Consider the setting in Section \ref{subsec:sim2}. For sampling latent variables only,
Figure \ref{fig:suppsv1} shows the average acceptance rates (red curves) and step sizes $\epsilon$ (black curves) during the burn-in period,
using the tuning procedure described in Section~\ref{subsec:tuning}.  The upper and lower thresholds
of acceptance rates for such adjustments are marked by the dashed lines. From Figure~\ref{fig:suppsv1}, our tuning procedure
seems effective in obtaining desirable acceptance rates for each algorithm.
Furthermore, larger step sizes are achieved for HAMS-A, HAMS-B, and pMALA* than other methods, while similar acceptance rates are obtained.
A possible explanation is that these three methods use coefficient $\frac{\epsilon^2}{1+\sqrt{1-\epsilon^2}}$ instead of $\frac{\epsilon^2}{2}$
for gradient updates and satisfy the rejection-free property (i.e., proposals are always accepted) for a normal target density with pre-specified variance.
Hence relatively large step sizes are allowed for these methods together with reasonable acceptance rates, when the target density is not far from such a normal density.
The differences in step sizes associated with the rejection-free property can be seen to underlie advantages of HAMS-A/B as well as improvement of pMALA* over pMALA in our results.

\begin{figure}[H]
  \begin{center}
  \includegraphics[height = 0.35\textheight]{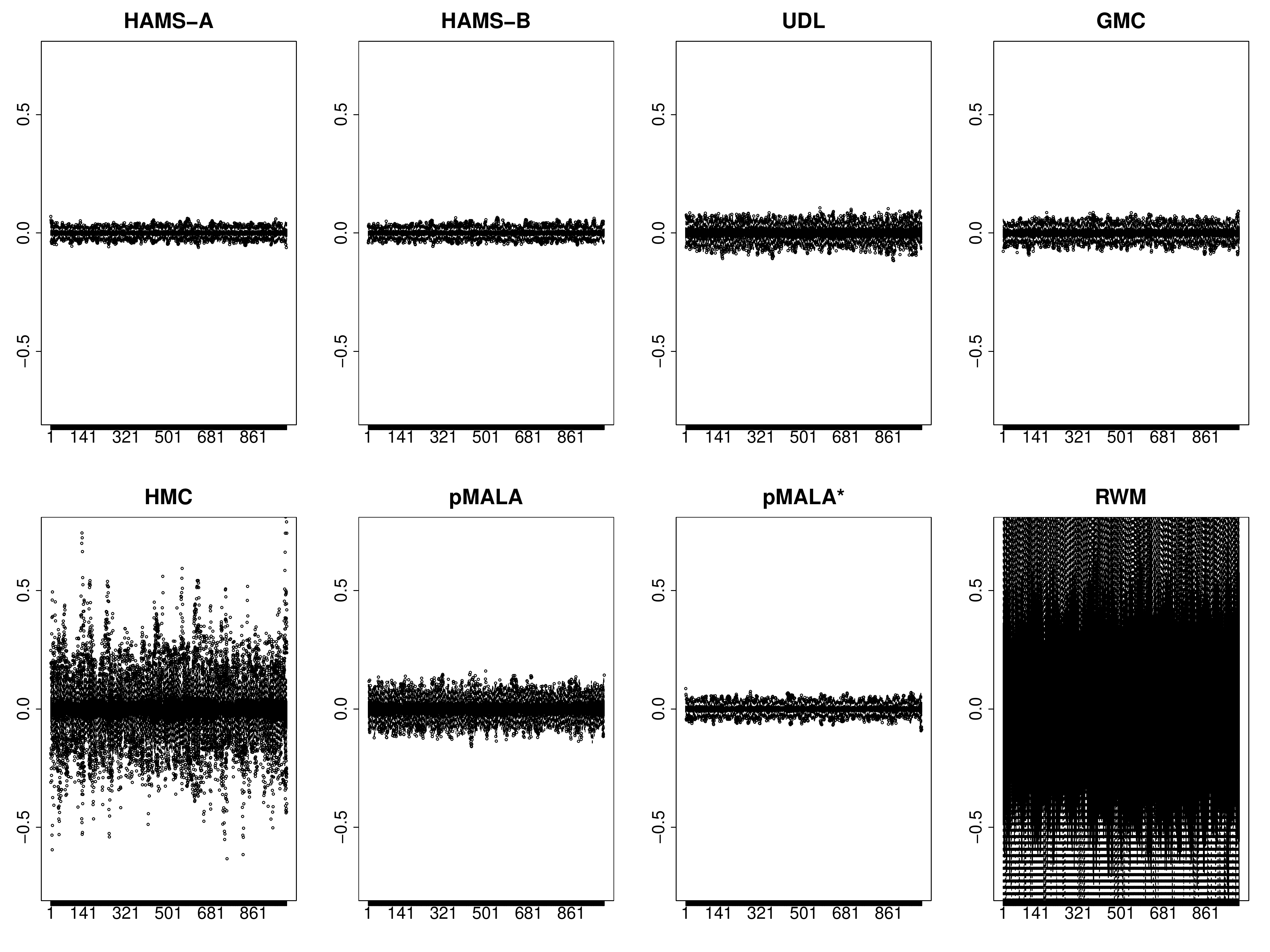}
  \end{center}
  \caption{Time-adjusted and centered boxplots of sample means of all latent variables over 50 repetitions for
sampling latent variables in the stochastic volatility model.
   \label{fig:suppsv2}}
\end{figure}

\begin{figure}[H]
  \begin{center}
  \includegraphics[height = 0.35\textheight]{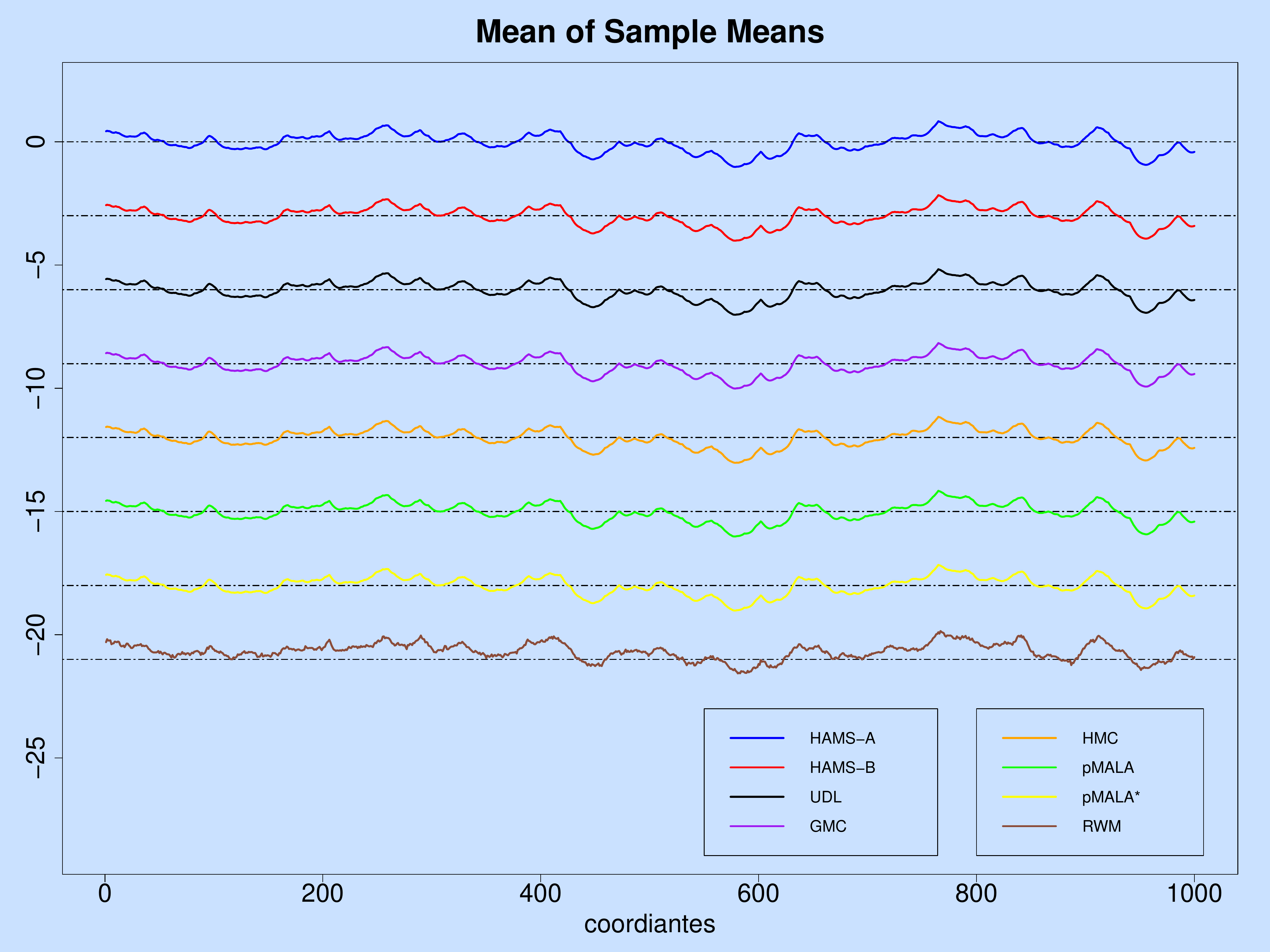}
  \end{center}
  \caption{Time-adjusted averages of sample means (shifted) of all latent variables over 50 repetitions for
sampling latent variables in the stochastic volatility model.
  \label{fig:suppsv3}}
\end{figure}

\begin{figure}[H]
  \begin{center}
  \includegraphics[height = 0.35\textheight]{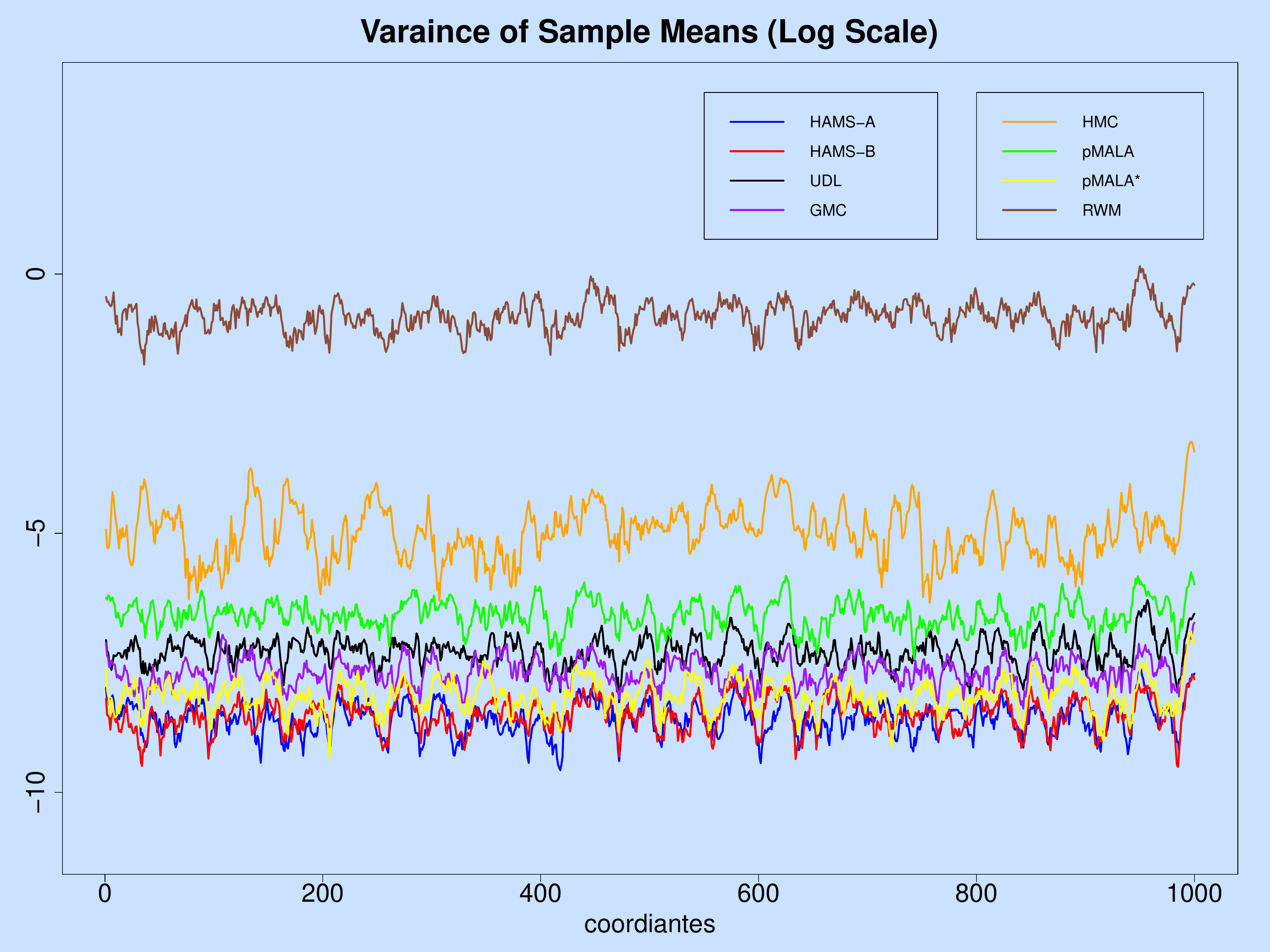}
  \end{center}
  \caption{Time-adjusted variances of sample means (log-scale) of all latent variables over 50 repetitions for
sampling latent variables in the stochastic volatility model.
  \label{fig:suppsv4}}
\end{figure}

\begin{figure}[H]
  \begin{center}
  \includegraphics[height = 0.35\textheight]{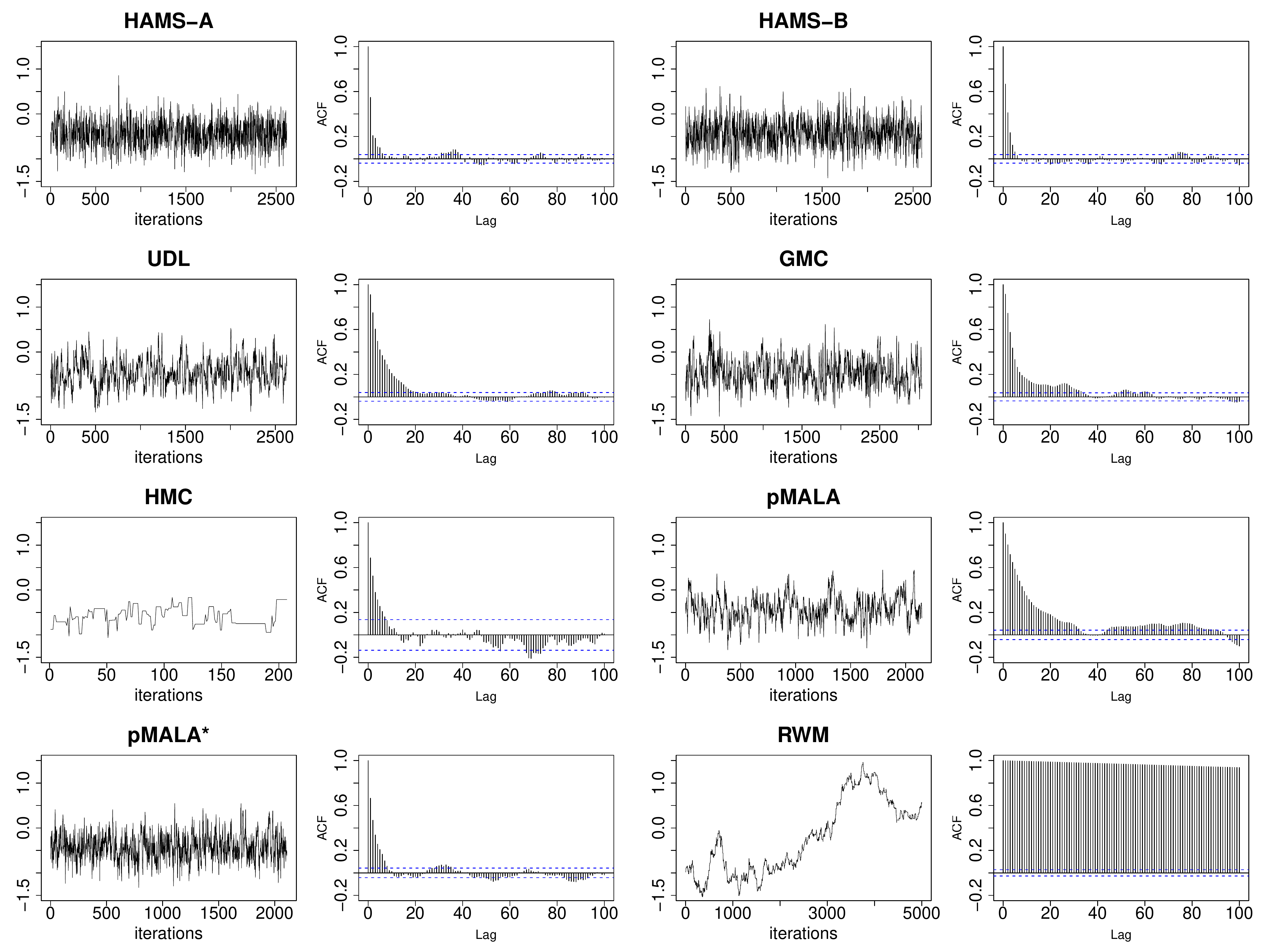}
  \end{center}
  \caption{Time-adjusted trace and ACF plots of one latent variable from an individual run
for sampling latent variables in the stochastic volatility model.
  \label{fig:suppsv5}}
\end{figure}

\begin{figure}[H]
  \begin{center}
  \includegraphics[height = 0.35\textheight]{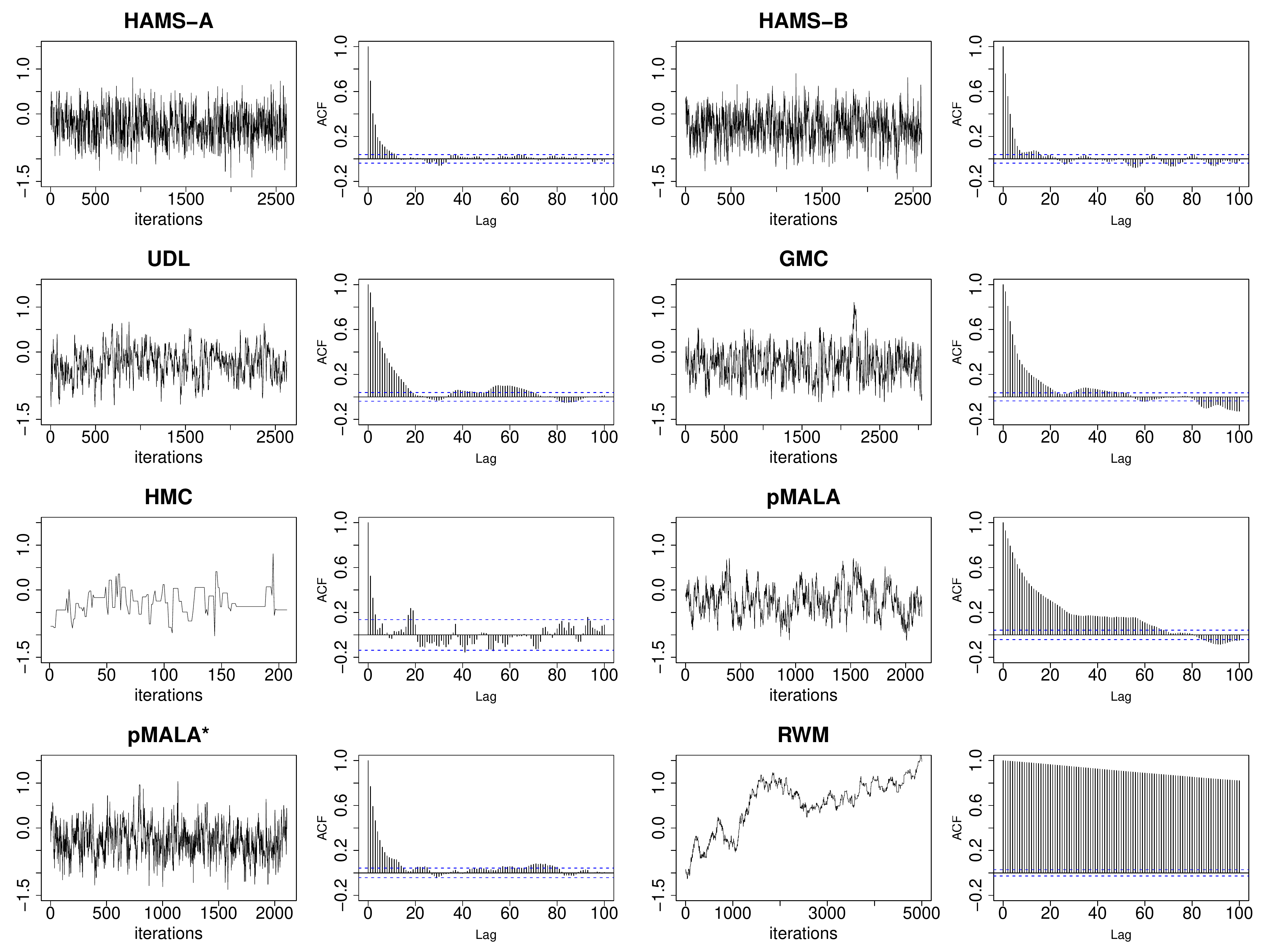}
  \end{center}
  \caption{Time-adjusted trace and ACF plots of one latent variable from an individual run
for sampling latent variables in the stochastic volatility model.
  \label{fig:suppsv6}}
\end{figure}

\begin{figure}[H]
  \begin{center}
  \includegraphics[height = 0.35\textheight]{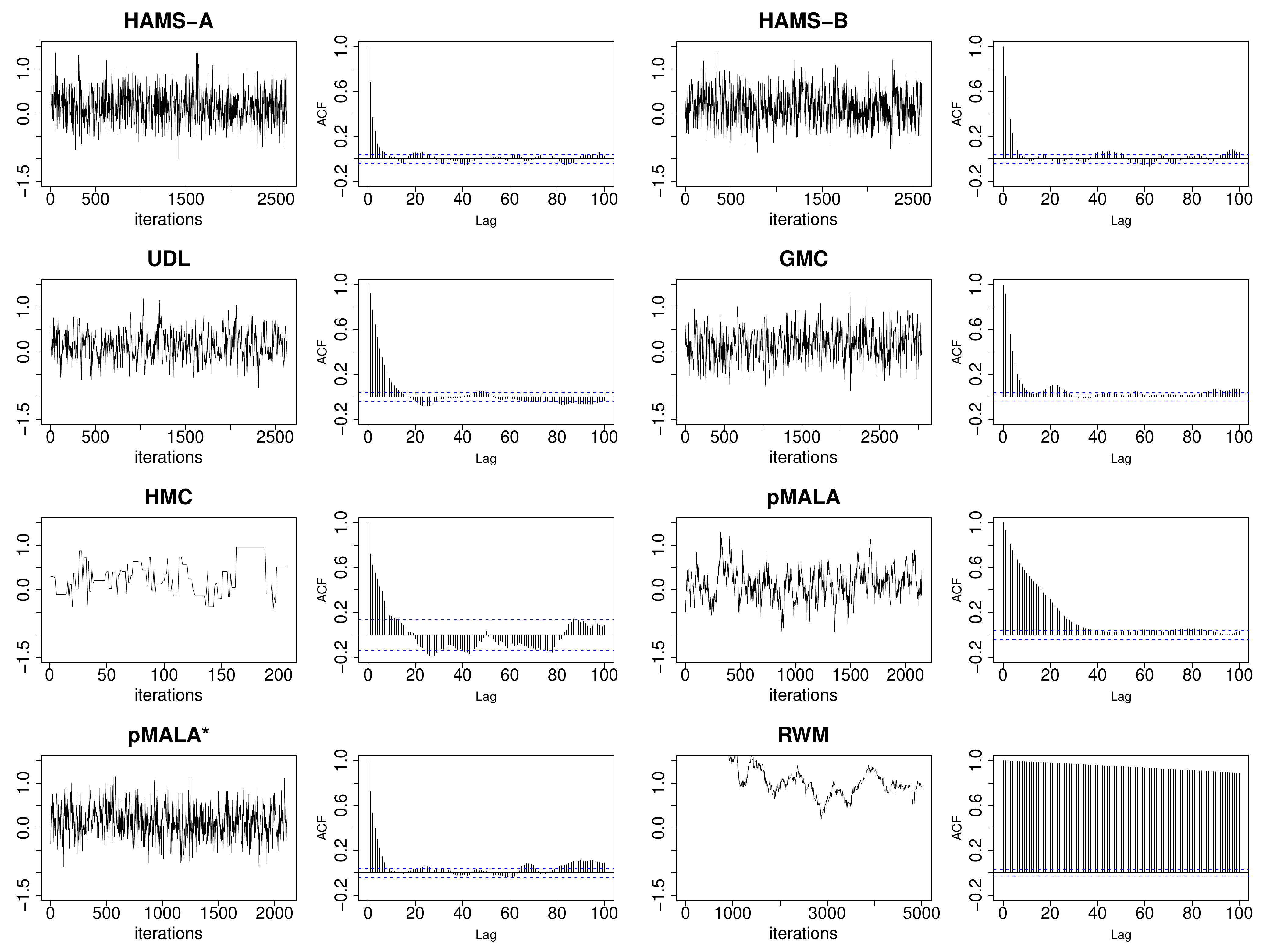}
  \end{center}
  \caption{Time-adjusted trace and ACF plots of one latent variable from an individual run
for sampling latent variables in the stochastic volatility model.
  \label{fig:suppsv7}}
\end{figure}

From Table \ref{tab:supptabone} (expanded from Table \ref{tab:tabone}), GMC has similar performance to UDL, while pMALA* improves upon pMALA considerably.
The time-adjusted centered boxplots of sample means in Figure \ref{fig:suppsv2} (expanded from Figire \ref{fig:svbox}) also confirm that pMALA* performs better than
the original pMALA. Figure \ref{fig:suppsv3} shows time-adjusted averages (over repeated runs) of sample means for all latent variables. The curves are
shifted (centered relative to the dashed lines) so that the overall shapes can be compared between methods. All methods yield
similar average sample means including RWM. Figure \ref{fig:suppsv4} shows time-adjusted variances in the log scale (over repeated runs) of sample means.
It is clear that HAMS-A and HAMS-B have the smallest variances, and hence are more consistent across repeated runs than other methods. pMALA* has
slightly larger variance, followed by GMC, UDL, pMALA, HMC and RWM.\ Additional
trace plots and ACFs are shown in Figures \ref{fig:suppsv5}  -- \ref{fig:suppsv7}, for different latent variables than in Figure~\ref{fig:svtrace}.

\begin{figure}[H]
  \centering
  \begin{subfigure}{.32\linewidth}
      \centering
      \includegraphics[width = \textwidth]{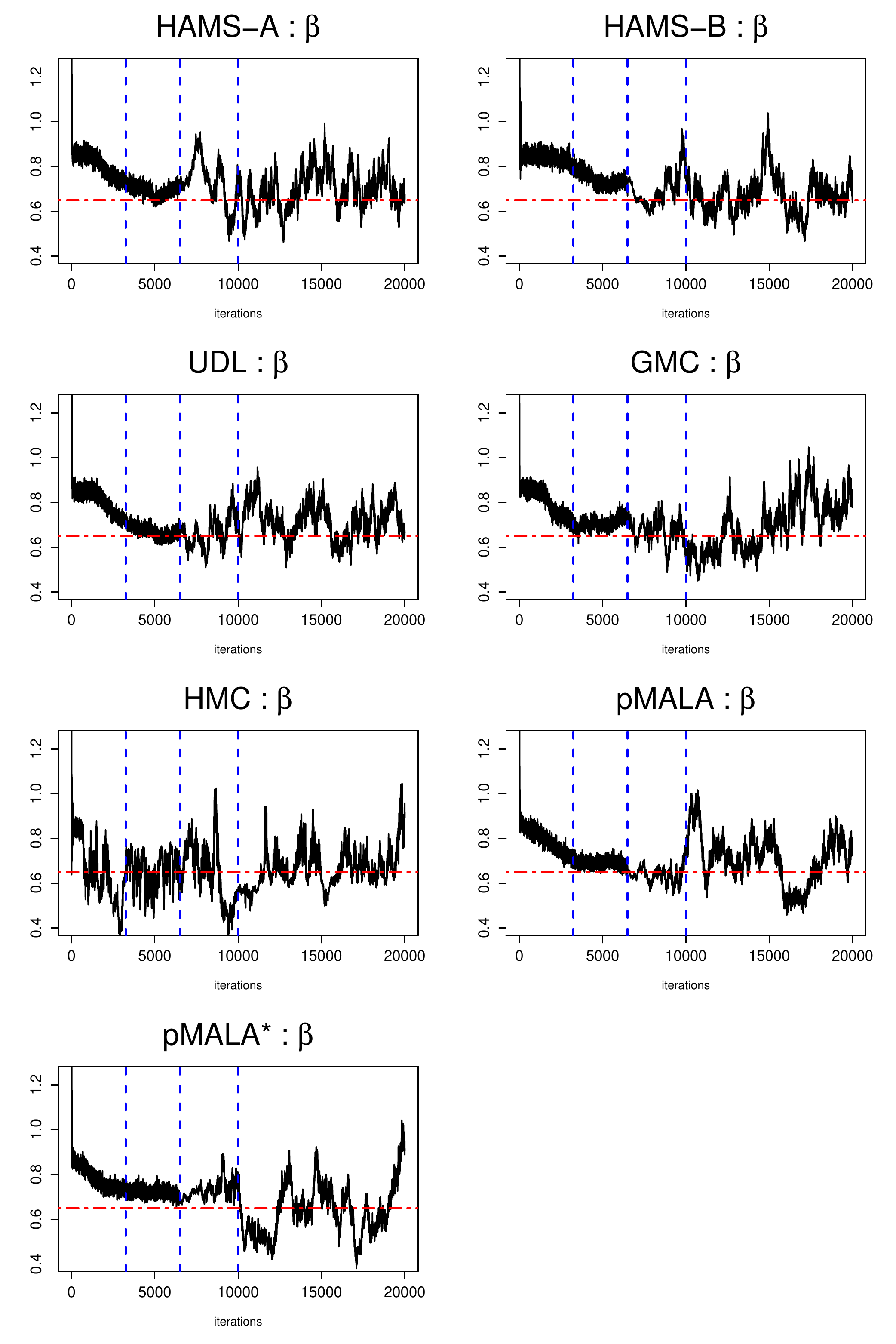}
      \caption{Trace plots of $\beta$}\label{fig:image1-SuppSV}
  \end{subfigure}
      \hfill
  \begin{subfigure}{.32\linewidth}
      \centering
      \includegraphics[width = \textwidth]{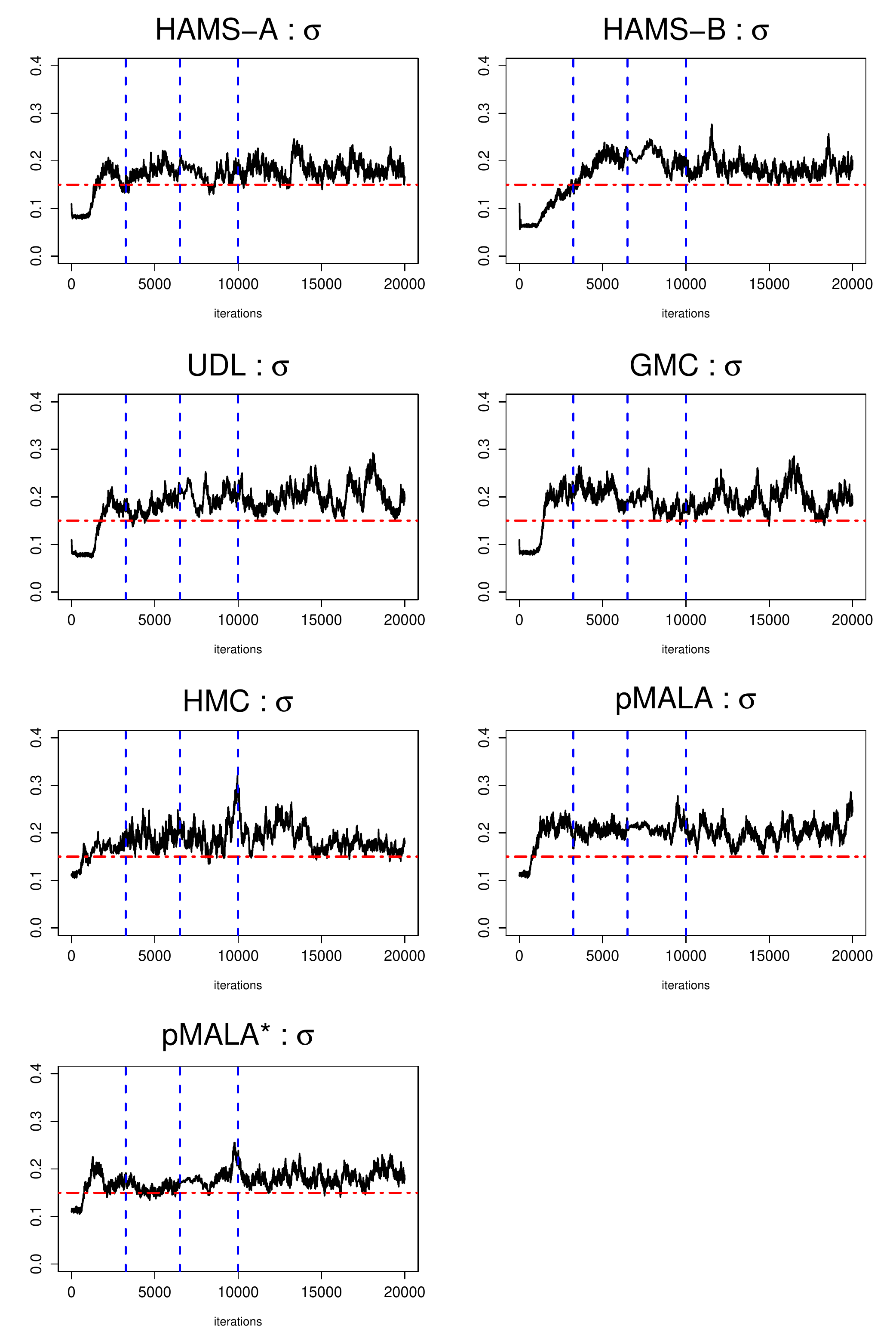}
      \caption{Trace plots of $\sigma$}\label{fig:image2-SuppSV}
  \end{subfigure}
  \hfill
  \begin{subfigure}{.32\linewidth}
      \centering
      \includegraphics[width = \textwidth]{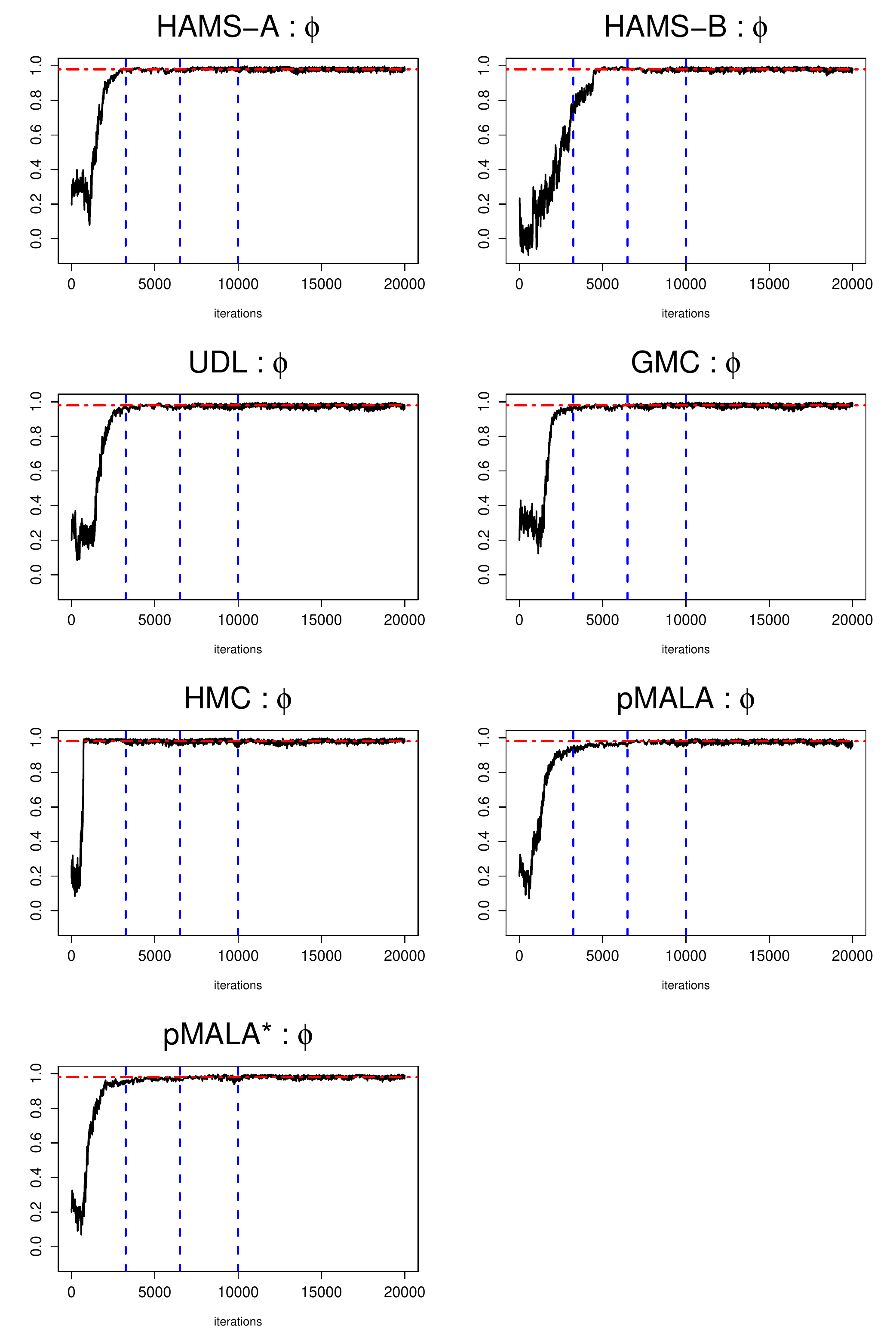}
      \caption{Trace plots of $\phi$}\label{fig:image3-SuppSV}
  \end{subfigure}

  \RawCaption{\caption{Trace plots from an individual run
for posterior sampling in the stochastic volatility model.
  Data generating parameter values are marked by red horizontal lines. There are four stages divided by blue vertical lines.
  The first two are without preconditioning, with 3250 iterations each. The last two are with preconditioning, with 3500 and 10000 iterations respectively.
  The first three stages are counted as burn-in.}
  \label{fig:suppsv9}}
  \end{figure}

Results of posterior sampling are presented in Table \ref{tab:supptabtwo} (expanded from Table~\ref{tab:tabtwo}). While pMALA* and GMC have reasonable sample means, they also have more variability than
our methods. pMALA* has large standard deviation in $\beta$ while GMC has large standard deviation in both $\beta$ and $\phi$.
Such behaviors are also observed in Figure \ref{fig:suppsv9}.

Finally, trace plots of each parameter from an individual run are
shown in Figure \ref{fig:suppsv9}. These trace plots are divided into four stages by blue vertical lines. In the
first stage, we apply no preconditioning and adjust step size $\epsilon$. In the second stage we fix $\epsilon$ and collect samples for
crude parameter estimates; we then evaluate preconditioning matrices using the sample means of parameters from the second stage
and fix them. In the third stage we apply preconditioning and adjust $\epsilon$. In the fourth stage, we fix $\epsilon$ and continue
applying preconditioning to collect working samples. \par

\begin{table}[H]
  \caption{Runtime and ESS comparison (including GMC and pMALA*) for sampling latent variables in the stochastic
  volatility model. Results are averaged over 50 repetitions. \label{tab:supptabone}}
  \centering \vspace{.4in}
  \begin{tabular}{|cccc|}
  \hline
  Method & Time (s) & \begin{tabular}[c]{@{}c@{}}ESS\\[-.1in] (min, median, max)\end{tabular} & $\frac{\mbox{minESS}}{\mbox{Time}}$  \\ \hline
  HAMS-A  & 98.7     & (2420, 3660, 6668)                                               & 24.51        \\
  HAMS-B  & 99.6     & (1915, 3404, 6229)                                            & 19.23      \\
  UDL    & 98.4     & (657, 1020, 1661)                                               & 6.68        \\
  GMC    & 85.0     & (752, 1249, 1914)                                               & 8.85        \\
  HMC    & 1250.1   & (1125, 3698, 11240)                                                & 0.90         \\
  pMALA  & 120.5    & (374, 610, 990)                                                 & 3.11       \\
  pMALA*    & 122.6     & (1740, 2879, 5429)                                               & 14.19        \\
  RWM    & 51.7     & (7, 12, 20)                                                      & 0.14         \\ \hline
\end{tabular}
\end{table}

\begin{table}[H]
  \caption{Comparison of posterior sampling (including GMC and pMALA*) in the stochastic volatility model. Standard
deviations of sample means are in parentheses. Results are averaged over 20 repetitions.  \label{tab:supptabtwo}}
  \centering
  \begin{tabular}{|ccccccc|}
  \hline
  Method & \multicolumn{1}{c|}{Time (s)} & \begin{tabular}[c]{@{}c@{}}\ \\[-.1in] $\beta$ (sd)\end{tabular} & \begin{tabular}[c]{@{}c@{}}Sample Mean\\[-.1in] $\sigma$ (sd)\end{tabular} & \multicolumn{1}{c|}{\begin{tabular}[c]{@{}c@{}}\ \\[-.1in] $\phi$ (sd)\end{tabular}} & \begin{tabular}[c]{@{}c@{}}ESS\\[-.1in] ($\beta,\sigma,\phi$)\end{tabular} & $\frac{\mbox{minESS}}{\mbox{Time}}$ \\ \hline
  HAMS-A  & 1951.3                        & 0.68 (0.034)                                                          & 0.19 (0.006)                                                      & 0.98 (0.001)                                                                              & (30, 73, 220)                                                 & 0.015      \\
  HAMS-B  & 1942.3                        & 0.68 (0.037)                                                          & 0.19 (0.007)                                                      & 0.98 (0.001)                                                                              & (25, 59, 188)                                                 & 0.013      \\
  UDL    & 1945.8                        & 0.68 (0.039)                                                          & 0.20 (0.008)                                                      & 0.98 (0.002)                                                                              & (29, 37, 87)                                                 & 0.015      \\
  GMC  & 1968.2                        & 0.67 (0.059)                                                          & 0.20 (0.007)                                                      & 0.98 (0.003)                                                                              & (35, 58, 169)                                                 & 0.018      \\
  HMC    & 20920.2                       & 0.69 (0.050)                                                          & 0.19 (0.014)                                                      & 0.98 (0.003)                                                                              & (19, 12, 78)                                                 & 0.001      \\
  pMALA  & 2013.0                        & 0.68 (0.040)                                                          & 0.20 (0.005)                                                      & 0.98 (0.001)                                                                              & (15, 30, 76)                                                  & 0.008      \\
  pMALA*  & 2015.2                        & 0.70 (0.054)                                                          & 0.19 (0.006)                                                      & 0.98 (0.001)                                                                              & (23, 53, 149)                                                 & 0.012      \\
  RWM    & 1311.1                        & 0.76 (0.050)                                                          & 0.47 (0.229)                                                      & 0.51 (0.149)                                                                              & (89, 12, 7)                                                   & 0.006      \\ \hline
  \end{tabular}
\end{table}

\begin{figure}[H]
  \begin{center}
  \includegraphics[height = 0.35\textheight]{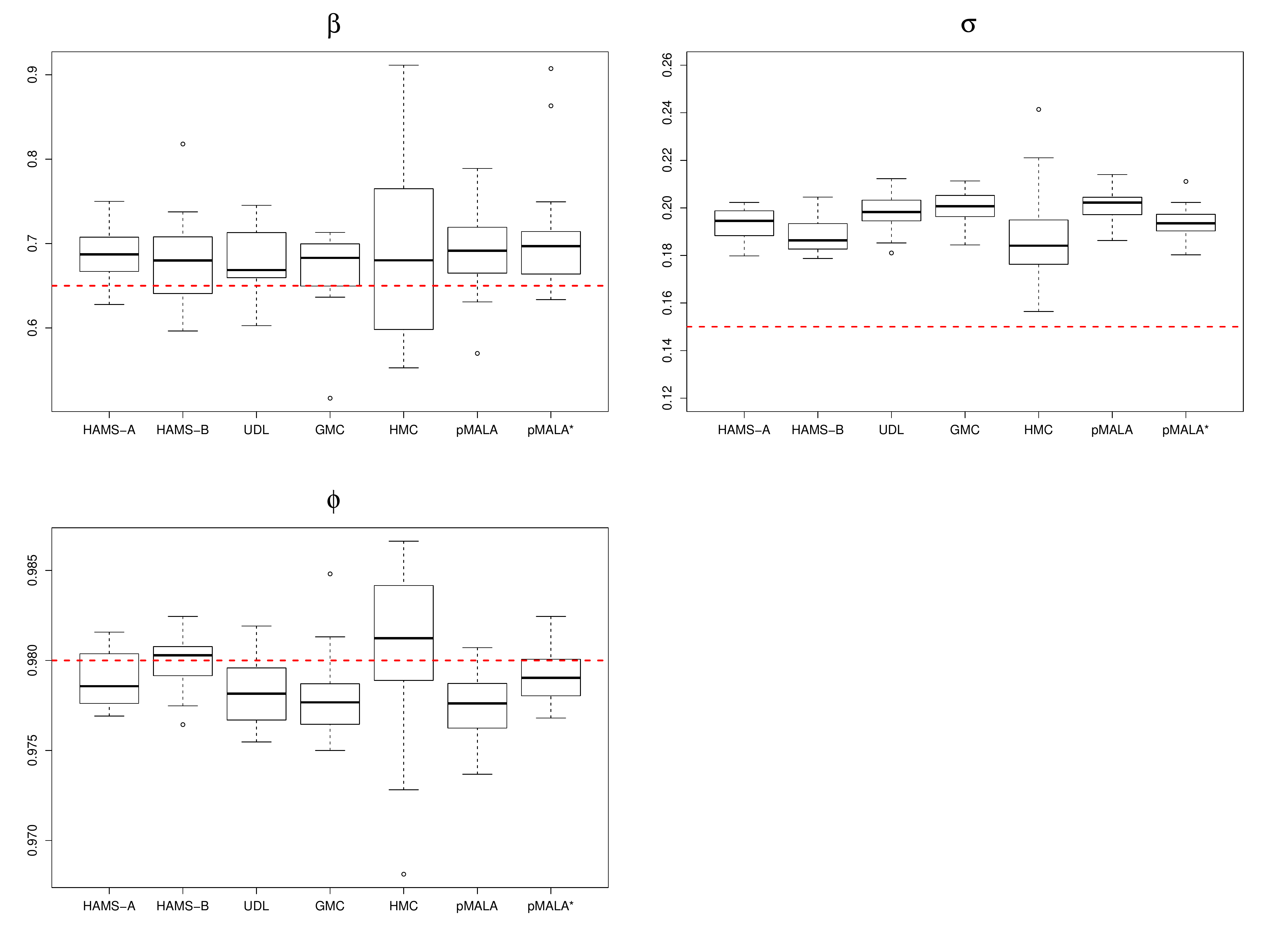}
  \end{center}
  \caption{Time-adjusted boxplots of sample means of parameters over 20 repetitions for posterior sampling in the stochastic volatility model.
  The data generating parameter values are marked by red lines.\label{fig:suppsv8}}
\end{figure}

\clearpage
%%%---------------------------------------------------------------------------------------
\subsection{Log-Gaussian Cox model}

\begin{figure}[H]
    \centering
    \begin{subfigure}{.4\linewidth}
        \centering
        \includegraphics[width = \textwidth]{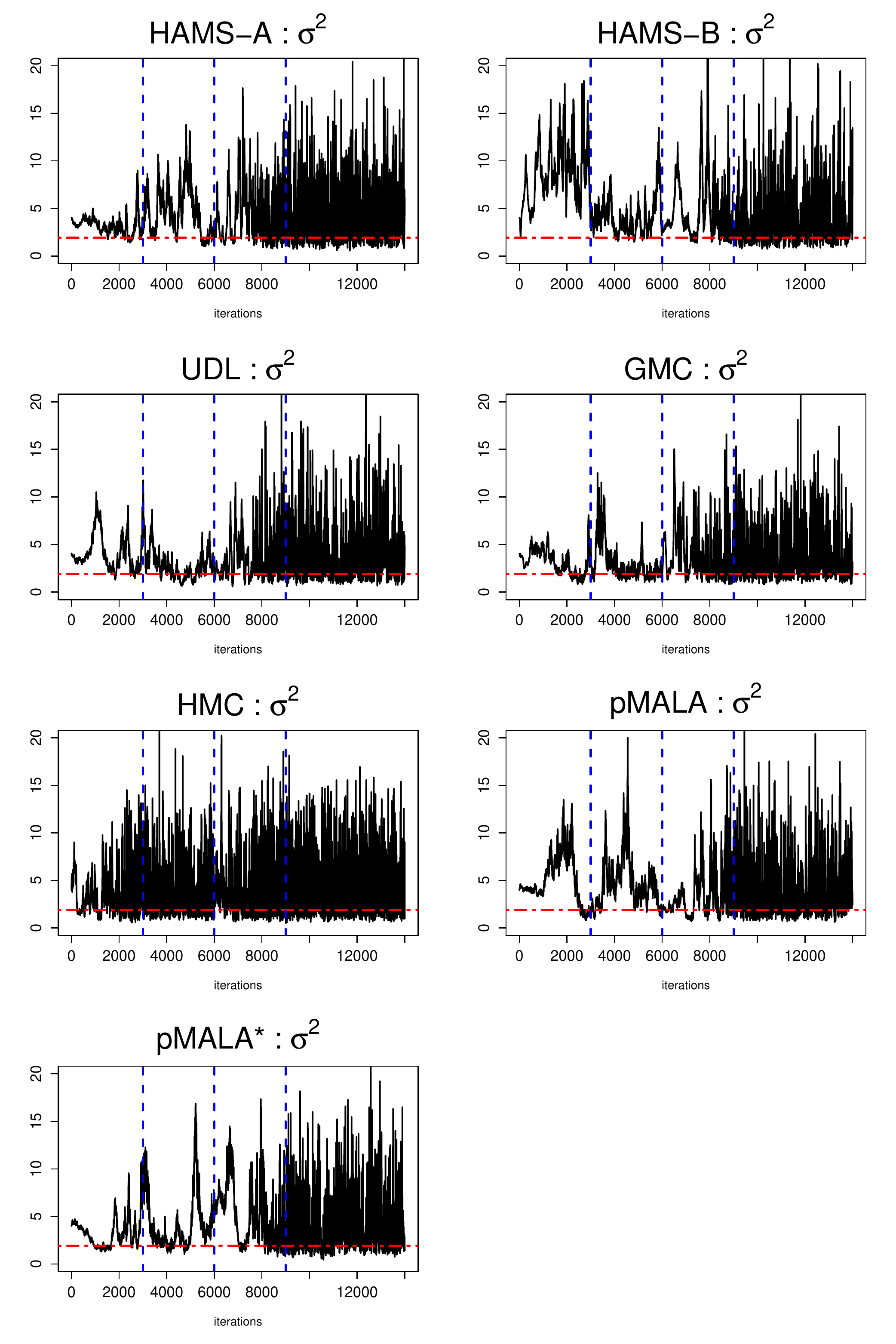}
        \caption{Trace plots of $\sigma^2$}\label{fig:image1-SuppCox}
    \end{subfigure}
        \hfill
    \begin{subfigure}{.4\linewidth}
        \centering
        \includegraphics[width = \textwidth]{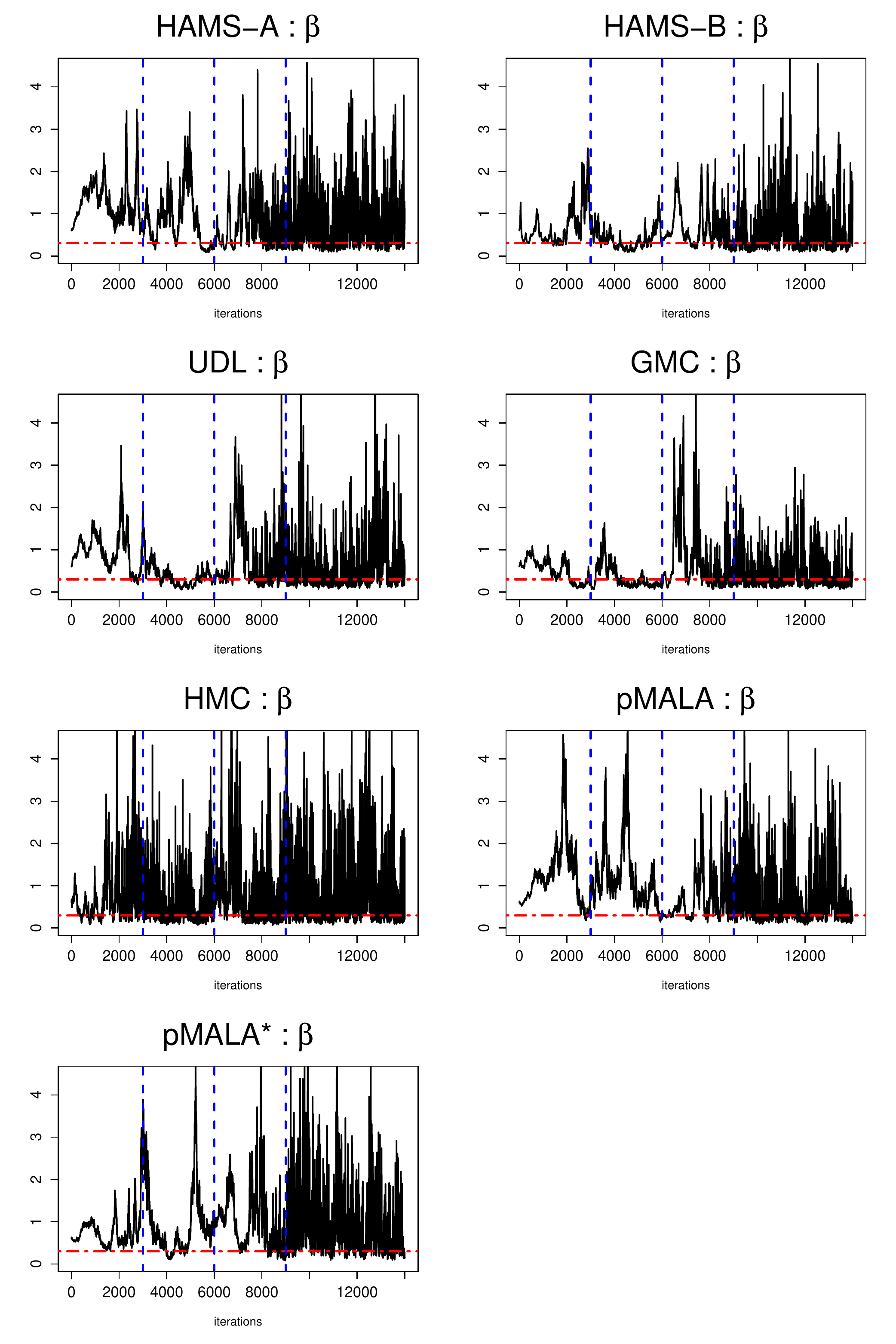}
        \caption{Trace plots of $\beta$}\label{fig:image2-SuppCox}
    \end{subfigure}

    \RawCaption{\caption{Trace plots from an individual run
for posterior sampling in the log-Gaussian Cox model ($n = 256$).
  Data generating parameter values are marked by red horizontal lines.There are four stages divided by blue vertical lines.
  The first two are without preconditioning, with 3000 iterations each. The last two are with preconditioning, with 3000 and 5000 iterations respectively.
  The first three stages are counted as burn-in.}
    \label{fig:suppcox9}}
    \end{figure}

We report additional simulation results for the log-Gaussian Cox model discussed in Section \ref{subsec:sim3}.
The overall conclusions remain similar as in the stochastic volatility model. When only sampling latent variables,
pMALA* improves upon the original pMALA, and GMC shows comparable performance to UDL. While all methods have similar average sample means,
HAMS-A and HAMS-B have the smallest variance. For posterior sampling results,
pMALA* inflates the standard deviations of sample means, but brings the estimates more aligned with HAMS.
Compared to the stochastic volatility model, the effect
of preconditioning can be seen more clearly from the trace plots in Figure \ref{fig:suppcox9}.

\begin{table}[H]
  \caption{Runtime and ESS comparison (including GMC and pMALA*) for sampling latent variables in the log-Gaussian Cox model ($n = 1024$).
   Results are averaged over 50 repetitions.  \label{tab:supptabthree}}
  \centering \vspace{.4in}
  \begin{tabular}{|cccc|}
  \hline
  Method & Time (s) & \begin{tabular}[c]{@{}c@{}}ESS\\[-.1in] (min, median, max)\end{tabular} & $\frac{\mbox{minESS}}{\mbox{Time}}$ \\ \hline
  HAMS-A  & 81.0    & (803, 1655, 5461)                                                 & 9.91         \\
  HAMS-B  & 78.8    & (619, 1376, 4831)                                              & 7.86        \\
  UDL    & 78.8    & (322, 622, 1761)                                                & 4.08         \\
  GMC    & 81.9    & (359, 742, 2081)                                                & 4.38         \\
  HMC    & 1285.9   & (935, 1621, 4523)                                              & 0.73         \\
  pMALA  & 116.4    & (184, 340, 1002)                                                 & 1.58         \\
  pMALA*    & 115.9    & (600, 1197, 4275)                                                & 5.17         \\
  RWM    & 51.1     & (8, 13, 22)                                                      & 0.16        \\ \hline
  \end{tabular}
\end{table}

\begin{figure}[H]
  \begin{center}
  \includegraphics[height = 0.35\textheight]{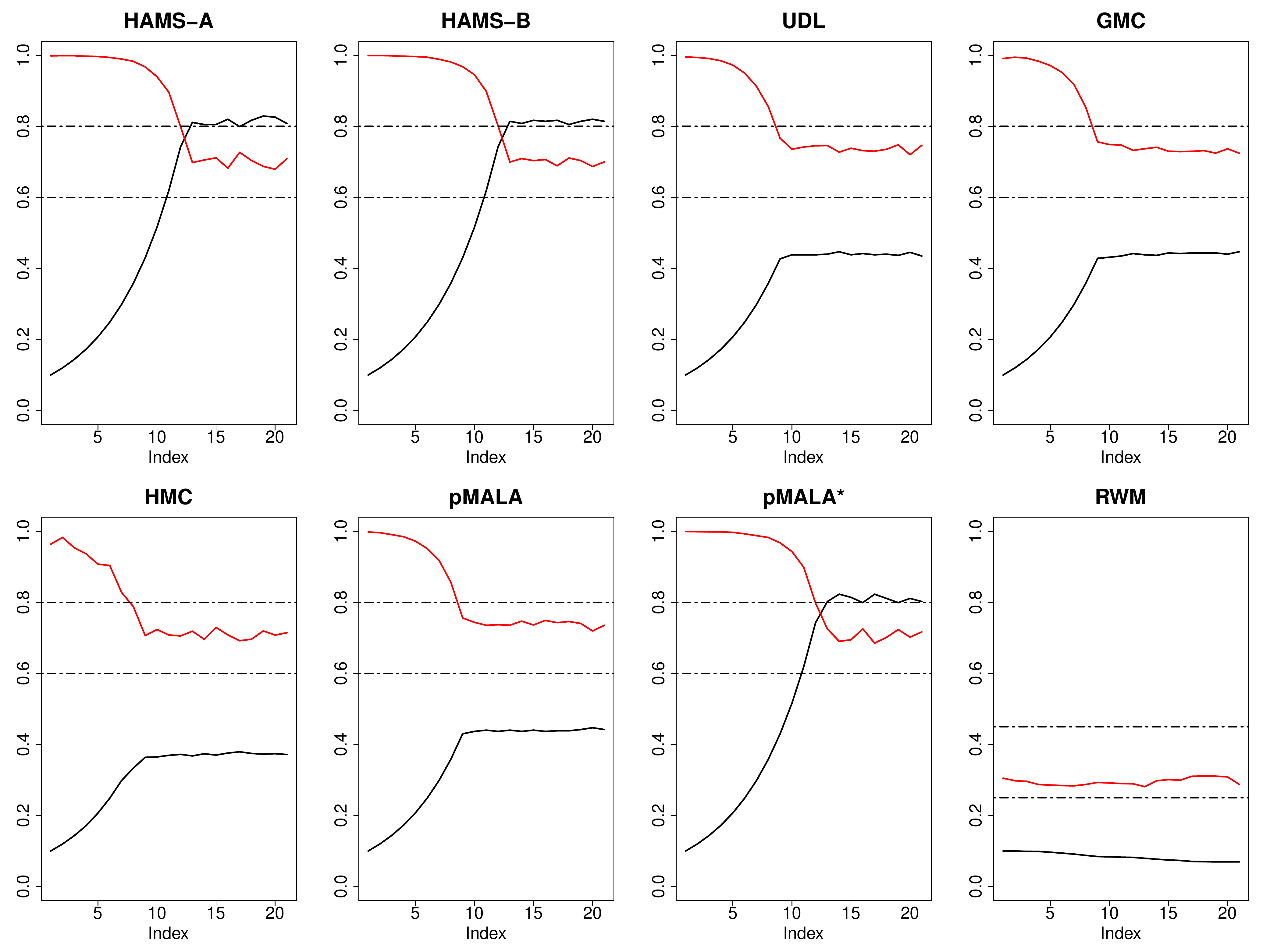}
  \end{center}
  \caption{Average step sizes (black) and acceptance rates (red) for sampling latent variables in the log-Gaussian Cox model ($n = 1024$).
  For every 250 iterations, acceptance rates are calculated and step sizes adjusted.
  Results are averaged over $50$ repetitions.  \label{fig:suppcox1}}
\end{figure}

\vspace{-.2in}

\begin{figure}[H]
  \begin{center}
  \includegraphics[height = 0.35\textheight]{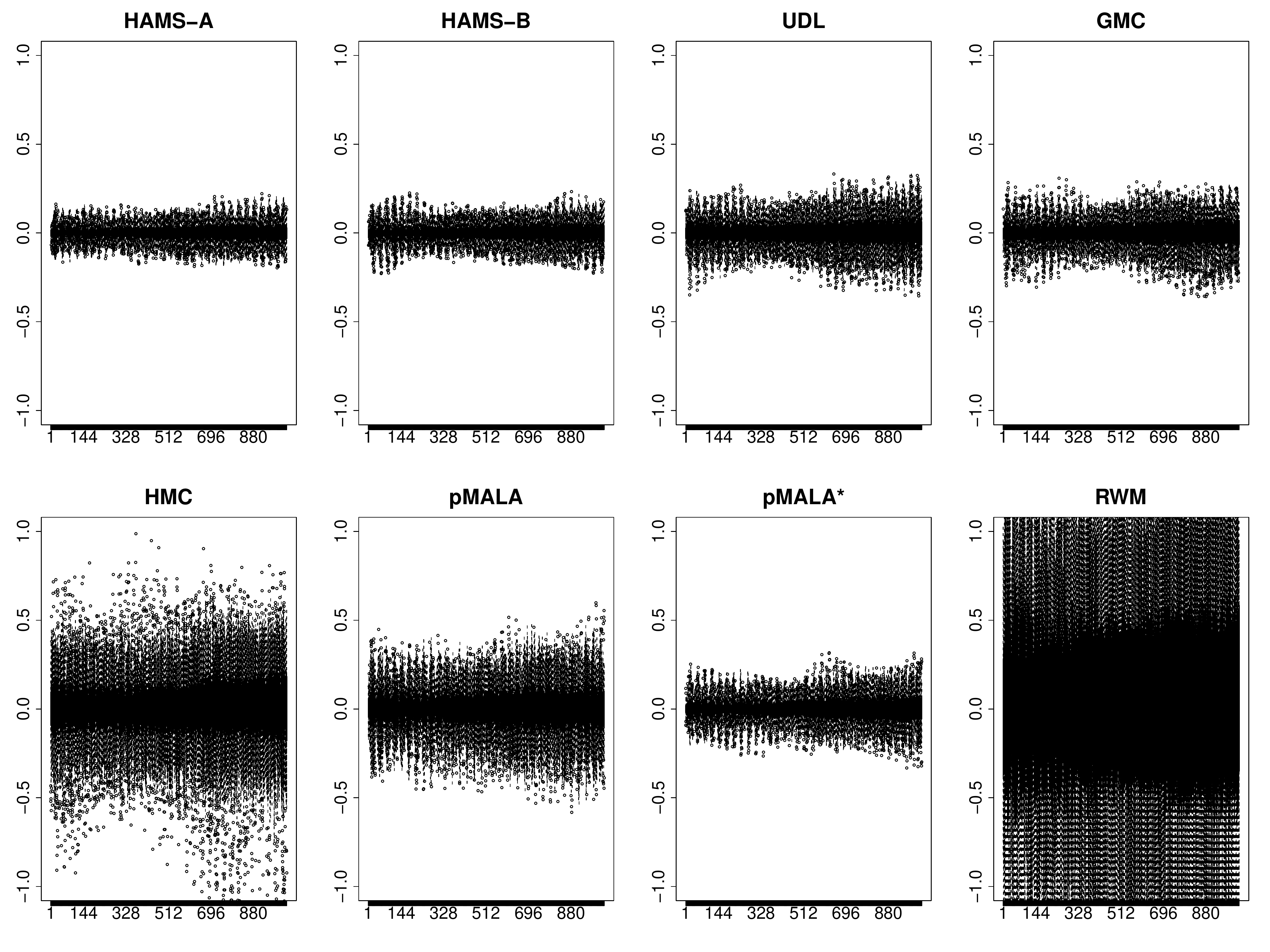}
  \end{center}
  \caption{Time-adjusted and centered boxplots of sample means of all latent variables over 50 repetitions for
sampling latent variables in the log-Gaussian Cox model ($n = 1024$).  \label{fig:suppcox2}}
\end{figure}

\begin{figure}[H]
  \begin{center}
  \includegraphics[height = 0.35\textheight]{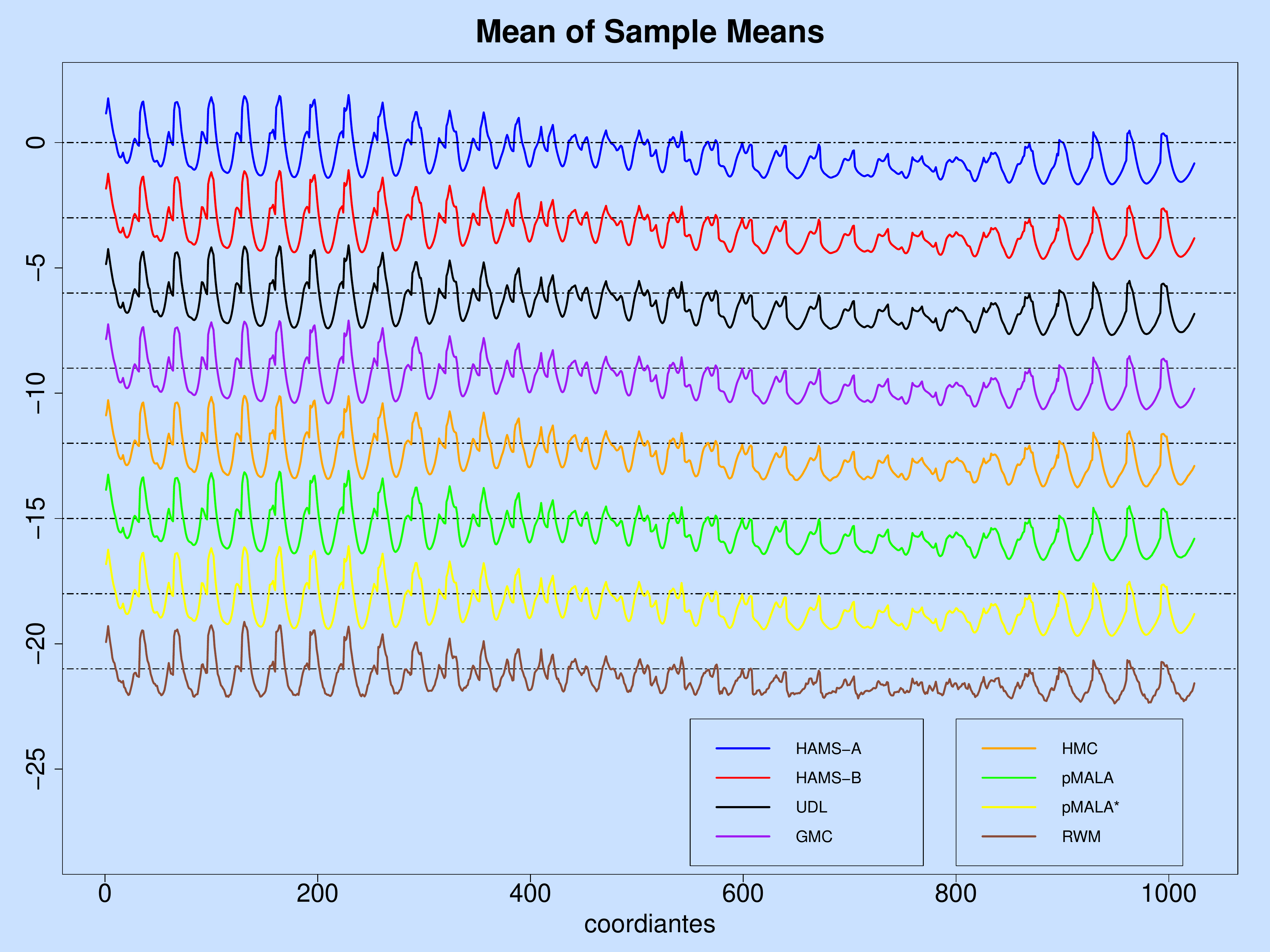}
  \end{center}
  \caption{Time-adjusted averages of sample means (shifted) of all latent variables over 50 repetitions for
sampling latent variables in the log-Gaussian Cox model ($n = 1024$).   \label{fig:suppcox3}}
\end{figure}

\begin{figure}[H]
  \begin{center}
  \includegraphics[height = 0.35\textheight]{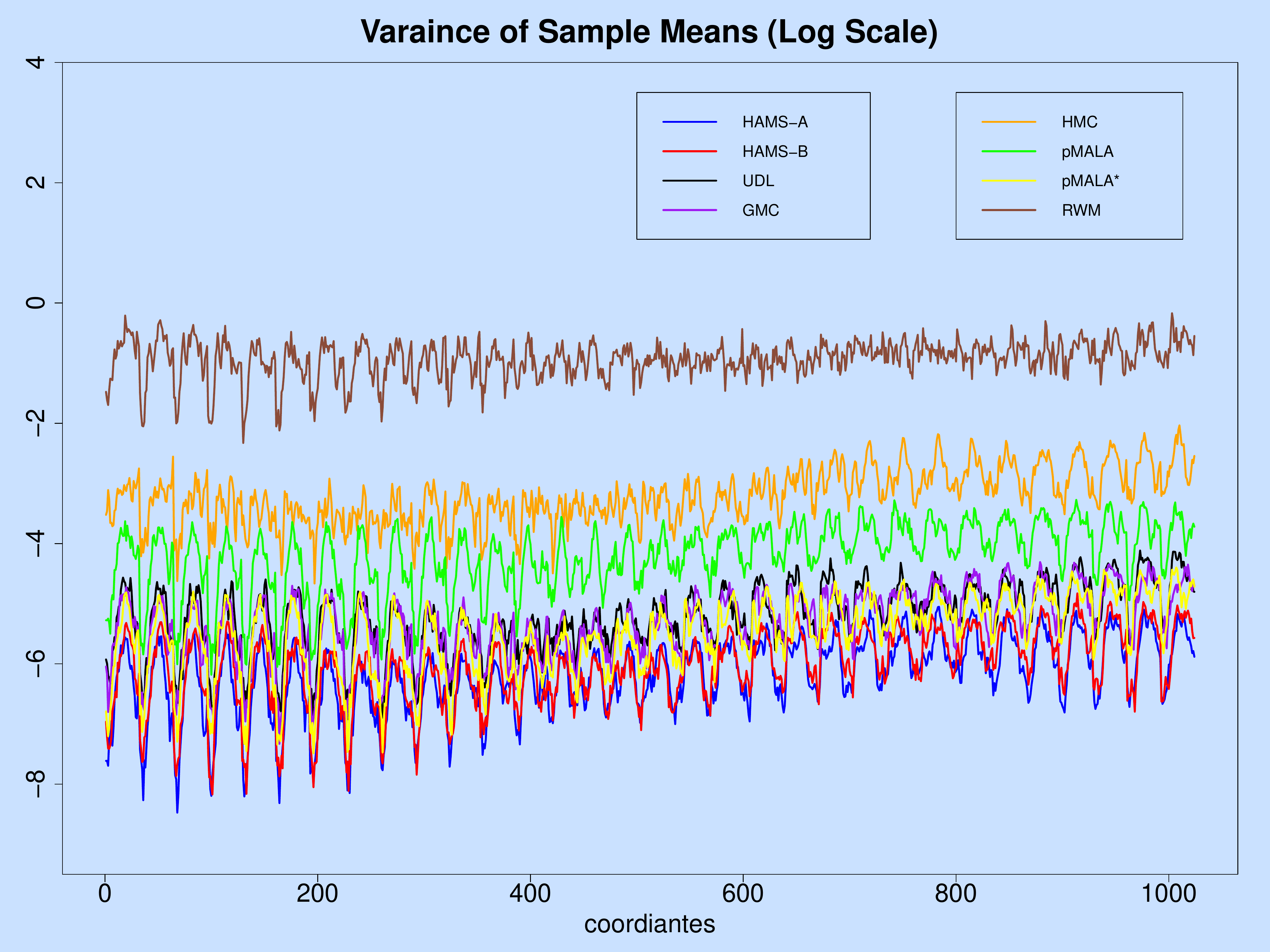}
  \end{center}
  \caption{Time-adjusted variances of sample means (log-scale) of all latent variables over 50 repetitions for
sampling latent variables in the log-Gaussian Cox model ($n = 1024$).
  \label{fig:suppcox4}}
\end{figure}

\begin{figure}[H]
  \begin{center}
  \includegraphics[height = 0.35\textheight]{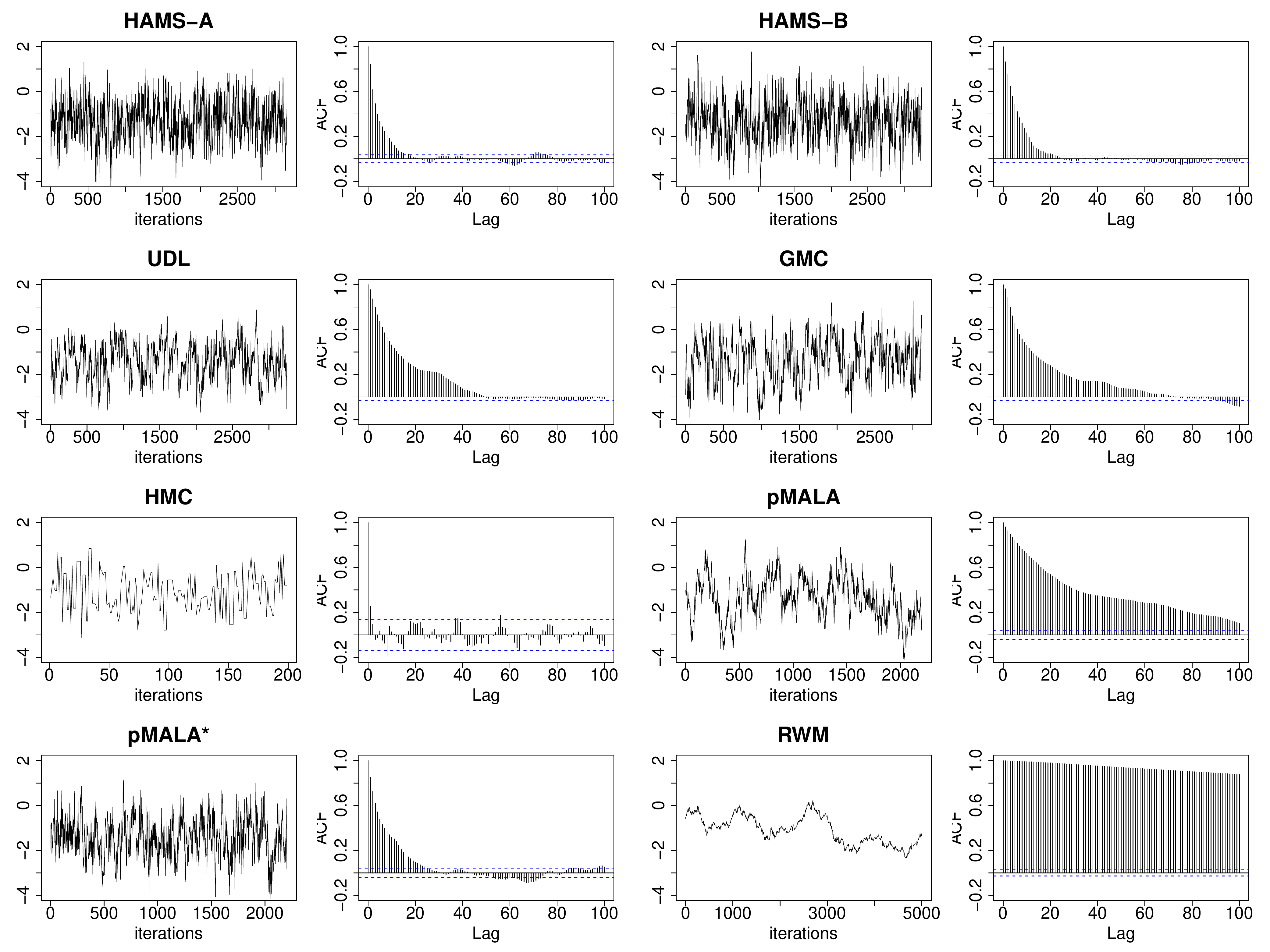}
  \end{center}
  \caption{Time-adjusted trace and ACF plots of one latent variable from an individual run
for sampling latent variables in the log-Gaussian Cox model ($n = 1024$).
  \label{fig:suppcox5}}
\end{figure}

\begin{figure}[H]
  \begin{center}
  \includegraphics[height = 0.35\textheight]{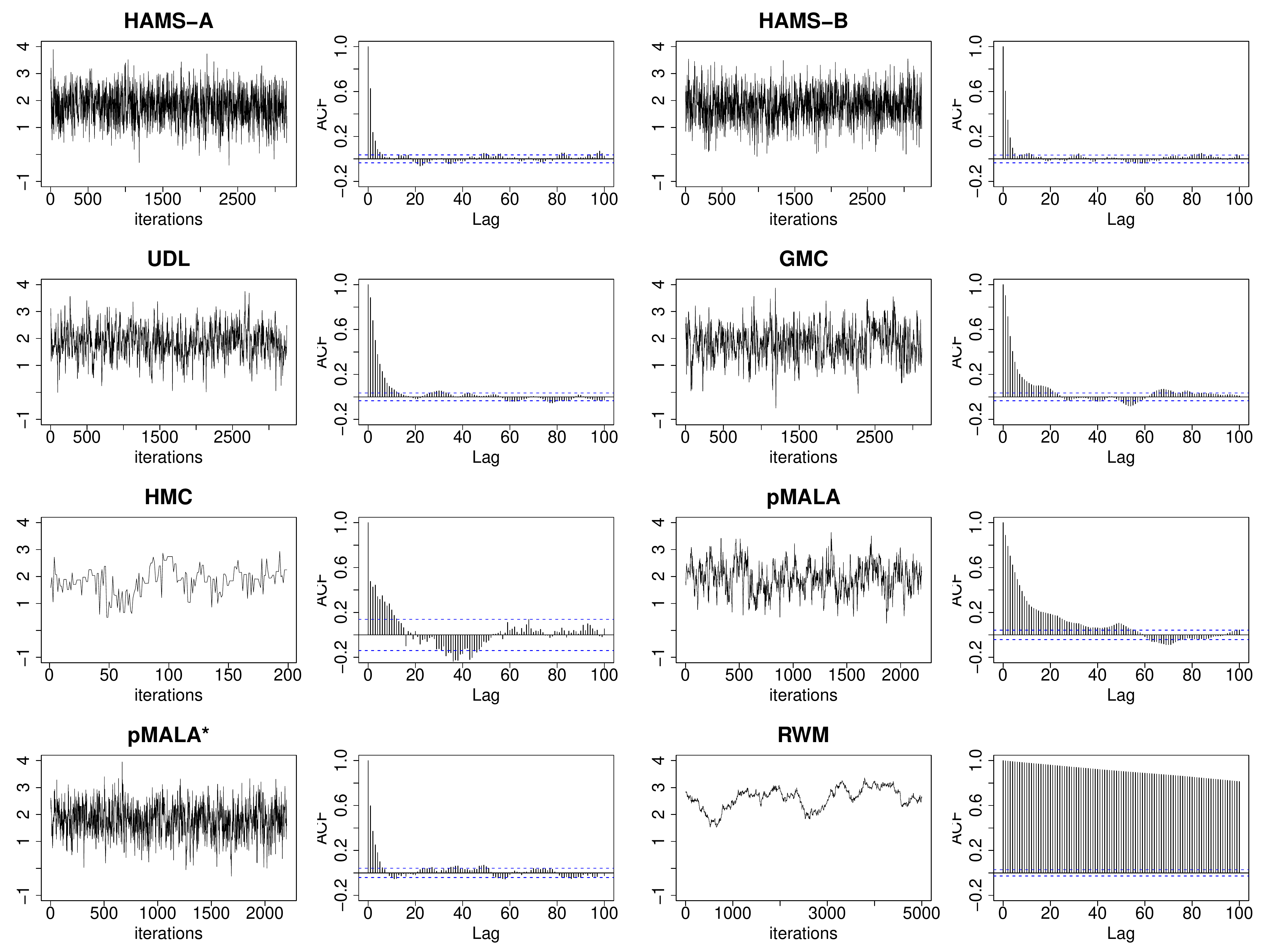}
  \end{center}
  \caption{Time-adjusted trace and ACF plots of one latent variable from an individual run
for sampling latent variables in the log-Gaussian Cox model ($n = 1024$).
  \label{fig:suppcox6}}
\end{figure}

\begin{figure}[H]
  \begin{center}
  \includegraphics[height = 0.35\textheight]{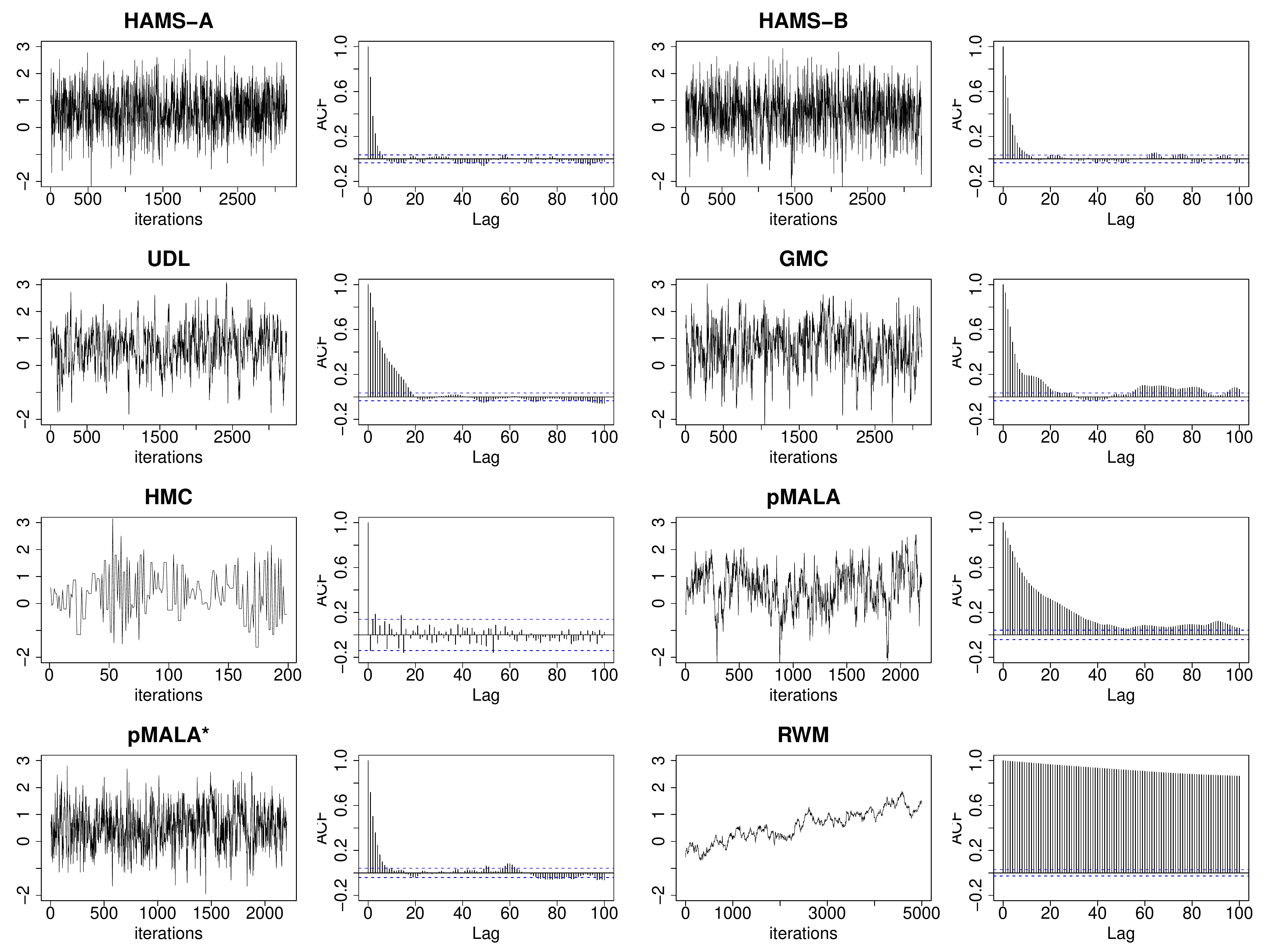}
  \end{center}
  \caption{Time-adjusted trace and ACF plots of one latent variable from an individual run
for sampling latent variables in the log-Gaussian Cox model ($n = 1024$).
  \label{fig:suppcox7}}
\end{figure}

\begin{table}[H]
  \caption{Comparison of posterior sampling (including GMC and pMALA*) in the log-Gaussian Cox model ($n = 256$). Standard
deviations of sample means are in parentheses. Results are averaged over 20 repetitions.   \label{tab:supptabfour}}
  \centering
  \begin{tabular}{|cccccc|}
  \hline
  Method & \multicolumn{1}{c|}{Time (s)} & \begin{tabular}[c]{@{}c@{}}Sample Mean\\[-.1in] $\sigma^2$ (sd)\end{tabular} & \multicolumn{1}{c|}{\begin{tabular}[c]{@{}c@{}}\ \\[-.1in] $\beta$ (sd)\end{tabular}} & \begin{tabular}[c]{@{}c@{}}ESS\\ ($\sigma^2$,$\beta$)\end{tabular} &$\frac{\mbox{minESS}}{\mbox{Time}}$ \\ \hline
  HAMS-A  & 2766.8                        & 3.90 (0.155)                                                          & 0.68 (0.073)                                                                                 & (978, 207)                                                     & 0.075         \\
  HAMS-B  & 2762.8                        & 3.93 (0.190)                                                          & 0.69 (0.106)                                                                                 & (838, 263)                                                     & 0.095         \\
  UDL    & 2759.1                        & 3.79 (0.171)                                                          & 0.59 (0.105)                                                                                 & (755, 246)                                                     & 0.089         \\
  GMC    & 2763.8                        & 3.81 (0.156)                                                          & 0.61 (0.132)                                                                                 & (884, 142)                                                     & 0.051         \\
  HMC    & 25386.0                       & 3.88 (0.084)                                                          & 0.75 (0.113)                                                                                 & (2253, 139)                                                    & 0.005         \\
  pMALA  & 2755.3                        & 3.76 (0.189)                                                          & 0.57 (0.101)                                                                                 & (528, 178)                                                     & 0.065         \\
  pMALA*    & 2758.4                        & 3.89 (0.223)                                                          & 0.69 (0.138)                                                                                 & (623, 182)                                                     & 0.066         \\
  RWM    & 1752.2                        & 3.70 (0.662)                                                          & 1.26 (1.434)                                                                                 & (226, 87)                                                      & 0.050         \\ \hline
  \end{tabular}
\end{table}

  \begin{figure}[H]
    \begin{center}
    \includegraphics[height = 0.35\textheight]{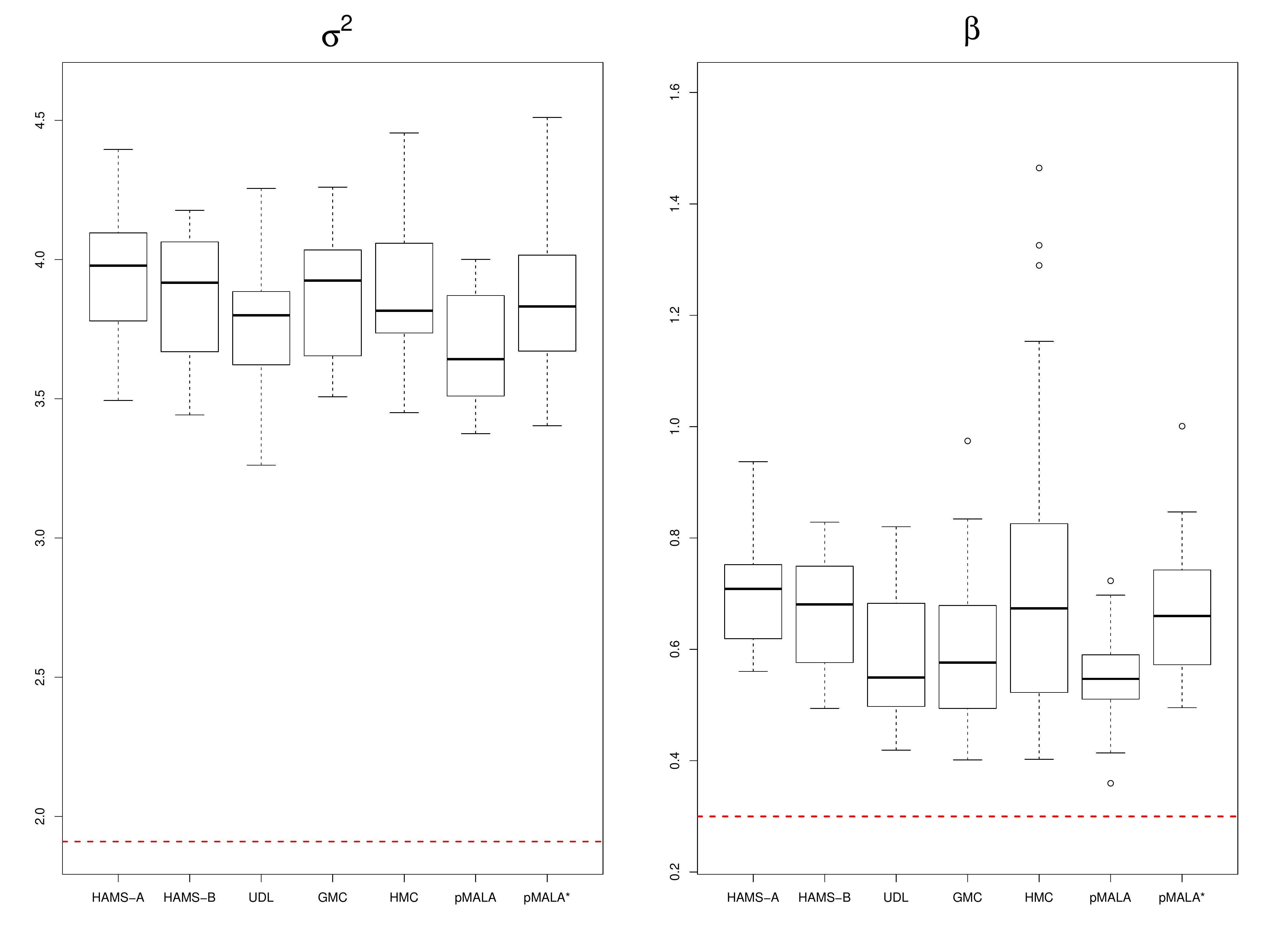}
    \end{center}
    \caption{Time-adjusted boxplots of sample means of parameters over 20 repetitions for posterior sampling in the log-Gaussian Cox model ($n = 256$).
  The data generating parameter values are marked by red lines. Due to skewness of posterior densities (Figure~\ref{fig:coxdensity}),
  the posterior modes are reasonably close to the true parameter values, while the posterior means are not, especially for $\sigma^2$. \label{fig:suppcox8}}
  \end{figure}

\end{document}